\documentclass[prd, preprint,amsmath,amssymb,showpacs,showkeys,superscriptaddress,nofootinbib, 10pt, twocolumn]{revtex4-2}
\usepackage{times}
\usepackage{amsmath}
\usepackage{graphicx}
\usepackage{epsfig}
\usepackage{dcolumn}
\usepackage{bm}
\usepackage{color}
\usepackage{longtable}
\usepackage{array}
\usepackage{multirow}
\usepackage{setspace}
\usepackage[mathlines]{lineno}
\usepackage{epsfig,graphics,subfigure,psfrag}
\usepackage[figuresright]{rotating}
\usepackage{makecell}
\usepackage{geometry}
\usepackage{epstopdf}
\usepackage{verbatim}
\usepackage{tabu}
\usepackage{booktabs}
\usepackage[figuresright]{rotating}

\usepackage[pdfstartview = FitH, colorlinks=red, urlcolor = blue,citecolor = blue, linkcolor = green]{hyperref}
\usepackage[hyphenbreaks]{breakurl}

\newcommand{\EEPPJ}{e^{+}e^{-} \rightarrow \pi^{+}\pi^{-}J/\psi}
\newcommand{\PPJ}{\pi^{+}\pi^{-}J/\psi}
\newcommand{\EE}{e^{+}e^{-}}
\newcommand{\MM}{\mu^{+}\mu^{-}}
\newcommand{\LL}{\ell^{+}\ell^{-}}
\newcommand{\PP}{\pi^{+}\pi^{-}}
\newcommand{\PZC}{\pi^{\pm}Z_{c}(3900)^{\mp}}
\newcommand{\ZCP}{Z_{c}(3900)^{\pm}}
\newcommand{\ZC}{Z_{c}(3900)}
\newcommand{\PPS}{(\pi^{+}\pi^{-})_{\rm{S\mbox{-}wave}}}
\newcommand{\BESIIIorcid}[1]{\href{https://orcid.org/#1}{\hspace*{0.1em}\raisebox{-0.45ex}{\includegraphics[width=1em]{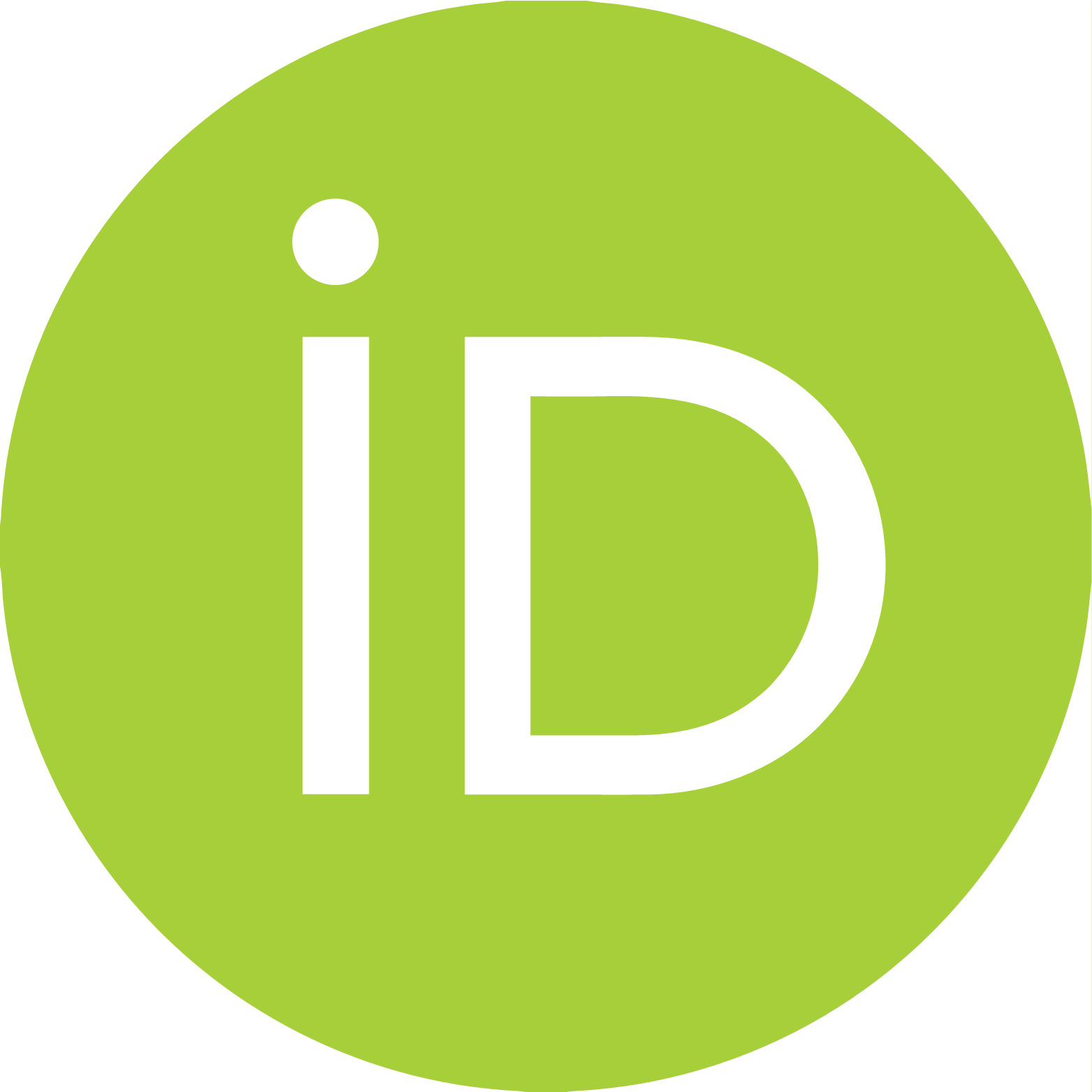}}}}

\emergencystretch=\maxdimen
\begin{document}
\title{\boldmath{Partial wave analysis of $e^{+}e^{-} \rightarrow \pi^{+}\pi^{-}J/\psi$ and cross section measurement of $e^{+}e^{-} \rightarrow \pi^{\pm}Z_{c}(3900)^{\mp}$ from 4.1271 to 4.3583 GeV}}
\author{
\author{Author list}
M.~Ablikim$^{1}$\BESIIIorcid{0000-0002-3935-619X},
M.~N.~Achasov$^{4,b}$\BESIIIorcid{0000-0002-9400-8622},
P.~Adlarson$^{76}$\BESIIIorcid{0000-0001-6280-3851},
O.~Afedulidis$^{3}$\BESIIIorcid{0009-0006-2899-9946},
X.~C.~Ai$^{81}$\BESIIIorcid{0000-0003-3856-2415},
R.~Aliberti$^{36}$\BESIIIorcid{0000-0003-3500-4012},
A.~Amoroso$^{75A,75C}$\BESIIIorcid{0000-0002-3095-8610},
Q.~An$^{72,59,\dagger}$,
Y.~Bai$^{58}$\BESIIIorcid{0000-0001-6593-5665},
O.~Bakina$^{37}$\BESIIIorcid{0009-0005-0719-7461}
% ...
\\
\vspace{0.2cm}
(BESIII Collaboration)\\
\vspace{0.2cm} {\it
$^{1}$ Institute of High Energy Physics, Beijing 100049, People's Republic of China\\
$^{2}$ Beihang University, Beijing 100191, People's Republic of China\\
$^{3}$ Bochum Ruhr-University, D-44780 Bochum, Germany\\
$^{4}$ Budker Institute of Nuclear Physics SB RAS (BINP), Novosibirsk 630090, Russia\\
$^{5}$ Carnegie Mellon University, Pittsburgh, Pennsylvania 15213, USA\\
$^{6}$ Central China Normal University, Wuhan 430079, People's Republic of China\\
$^{7}$ Central South University, Changsha 410083, People's Republic of China\\
$^{8}$ China Center of Advanced Science and Technology, Beijing 100190, People's Republic of China\\
$^{9}$ China University of Geosciences, Wuhan 430074, People's Republic of China\\
$^{10}$ Chung-Ang University, (A)Seoul, 06974, Republic of Korea; (B)84 Heukseok-ro, Dongjak-gu, Seoul, 06974, Republic of Korea\\
$^{11}$ COMSATS University Islamabad, Lahore Campus, Defence Road, Off Raiwind Road, 54000 Lahore, Pakistan\\
$^{12}$ Fudan University, Shanghai 200433, People's Republic of China\\
$^{13}$ GSI Helmholtzcentre for Heavy Ion Research GmbH, D-64291 Darmstadt, Germany\\
$^{14}$ Guangxi Normal University, Guilin 541004, People's Republic of China\\
$^{15}$ Guangxi University, Nanning 530004, People's Republic of China\\
$^{16}$ Hangzhou Normal University, Hangzhou 310036, People's Republic of China\\
$^{17}$ Hebei University, Baoding 071002, People's Republic of China\\
$^{18}$ Helmholtz Institute Mainz, Staudinger Weg 18, D-55099 Mainz, Germany\\
$^{19}$ Henan Normal University, Xinxiang 453007, People's Republic of China\\
$^{20}$ Henan University, Kaifeng 475004, People's Republic of China\\
$^{21}$ Henan University of Science and Technology, Luoyang 471003, People's Republic of China\\
$^{22}$ Henan University of Technology, Zhengzhou 450001, People's Republic of China\\
$^{23}$ Huangshan College, Huangshan 245000, People's Republic of China\\
$^{24}$ Hubei University of Automotive Technology, Shiyan 442002, People's Republic of China\\
$^{25}$ Hunan Normal University, Changsha 410081, People's Republic of China\\
$^{26}$ Hunan University, Changsha 410082, People's Republic of China\\
$^{27}$ Indian Institute of Technology Madras, Chennai 600036, India\\
$^{28}$ Indiana University, Bloomington, Indiana 47405, USA\\
$^{29}$ INFN Laboratori Nazionali di Frascati, (A)INFN Laboratori Nazionali di Frascati, I-00044, Frascati, Italy; (B)INFN Sezione di Perugia, I-06100, Perugia, Italy; (C)University of Perugia, I-06100, Perugia, Italy\\
$^{30}$ INFN Sezione di Ferrara, (A)INFN Sezione di Ferrara, I-44122, Ferrara, Italy; (B)University of Ferrara, I-44122, Ferrara, Italy\\
$^{31}$ Inner Mongolia University, Hohhot 010021, People's Republic of China\\
$^{32}$ Institute of Modern Physics, Lanzhou 730000, People's Republic of China\\
$^{33}$ Institute of Physics and Technology, Peace Avenue 54B, Ulaanbaatar 13330, Mongolia\\
$^{34}$ Instituto de Alta Investigaci\'on, Universidad de Tarapac\'a, Casilla 7D, Arica 1000000, Chile\\
$^{35}$ Jilin University, Changchun 130012, People's Republic of China\\
$^{36}$ Johannes Gutenberg University of Mainz, Johann-Joachim-Becher-Weg 45, D-55099 Mainz, Germany\\
$^{37}$ Joint Institute for Nuclear Research, 141980 Dubna, Moscow region, Russia\\
$^{38}$ Justus-Liebig-Universitaet Giessen, II. Physikalisches Institut, Heinrich-Buff-Ring 16, D-35392 Giessen, Germany\\
$^{39}$ Lanzhou University, Lanzhou 730000, People's Republic of China\\
$^{40}$ Liaoning Normal University, Dalian 116029, People's Republic of China\\
$^{41}$ Liaoning University, Shenyang 110036, People's Republic of China\\
$^{42}$ Nanjing Normal University, Nanjing 210023, People's Republic of China\\
$^{43}$ Nanjing University, Nanjing 210093, People's Republic of China\\
$^{44}$ Nankai University, Tianjin 300071, People's Republic of China\\
$^{45}$ National Centre for Nuclear Research, Warsaw 02-093, Poland\\
$^{46}$ North China Electric Power University, Beijing 102206, People's Republic of China\\
$^{47}$ Peking University, Beijing 100871, People's Republic of China\\
$^{48}$ Qufu Normal University, Qufu 273165, People's Republic of China\\
$^{49}$ Renmin University of China, Beijing 100872, People's Republic of China\\
$^{50}$ Shandong Normal University, Jinan 250014, People's Republic of China\\
$^{51}$ Shandong University, Jinan 250100, People's Republic of China\\
$^{52}$ Shanghai Jiao Tong University, Shanghai 200240, People's Republic of China\\
$^{53}$ Shanxi Normal University, Linfen 041004, People's Republic of China\\
$^{54}$ Shanxi University, Taiyuan 030006, People's Republic of China\\
$^{55}$ Sichuan University, Chengdu 610064, People's Republic of China\\
$^{56}$ Soochow University, Suzhou 215006, People's Republic of China\\
$^{57}$ South China Normal University, Guangzhou 510006, People's Republic of China\\
$^{58}$ Southeast University, Nanjing 211100, People's Republic of China\\
$^{59}$ State Key Laboratory of Particle Detection and Electronics, Beijing 100049, Hefei 230026, People's Republic of China\\
$^{60}$ Sun Yat-Sen University, Guangzhou 510275, People's Republic of China\\
$^{61}$ Suranaree University of Technology, University Avenue 111, Nakhon Ratchasima 30000, Thailand\\
$^{62}$ Tsinghua University, Beijing 100084, People's Republic of China\\
$^{63}$ Turkish Accelerator Center Particle Factory Group, (A)Istinye University, 34010, Istanbul, Turkey; (B)Near East University, Nicosia, North Cyprus, 99138, Mersin 10, Turkey\\
$^{64}$ University of Chinese Academy of Sciences, Beijing 100049, People's Republic of China\\
$^{65}$ University of Groningen, NL-9747 AA Groningen, The Netherlands\\
$^{66}$ University of Hawaii, Honolulu, Hawaii 96822, USA\\
$^{67}$ University of Jinan, Jinan 250022, People's Republic of China\\
$^{68}$ University of Manchester, Oxford Road, Manchester, M13 9PL, United Kingdom\\
$^{69}$ University of Muenster, Wilhelm-Klemm-Strasse 9, 48149 Muenster, Germany\\
$^{70}$ University of Oxford, Keble Road, Oxford OX13RH, United Kingdom\\
$^{71}$ University of Science and Technology Liaoning, Anshan 114051, People's Republic of China\\
$^{72}$ University of Science and Technology of China, Hefei 230026, People's Republic of China\\
$^{73}$ University of South China, Hengyang 421001, People's Republic of China\\
$^{74}$ University of the Punjab, Lahore-54590, Pakistan\\
$^{75}$ University of Turin and INFN, (A)University of Turin, I-10125, Turin, Italy; (B)University of Eastern Piedmont, I-15121, Alessandria, Italy; (C)INFN, I-10125, Turin, Italy\\
$^{76}$ Uppsala University, Box 516, SE-75120 Uppsala, Sweden\\
$^{77}$ Wuhan University, Wuhan 430072, People's Republic of China\\
$^{78}$ Yantai University, Yantai 264005, People's Republic of China\\
$^{79}$ Yunnan University, Kunming 650500, People's Republic of China\\
$^{80}$ Zhejiang University, Hangzhou 310027, People's Republic of China\\
$^{81}$ Zhengzhou University, Zhengzhou 450001, People's Republic of China\\
\vspace{0.2cm}
$^{\dagger}$ Deceased\\
$^{a}$ Also at the Moscow Institute of Physics and Technology, Moscow 141700, Russia\\
$^{b}$ Also at the Novosibirsk State University, Novosibirsk, 630090, Russia\\
$^{c}$ Also at the NRC "Kurchatov Institute", PNPI, 188300, Gatchina, Russia\\
$^{d}$ Also at Goethe University Frankfurt, 60323 Frankfurt am Main, Germany\\
$^{e}$ Also at Key Laboratory for Particle Physics, Astrophysics and Cosmology, Ministry of Education; Shanghai Key Laboratory for Particle Physics and Cosmology; Institute of Nuclear and Particle Physics, Shanghai 200240, People's Republic of China\\
$^{f}$ Also at Key Laboratory of Nuclear Physics and Ion-beam Application (MOE) and Institute of Modern Physics, Fudan University, Shanghai 200443, People's Republic of China\\
$^{g}$ Also at State Key Laboratory of Nuclear Physics and Technology, Peking University, Beijing 100871, People's Republic of China\\
$^{h}$ Also at School of Physics and Electronics, Hunan University, Changsha 410082, China\\
$^{i}$ Also at Guangdong Provincial Key Laboratory of Nuclear Science, Institute of Quantum Matter, South China Normal University, Guangzhou 510006, China\\
$^{j}$ Also at MOE Frontiers Science Center for Rare Isotopes, Lanzhou University, Lanzhou 730000, People's Republic of China\\
$^{k}$ Also at Lanzhou Center for Theoretical Physics, Lanzhou University, Lanzhou 730000, People's Republic of China\\
$^{l}$ Also at the Department of Mathematical Sciences, IBA, Karachi 75270, Pakistan\\
$^{m}$ Also at Ecole Polytechnique Federale de Lausanne (EPFL), CH-1015 Lausanne, Switzerland\\
$^{n}$ Also at Helmholtz Institute Mainz, Staudinger Weg 18, D-55099 Mainz, Germany\\
$^{o}$ Also at School of Physics, Beihang University, Beijing 100191, China\\
$^{p}$ Also at Hubei key laboratory of energy storage and power battery, Hubei University of Automotive Technology, Shiyan 442002, China \\
}
}




\begin{abstract}
  \par Based on 12.0 $\mathrm{fb^{-1}}$ of $\EE$ collision data samples collected by the BESIII detector at center-of-mass energies from 4.1271 to 4.3583 GeV, a partial wave analysis is performed for the process $e^{+}e^{-} \rightarrow \pi^{+}\pi^{-}J/\psi$. The cross sections for the subprocesses ${e^{+}e^{-}\rightarrow\pi^{+}Z_{c}(3900)^{-}+c.c.\rightarrow\pi^{+}\pi^{-}J/\psi}$, $f_{0}(980)(\rightarrow\pi^{+}\pi^{-})J/\psi$, and $(\pi^{+}\pi^{-})_{\rm{S\mbox{-}wave}} J/\psi$ are measured for the first time. The mass and width of the $Z_{c}(3900)^{\pm}$ are determined to be $3884.6\pm0.7\pm3.3$~MeV/$c^{2}$ and $37.2\pm1.3\pm6.6$~MeV, respectively. The first errors are statistical and the second systematic. The final state $(\pi^{+}\pi^{-})_{\rm{S\mbox{-}wave}} J/\psi$ dominates the process $e^{+}e^{-} \rightarrow \pi^{+}\pi^{-}J/\psi$. By analyzing the cross sections of $\pi^{\pm}Z_{c}(3900)^{\mp}$ and $f_{0}(980)J/\psi$, $Y(4220)$ has been observed. Its mass and width are determined to be $4225.7\pm4.1\pm3.4$~MeV/$c^{2}$ and $57.5\pm9.4\pm12.1$~MeV, respectively.
\end{abstract}

\maketitle

\oddsidemargin  -0.2cm
\evensidemargin -0.2cm

\section{Introduction}
\par Over the past twenty years several unpredicted charmoniumlike states, usually referred to as $XYZ$ mesons, have been discovered. The $\ZCP$ state was simultaneously observed by BESIII and Belle Collaborations in the $\EEPPJ$ process~\cite{BESIII:2013ris, Belle:2013yex} and confirmed by CLEO-c Collaboration~\cite{Xiao:2013iha}. The mass and width of the $\ZCP$ were determined by directly fitting the invariant mass distribution of $\pi^{\pm}J/\psi$, as well as by partial wave analysis (PWA)~\cite{BESIII:2017bua,Pilloni:2016obd,Danilkin:2020kce}, with consistent values among the different experiments. Although there are many theoretical explanations for the $\ZCP$, its internal structure is still unclear. Currently, it is thought to be either a tetra-quark or molecular state~\cite{Wang:2013cya, Li:2013xia, Cui:2013yva, Voloshin:2013dpa, Braaten:2013boa, Wilbring:2013cha,Guo:2017jvc, Wang:2013vex}. 

\par According to the interpretation of the $\ZC$ as a molecular or a tetra-quark state, there is a strong correlation between the $Y(4230)$ and $\pi\ZC$ spectrum~\cite{Cleven:2013mka, Giron:2020fvd}. The latest results of BESIII regarding the neutral process $\EE\rightarrow\pi^{0}\ZC^{0}(\rightarrow\pi^{0}J/\psi)$~\cite{Ablikim:2020pzw} show that the cross section line shape is consistent with that of the $Y(4230)$. Compared to the neutral mode, the charged process $\EE \rightarrow \pi^{\pm}Z_{c}(3900)^{\mp}(\rightarrow\pi^{\mp}J/\psi)$ has larger signal yield and a lower background level. It is, therefore, a more promising channel to study the relationship between the $\ZCP$ and $Y$ states. In addition, there might also be similar correlation between $Y$ states and resonances of the $\PP$ system. For example, Ref.~\cite{MartinezTorres:2009xb} predicted a possible peak around 4150 MeV/$c^{2}$ with a width of 90 MeV, which may have a strong coupling to $f_{0}(980)J/\psi$. In Ref.~\cite{Coito:2019cts}, it is claimed that Y(4230) comes from the mass shift of the loop amplitude of the decay $\psi(4160)\rightarrow D_{s}^{\ast}\bar{D}_{s}^{\ast}\rightarrow f_{0}(980)J/\psi\rightarrow\PPJ$. All these models require a measurement of the cross section of $\EE\rightarrow f_{0}(980)J/\psi$. Besides, providing the invariant mass/Dalitz plot distributions at different center-of-mass (c.m.) energies would be helpful for theorists to explain the ``triangle singularity"~\cite{Gong:2016jzb} with $\ZC$.

\par In this paper, we measure the cross sections of ${\EE\rightarrow\pi^{\pm}Z_{c}(3900)^{\mp}(\rightarrow\pi^{\mp}J/\psi)}$, ${f_{0}(980)(\rightarrow\PP)J/\psi}$, and $\PPS J/\psi$ [the ensemble of $f_{0}(500)$, $f_{0}(980)$, and $f_{0}(1370)$] through PWA of 17 data samples with c.m. energies $\sqrt{s}$ from 4.1271 to 4.3583 GeV. Two different models are used to parametrize $\PPS$ to obtain the cross sections: one is the Breit-Wigner (BW) function, the other is the K-matrix~\cite{Wigner:1946zz,Aitchison:1972ay,Anisovich:2002ij,Peters:2004qw}. The parameters of $\ZCP$ and the fractions of the subprocesses are given by PWA. Based on the precise cross section measurement of the $\EEPPJ$ process reported in Ref.~\cite{BESIII:2022qal}, the cross sections of the subprocesses are given. The data sample with $\sqrt{s} = 4.1780$ GeV is used as an example to illustrate the analysis method.

\section{The BESIII detector and the datasets}
\label{datasets}

The BESIII detector~\cite{bes3} records $\EE$ collisions provided by the BEPCII storage ring~\cite{bepcii} in a c.m. energy range from 2.00 to 4.95 GeV. The cylindrical core of the detector covers 93\% of the full solid angle and consists of a helium-based multilayer drift chamber~(MDC), a plastic scintillator time-of-flight system~(TOF), and a CsI(Tl) electromagnetic calorimeter~(EMC), which are all enclosed in a superconducting solenoidal magnet providing a 1.0~T magnetic field. The solenoid is supported by an octagonal flux-return yoke with resistive plate counter muon identification modules interleaved with steel (MUC). The charged-particle momentum resolution at $1~{\rm GeV}/c$ is $0.5\%$, and the $\mathrm{d}E/\mathrm{d}x$ resolution is $6\%$ for electrons from Bhabha scattering. The EMC measures photon energies with a resolution of $2.5\%$ ($5\%$) at $1$~GeV in the barrel (end cap) region. The time resolution in the TOF barrel region is 68~ps, while that in the end cap region is 110~ps.  The end cap TOF system was upgraded in 2015 using multigap resistive plate chamber technology, providing a time resolution of 60~ps~\cite{etof}.

\par The data samples in this work are listed in Table~\ref{totCrosssection}. There are 17 energy points with c.m. energies from 4.1271 to 4.3583~GeV with an integrated luminosity more than 400~$\rm pb^{-1}$ at each point. The luminosity and the energy are measured using Bhabha scattering~\cite{BESIII:2022xii} and radiative di-muon production~\cite{BESIII:2020eyu}, respectively.

\par The {\sc geant4}-based~\cite{GEANT4} Monte Carlo (MC) simulation software packages {\sc boost}~\cite{boost} and  {\sc evtgen}~\cite{EvtGen} are used to determine detection efficiencies and to estimate the background contributions. The generator {\sc kkmc}~\cite{Jadach:2000ir} is used to model the beam energy spread and the initial state radiation emission in $\EE$ annihilations. Final state radiation from charged particles is incorporated using the {\sc photos} package~\cite{photos}. The phase-space (PHSP) MC samples of ${4\times10^{5}}$ $\EEPPJ$, ${J/\psi \rightarrow \LL~(\ell = e,~\mu)}$ events are generated for each energy point, and about half of them that survived after reconstruction and event selection are used for MC integration of PWA. The background MC samples of $\EE\rightarrow\EE\rho^{0}(\rightarrow\PP)$ (PHSP), $\EE\rightarrow\PP\rho^{0}(\rightarrow\PP)$ (PHSP), and the two-photon process $\EE \rightarrow \EE\MM$ ~\cite{Berends:1986if,Berends:1986ig} are generated to estimate the effect of corresponding background channels.

\section{Event selection}
\par The preliminary event selection criteria are the same as those used in Ref.~\cite{BESIII:2022qal}. The charged tracks need to meet the vertex fit and geometric acceptance criteria of the detector and the final event candidate requires four charged tracks with a total charge of zero. The momentum measured by the MDC, the energy deposited in the EMC of the charged tracks, and the combined lepton pair invariant mass are used to distinguish leptons and pions. The angular distribution information, $\mathrm{d}E/\mathrm{d}x$ information, and boosted decision tree method are used to suppress background events of gamma conversion events, low-momentum $e/\pi$ misidentification events, and two-photon process events. A four-constraint fit imposing energy-momentum conservation with the hypothesis of $\EE \rightarrow \PP\LL$ is applied to the final charged tracks to suppress the background contribution and to improve the energy and momentum resolution.

\par In addition, we further require that at least one of the muon tracks in the $J/\psi\rightarrow\MM$ mode has a MUC hit depth to eliminate the background of the $2(\PP)$ process, and the selection criterion is optimized by $S/\sqrt{S+B}$ with MC samples, where $S$ and $B$ indicate the numbers of events estimated from the signal and background regions, respectively. We choose $3.09<m(\ell^{+}\ell^{-})<3.11$ GeV/$c^{2}$ as the signal region and $3.02<m(\ell^{+}\ell^{-})<3.07$ GeV/$c^{2}$ and $3.13<m(\ell^{+}\ell^{-})<3.16$ GeV/$c^{2}$ as sideband regions to estimate the background contributions. The background level of most samples is less than 10\% after event selection. Finally, a five-constraint fit is implied with the invariant mass of the lepton pairs constrained to the known $J/\psi$ mass~\cite{PDG}.

\section{Amplitude construction and the PWA result}
\par Amplitudes of the PWA are constructed with the helicity-covariant method~\cite{Chung:1997jn, Chung:2007nn} and include the cascading processes $\EE\rightarrow\pi^{\pm}Z_{c}(3900)^{\mp}(\rightarrow\pi^{\mp}J/\psi)$ and ${\EE\rightarrow f_{j}(\rightarrow\PP) J/\psi}$, where $f_{j}$ represents the $f_{0}(500)$, $f_{0}(980)$, $f_{0}(1370)$, or $f_{2}(1270)$. The construction method of the amplitude is the same as that in Refs.~\cite{BESIII:2017bua, Ablikim:2020pzw, LHCb:2015yax}. The PWA is performed using the AmpTool package~\cite{pwa}.

\par For a decay $a(J_{a}, M) \rightarrow b(J_{b}, \lambda_{b}) + c(J_{c}, \lambda_{c}$) of particle $a$ into the pair $b+c$, where spin and helicity are indicated in the parentheses, the amplitude is given by
\begin{equation}
    A_{\lambda_{b},\lambda_{c}}^{J_{a}}(\theta,\phi) =  N_{J_{a}}F_{\lambda_{b},\lambda_{c}}^{J_{a}}D_{M,\lambda}^{J_{a}\ast}(\phi,\theta,0),
\end{equation}
\\where $N_{J_{a}}$ is the normalization factor, $\lambda = \lambda_{b}-\lambda_{c}$ and $F_{\lambda_{b},\lambda_{c}}^{J_{a}}$ is the helicity amplitude, which is constrained by parity conservation. The $D_{M,\lambda}^{J_{a}}(\phi,\theta,0)$ is the Wigner-$D$ function which describes the angular distribution of the final state particle with its polar ($\theta$) and azimuthal ($\phi$) angles in the rest frame of the mother particle. The amplitude $F_{\lambda_{b},\lambda_{c}}^{J_{a}}$ is given by Chung's formula~\cite{Chung:1997jn},
\begin{equation}
\label{hcamp}
\begin{aligned}
    F_{\lambda_{b},\lambda_{c}}^{J_{a}} = &\sum_{LS}G_{LS}^{J_{a}}\sqrt{\frac{2L+1}{2J_{a}+1}}
    \langle L,0,S,\lambda|J_{a},\lambda \rangle \\
     &\langle S_{b},\lambda_{b},S_{c},-
    \lambda_{c}|S,\lambda\rangle p^{L}B_{L}(p,r),\\
\end{aligned}
\end{equation}
\\where $G_{LS}$ is the coupling constant in the $LS$ coupling scheme, $L$ is the orbital angular momentum between $b$ and $c$, and $S$ is the coupled spin angular momentum of $b$ and $c$. The angular brackets denote Clebsch-Gordan coefficients, and the orbital angular momentum barrier factor $p^{L}B_{L}(p,r)$ involves the Blatt-Weisskopf functions~\cite{Chung:1997jn}.

The total differential cross section is given by

\begin{widetext}
\begin{equation}
\label{DiffCs}
\begin{array}{l}
\begin{aligned}
    T
    &= \sum_{\lambda_{Y},\Delta\lambda_{\ell}}\biggl|\sum_{f_{j},\lambda_{J/\psi},\lambda_{f_{j}}}A^{Y \rightarrow f_{j}J/\psi}_{\lambda_{f_{j}},\lambda_{J/\psi}}(\theta_{f_{j}},\phi_{f_{j}}) BW(m(\PP)) A^{f_{j}\rightarrow\PP}_{0,0}(\theta_{\pi^{+}},\phi_{\pi^{+}})A^{J/\psi\rightarrow \LL }_{\lambda_{\ell^{+}},\lambda_{\ell^{-}}}(\theta_{\ell^{+}},\phi_{\ell^{+}})  \\
    & + e^{i\Delta\lambda_{\ell}\alpha_{\ell}}\sum_{Z_{c},\lambda_{Z_{c}},\lambda_{J/\psi}}A^{Y \rightarrow \pi Z_{c}}_{\lambda_{Z_{c}},0}(\theta_{Z_{c}},\phi_{Z_{c}}) BW(m(\pi J/\psi)) A^{Z_{c}\rightarrow\pi J/\psi}_{\lambda_{J/\psi},\lambda_{\pi}}(\theta_{J/\psi},\phi_{J/\psi})A^{J/\psi\rightarrow \LL }_{\lambda_{\ell^{+}},\lambda_{\ell^{-}}}(\theta_{\ell^{+}},\phi_{\ell^{+}})
    \biggr|^{2},
\end{aligned}
\end{array}
\end{equation}
\end{widetext}
where $\lambda_{P}$ is the helicity of the particle $P$ ($\ell^{\pm}$, $Z_{c}(3900)^{\pm}$, $J/\psi$, and $f_{j}$), $Z_{c}$ ($\pi$) represents the $Z_{c}(3900)^{+}$ ($\pi^{-}$) and $Z_{c}(3900)^{-}$ ($\pi^{+}$), and $(\theta_{P}, \phi_{P})$ are the polar and azimuthal angles of particle $P$ in the helicity frame of the cascading process. The summation is performed for the virtual photon ($\lambda_{Y}$), and the lepton pairs in the final state ($\Delta \lambda_{l}$), then $\lambda_{Y}=\pm1$ and $\Delta \lambda_{l} = \pm1$. The $m(\PP)$ and $m(\pi J/\psi)$ are the invariant masses of $\PP$ and $\pi^{\pm} J/\psi$. The amplitude of $J/\psi\rightarrow \LL$ is considered, and an additional term $e^{i\Delta\lambda_{\ell}\alpha_{\ell}}$ is introduced to align the helicity frames in two different decay chains~\cite{LHCb:2015yax}. (If we make a one-to-one correspondence for the final states of $e^{+}e^{-} \rightarrow \pi^{+}\pi^{-}J/\psi$ and the process $\Lambda_b^0\to p K^-J/\psi$ studied in Ref.~\cite{LHCb:2015yax} , with $\pi^+\leftrightarrow p$, $\pi^-\leftrightarrow K^-$ and $J/\psi\leftrightarrow J/\psi$, the helicity frames for these two processes are identical and so the calculation methods for the helicity angles and $\alpha_l$ are also the same).

\par The subprocesses, including $Z_{c}(3900)^{\pm}$, $f_{0}(500)$, $f_{0}(1370)$, and $f_{2}(1270)$, are parametrized with constant-width relativistic BW functions. The $f_{0}(980)$ is described by a Flatt\'{e} formula with parameters taken from Ref.~\cite{BES:2004twe}. The energy-dependent width of the $f_{0}(500)$ is parametrized as $\sqrt{1-\frac{4m_{\pi}^{2}}{s}}\Gamma$~\cite{BESIII:2017bua, Ablikim:2020pzw}. The resonant parameters of these well-known mesons are taken from the world averaged values (Particle Data Group, PDG)~\cite{PDG}. We also use the K-matrix method~\cite{Peters:2004qw} to parametrize $\PPS$, where its parameters are fixed to the values obtained from the scattering experiments~\cite{BaBar:2008inr, Anisovich:2002ij}. The relative magnitudes and phases of the individual subprocesses are determined by performing an unbinned extended maximum likelihood (EML) fit at each energy point. The likelihood function is
\begin{equation}
\label{Likehoodmx}
    \mathcal{L} = \frac{e^{-\mu}\mu^{N}}{N!}\prod^{N}_{i=1} \frac{I(\omega_{i},\alpha)\epsilon(\omega_{i})}{\mu},
\end{equation}
where $I(\omega_{i},\alpha) \equiv T_{i}$ is the total differential cross section as defined in Eq.(~\ref{DiffCs}) with a set of four-momenta $\omega_{i} = (p_{\pi^{+}},p_{\pi^{-}}, p_{\ell^{+}}, p_{\ell^{-}} )_{i}$ and float parameters $\alpha$ [the coupling constants $G_{LS}$ and the resonance parameters of $Z_c(3900)^{\pm}$]. The $\epsilon(\omega_{i})$ is the event selection efficiency, which is precalculated and does not affect fit parameters. The $N$ is the number of observed events. The normalization factor ${\mu = \int I(\omega_{i},\alpha) \epsilon(\omega_{i}) d\omega_{i} = \frac{1}{N_{mc}}\sum_{j=1}^{N_{mc}}I(\omega_{j},\alpha)}$. $N_{mc}$ is the $\EEPPJ$ events generated with the PHSP model that pass the detector simulation and event selection. We can see that $\mu$ is the expected value of the Poisson term in Eq.(~\ref{Likehoodmx}), and it will converge to the number of signal events in the EML fit. Thus, the magnitude of the coupling constants will converge to suitable values automatically.  However, to remove the phase ambiguity, we fix the phase of the coupling constant of the first amplitude of the $Y\to \pi Z_c(3900)$ process to 0.  Technically, rather than maximizing $\mathcal{L}$, the log-likelihood $-\mathrm{ln}\mathcal{L}$ is minimized. The background contribution to the overall likelihood value is calculated and subtracted with the events in the $J/\psi$ sideband region.

After the fitting, the fraction of a process $i$ is calculated using
\begin{equation}
    F_i=\frac{\int |A_i|^2 d\phi}{\int|\sum_k A_k|^2 d\phi}=\frac{\sum_{j\in \rm{PHSP}}|A_j(\omega_j,\alpha)|^2}{\sum_{j\in \rm{PHSP}}|\sum_k A_k(\omega_j,\alpha)|^2},
\end{equation}
where the numerator is the amplitude of process $i$ and the denominator is the sum of all the amplitudes included in the fit. The PHSP MC without passing the detector reconstruction and event selection is used to perform the calculation, so what we get is the fraction at the physical level, and there is no need to consider the efficiency and so on. The interference between process $i$ and $j$ is
\begin{equation}
    Inter_{i,j}=\frac{\int |A_i+A_j|^2-|A_i|^2-|A_j|^2 d\phi}{\int|\sum_k A_k|^2 d\phi},
\end{equation}
which is similar to the calculation of fractions.

\par In order to reduce the number of unnecessary parameters and subprocesses with low significance (less than 5$\sigma$), we perform a significance check based on the subprocesses used in Ref.~\cite{BESIII:2017bua}. We remove the PHSP process and only keep the $(L, S)=(0, 1)$ wave of $f_{2}(1270)$, where $L$ is the orbital angular momentum between $J/\psi$ and $f_{2}(1270)$, and $S$ is the spin of the [$J/\psi,~f_{2}(1270)$] system. The significance of $\pi^{\pm}Z_{c}(4020)^{\mp}$ in this energy range is less than 3.0$\sigma$. A simultaneous fit to four adjacent energy points is used to obtain the parameters of $\ZCP$, and then give the fractions of the subprocesses by a single-point fit with the $\ZCP$ parameters fixed. The masses and widths of the subprocesses are shared as common parameters in the simultaneous fit, while the production magnitudes and the relative phases are independent. The mass and width of the $\ZCP$ obtained by the simultaneous fit in different energy regions are listed in Table~\ref{SimultFitMW}, and the average values of the fitted results of four sets of data samples are taken as the final parameters of $\ZCP$, which are ($M,~\Gamma$) = ($3884.6\pm0.7\pm3.3$~MeV/$c^{2}$, $37.2\pm1.3\pm6.6$~MeV).

\par When performing the fit to get the fraction of intermediate states, we meet some local minimum and multiple solution problems. At each energy point, we perform 10,000 fits with random initial parameter values. The cross sections and fractions with different FCNs (target function, here it is $-2\rm{ln}\mathcal{L}$ in the extended maximum likelihood fit) are counted (only the solutions that appear more than 100 times are considered), as shown in Fig.~\ref{Mult_Solution_CS_Fractions}. The cumulative number of times and the fractions of subprocesses for different FCNs are listed in Tables~\ref{FCNallBW1} and~\ref{FCNallBW2}. We can see that there are some solutions with quite different values but generally close FCNs. When looking at some single energy points, it might be difficult to choose the physical solution. The result in Ref.~\cite{BESIII:2017bua} can be repeated within the 0.5\% FCN difference range, but it does not correspond to the solution with minimum FCN in our fit. By our study, many factors such as the statistics of PHSP MC and the line shape of $Z_{c}(3900)$, may cause a fluctuation in the fit and the result may converge to a different solution. This explains the difference about the cross section of $\EE \to \PZC$ between this work and the previous BESIII result~\cite{BESIII:2017bua}. We have the result at many adjacent c.m. energy points which can help us to select a set of reasonable solutions. We found at most energy points that choosing the solution with minimum FCN would give a reasonable set of solutions. The fraction and cross section of intermediate states of adjacent c.m. energies are smooth, and there are no abnormal jumps/dips. There is only one exception at $\sqrt{s}=4.2776$ GeV. The solution with minimum FCN seems to cause a jump on the cross section of both $e^+e^-\to\pi^{\pm} Z_c(3900)^{\mp}$ and $f_0(980)J/\psi$. However, since the uncertainty at this energy point is large and it is far away from the peak of $Y(4220)$, we just keep the solution with minimum FCN as the nominal result at all energy points. All the following discussion and results are based on this choice.

\par
\begin{figure*}
\centering
  \includegraphics[width=0.9\textwidth, height=0.4\textwidth]{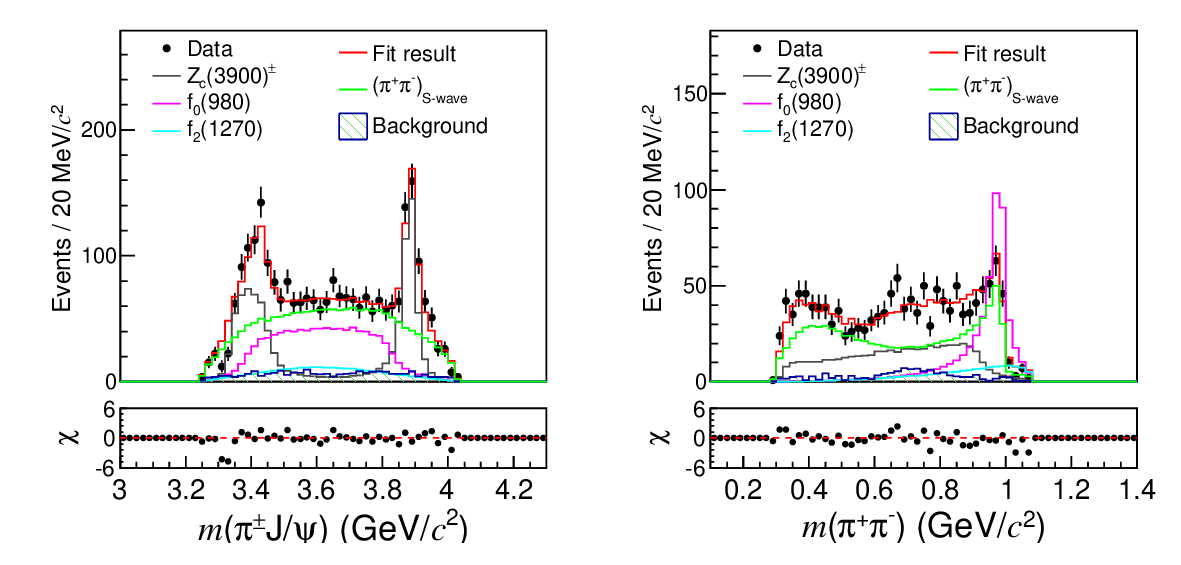}
  \put(-280, 130){(a)}
  \put(-50,  130){(b)}

  \includegraphics[width=0.9\textwidth, height=0.4\textwidth]{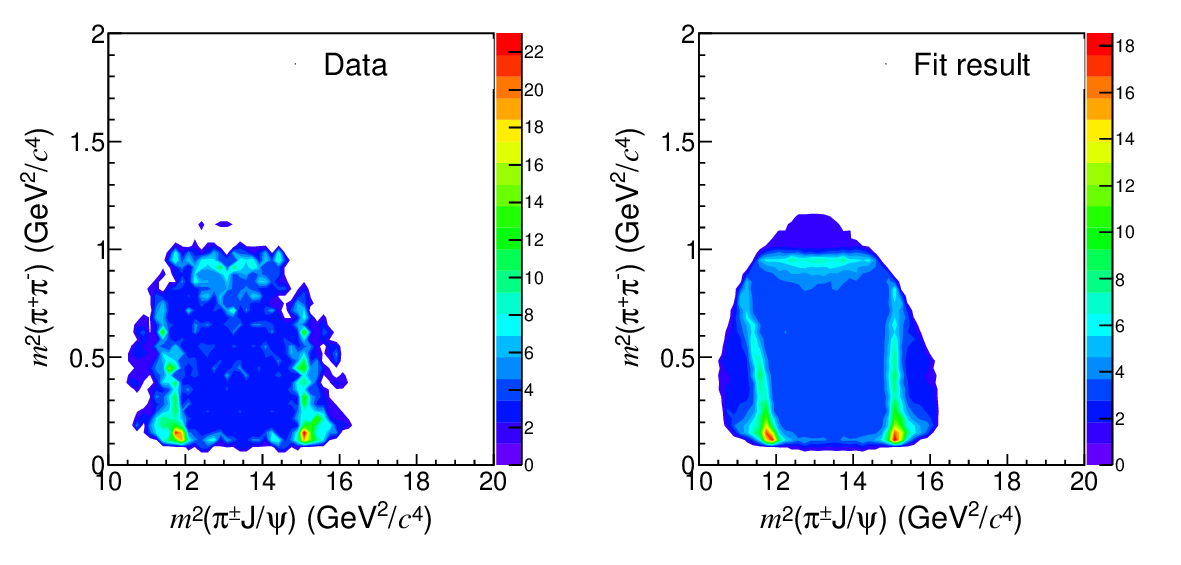}
  \put(-290, 140){(c)}
  \put(-60,  140){(d)}

  \includegraphics[width=0.9\textwidth, height=0.4\textwidth]{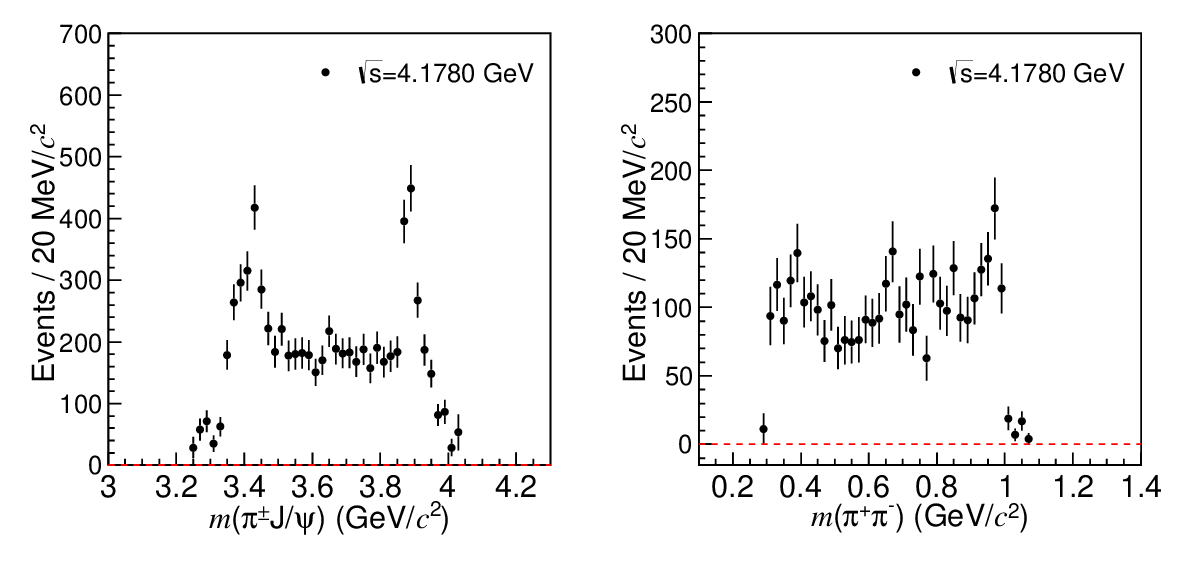}
  \put(-280, 130){(e)}
  \put(-50,  130){(f)}

\caption{\footnotesize{The PWA fit result of $\sqrt{s}=4.1780$ GeV sample.} Figures (a) and (b) show the distributions of $m(\pi^{\pm}J/\psi)$ and $m(\pi^{+}\pi^{-})$, respectively. The top plots come from the fit result of PWA, and the bottom parts are the pull distribution of the total fit result. The goodness of fits ($\chi^{2}$/ndf) of (a) and (b) are 1.48 and 1.22, respectively. In (a) and (b), the $\PPS$ includes $f_{0}(500)$, $f_{0}(980)$, and $f_{0}(1370)$. Figures (c) and (d) are the Dalitz plots for $\pi^{\pm}J/\psi$ and $\pi^{+}\pi^{-}$, (c) is from data, and (d) is the fit result, where the $\chi^{2}$/ndf is 1.71. Figures (e) and (f) are the distributions of $m(\pi^{\pm}J/\psi)$ and $m(\pi^{+}\pi^{-})$ after background deduction and efficiency correction. In (e) and (f), the sideband events are used as an estimate of the background events and subtracted from events in the signal region.}
\label{PWAresult}
\end{figure*}

\begin{figure*}
\centering
  \includegraphics[width=0.9\textwidth, height=0.4\textwidth]{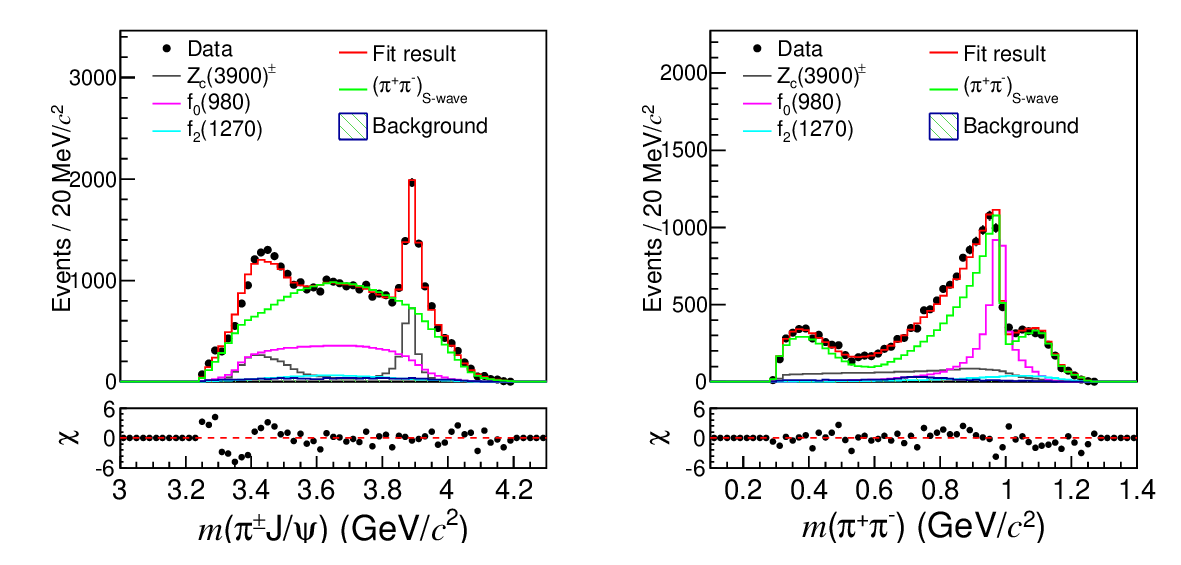}
  \put(-280, 130){(a)}
  \put(-50,  130){(b)}

  \includegraphics[width=0.9\textwidth, height=0.4\textwidth]{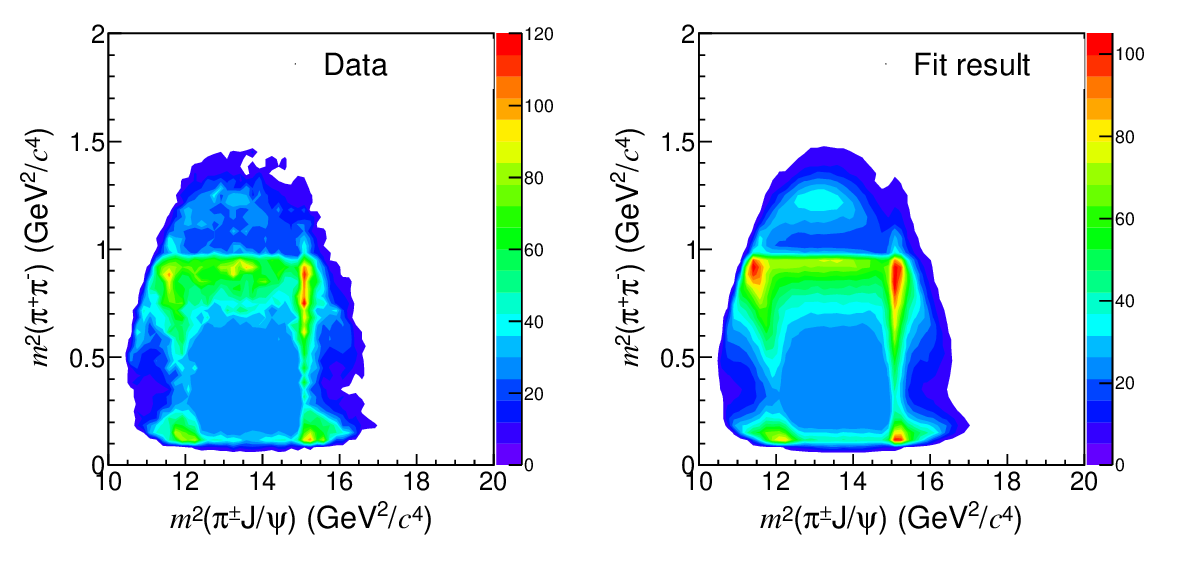}
  \put(-290, 140){(c)}
  \put(-60,  140){(d)}

\caption{\footnotesize{Fitting results for all data samples are merged together.} Figures (a) and (b) show the distributions of $m(\pi^{\pm}J/\psi)$ and $m(\pi^{+}\pi^{-})$, respectively. The top plots come from the fit result of PWA, and the bottom parts are the pull distribution of the total fit result. The fit of $m(\pi^{\pm}J/\psi)$ does not agree well with the data near 3.3 GeV/$c^{2}$. Figures (c) and (d) are the Dalitz plots for $\pi^{\pm}J/\psi$ and $\pi^{+}\pi^{-}$, (c) is from data and (d) is the fit result.}
\label{PWAresultTotal}
\end{figure*}

 The invariant mass distributions of $\pi^{\pm}J/\psi$ and $\PP$ as well as the Dalitz plots at $\sqrt{s} = 4.1780$ GeV are shown in Figs.~\ref{PWAresult}(a)-\ref{PWAresult}(d) (the remaining energy points are shown in Figs.~\ref{PWAresultTotal1},~\ref{PWAresultTotal2},~\ref{PWAresultTotal3} ,and~\ref{PWAresultTotal4} in the Appendix), where the $m(\pi^{\pm}J/\psi)$ is the sum of $m(\pi^{+}J/\psi)$ and $m(\pi^{-}J/\psi)$. The goodness of fit ($\chi^{2}/ndf$) of the $m(\pi^{\pm}J/\psi)$, $m(\pi^{+}\pi^{-})$, and Dalitz distributions is listed in Table~\ref{chisqvsndf1} in the Appendix. For the Dalitz distributions, we have used the adaptive binning scheme~\cite{LHCb:2024yzj} to make sure the bin content in each bin is not less than 10. In addition, the distributions of $m(\pi^{\pm} J/\psi)$ and $m(\PP)$ after background deduction and efficiency correction are also given, which are shown in Figs.~\ref{PWAresult} (e) and (f). At multiple c.m. energies and the combination of all data samples (in Fig.~\ref{PWAresultTotal}), we note that systematically the data are not described very well by our fits for $\pi^{\pm} J/\psi$ invariant masses in the region around 3.4 GeV, which is the reflection of $Z_c(3900)$. We have tried to solve  this problem. The first attempt is adding the contribution of $Z_{c}(4020)^{\pm}$, however, the significance of $Z_c(4020)$ is too small and cannot improve the fitting quality significantly. Changing the parameters of the $\PPS$ to the K-matrix or using the Flatt{\'e} formula used in ~\cite{BESIII:2017bua} to parametrize $Z_{c}(3900)^{\pm}$ cannot solve this problem either. The efficiency corrected distributions of $m(\pi^{\pm} J/\psi)$ (Fig.~\ref{CorrectMasspiJpsi}) and $m(\PP)$ (Fig.~\ref{CorrectMasspipi}) are placed in the Appendix to facilitate the theorists who can make further studies. The Argand plots~\cite{PDG} of energy points with larger statistics are shown in Fig.~\ref{Agrandplot}. Although the uncertainties are large, they show the features of a resonance.

\par In this paper, we only report the result below $\sqrt{s}=4.3583$ GeV, due to the decrease in the $Z_{c}(3900)^{\pm}$ production cross section with increasing energy, and the second $Z_c$ states might be needed. However, after adding a new subprocess $\pi^{\pm}Z_{c}(4020)^{\mp}$ which is parametrized with a BW function with its mass and width fixed to PDG values, the fit results with the samples at $\sqrt{s}=$ 4.3768, 4.3954, 4.4156, and 4.4359 GeV are obtained and shown in Fig.~\ref{PWAabove4360}. The $m(\pi^{\pm}J/\psi)$ is not well fitted, and the line shape of $Z_{c}(3900)^{\pm}$ or $Z_{c}(4020)^{\pm}$ may affect the fit. Detailed study of these data samples is ongoing by BESIII and will be published in another paper.

\begin{table*}
\caption{The mass ($M$) and width ($\Gamma$) of $\ZCP$ obtained by the simultaneous fit in different energy regions. The first uncertainty is statistical, and the second uncertainty for average values is systematic.}
\centering
\renewcommand\arraystretch{1.2}
\renewcommand\tabcolsep{8.0pt}
\scalebox{1.0}{
\begin{tabular}{lll}
\hline
\hline
  Sample           &      $M$~(MeV/$c^{2}$)                &   $\Gamma$~(MeV)            \\
\hline
  $4.1567-4.1989$  &     $3883.5\pm1.6$    &  $38.6\pm3.6$    \\
  $4.2091-4.2357$  &     $3884.0\pm1.0$    &  $37.8\pm1.6$    \\
  $4.2438-4.2776$  &     $3884.9\pm1.8$    &  $34.2\pm3.3$    \\
  $4.2866-4.3583$  &     $3890.0\pm2.3$    &  $36.1\pm4.2$    \\
\hline
  Average          &     $3884.6\pm0.7\pm3.3$    &  $37.2\pm1.3\pm6.6$    \\
\hline
\hline
\end{tabular}}
\label{SimultFitMW}
\end{table*}

\par The influence of the statistics of the MC samples used for integration and the reconstruction resolution are also taken into consideration, and a correction independent of c.m. energy is given according to the MC-based input and output (I/O). The correction values are listed in Table~\ref{MWCorrection}. After obtaining the fractions of the subprocesses (listed in the Appendix), we extract the cross section of each subprocess according to the total cross section of the $\EEPPJ$ process given in Ref.~\cite{BESIII:2022qal}. The results are listed in Table~\ref{totCrosssection} and the corresponding cross section plots are shown in Fig.~\ref{FitCs}. The interference magnitudes between each subprocess at $\sqrt{s} = 4.2263$ and 4.2580 GeV are listed in Table~\ref{InterferenceTerm4230}. It can be seen that the interference between $\PPS$ is strong, therefore, we only present the results for the narrower structures and $\PPS$ as a whole.

\begin{table}[!htbp]
\caption{The mass ($M^{\rm Z}$ in MeV/$c^{2}$) and width ($W^{\rm Z}$ in MeV) correction values of $\ZCP$. The fraction correction values of $\pi^{\pm}Z_{c}(3900)^{\mp}$ ($F^{Z}$ in \%), $f_{0}(980)J/\psi$ ($F^{980}$ in \%), and $(\pi^{+}\pi^{-})S_{\rm wave}J/\psi$ ($F^{\rm S}$ in \%) .}
\centering
\renewcommand\arraystretch{1.2}
\scalebox{0.9}{
\begin{tabular}{llllll}
\hline
\hline
  Parameter    &   $M^{Z}$     &   $W^{Z}$   &   $F^{Z}$ & $F^{980}$  &   $F^{S}$ \\
\hline
  Value         &  $0.9\pm0.1$  &  $0.6\pm0.1$    &   $-0.2\pm0.1$    &  $0.3\pm0.1$   &   $0.2\pm0.1$       \\
\hline
\hline
\end{tabular}}
\label{MWCorrection}
\end{table}

\section{Systematic uncertainties}

\par The systematic uncertainties for the $\ZCP$ parameters and the cross sections of the subprocesses come from the amplitude modeling, the background estimation, reconstruction resolution, and the statistics of the PHSP MC samples in the PWA. In addition, since we first use the four-samples simultaneous fit to obtain the parameters of the $\ZCP$, and then give the fractions and cross sections of the subprocesses with the parameters fixed, the uncertainties of the cross sections of the subprocesses also include the uncertainty due to the $\ZCP$ parameters and the uncertainty from the total cross section of the process $\EEPPJ$.

\par The uncertainties associated with the amplitude modeling in PWA arise from the parametrizations of the subprocesses, the radius of the angular momentum barrier factor, and the possible missing components. Each term of the uncertainties associated with the parametrizations of subprocesses and barrier factors is studied by varying the parameters within their possible value range and uncertainty~\cite{BESIII:2017bua,BES:2004twe,PDG}, and the largest difference between the results and the nominal result is regarded as the uncertainty. For the BW function of propagators, widths of the resonances are replaced by the energy-dependent widths to estimate the uncertainty, and the difference between the nominal and new results is taken as the systematic uncertainty.

\par The uncertainty related to the background estimation is studied by varying the $m(\ell^{+}\ell^{-})$ signal region and using the background MC sample instead of sideband events, and the difference is taken as the system uncertainty. Since the significance of the three-body PHSP process is lower than 5$\sigma$ in the simultaneous fit, it is not considered in the nominal fit. At the same time, due to the limited MC statistics, the very wide $f_{0}(1370) J/\psi$ and PHSP processes cannot be distinguished in the fit, that is, $f_{0}(1370)$ actually represents a class of wider states. Therefore, we do not consider the possible effect caused by missing the PHSP process in the fit. For the MC sample statistics and reconstruction resolution, we take the maximum difference between the I/O bias value and the correction value as the systematic uncertainty.

\par Assuming all the sources are independent, they are added in quadrature. The total systematic uncertainties are listed in Table~\ref{totCrosssection}. In parametrizing $\PPS$, the K-matrix method is used as a comparison to the BW method, and the difference is taken as the systematic uncertainty of the S-wave parametrization.

\section{Fit the cross section}
\par
\begin{figure*}
\centering
  \includegraphics[width=0.45\textwidth, height=0.35\textwidth]{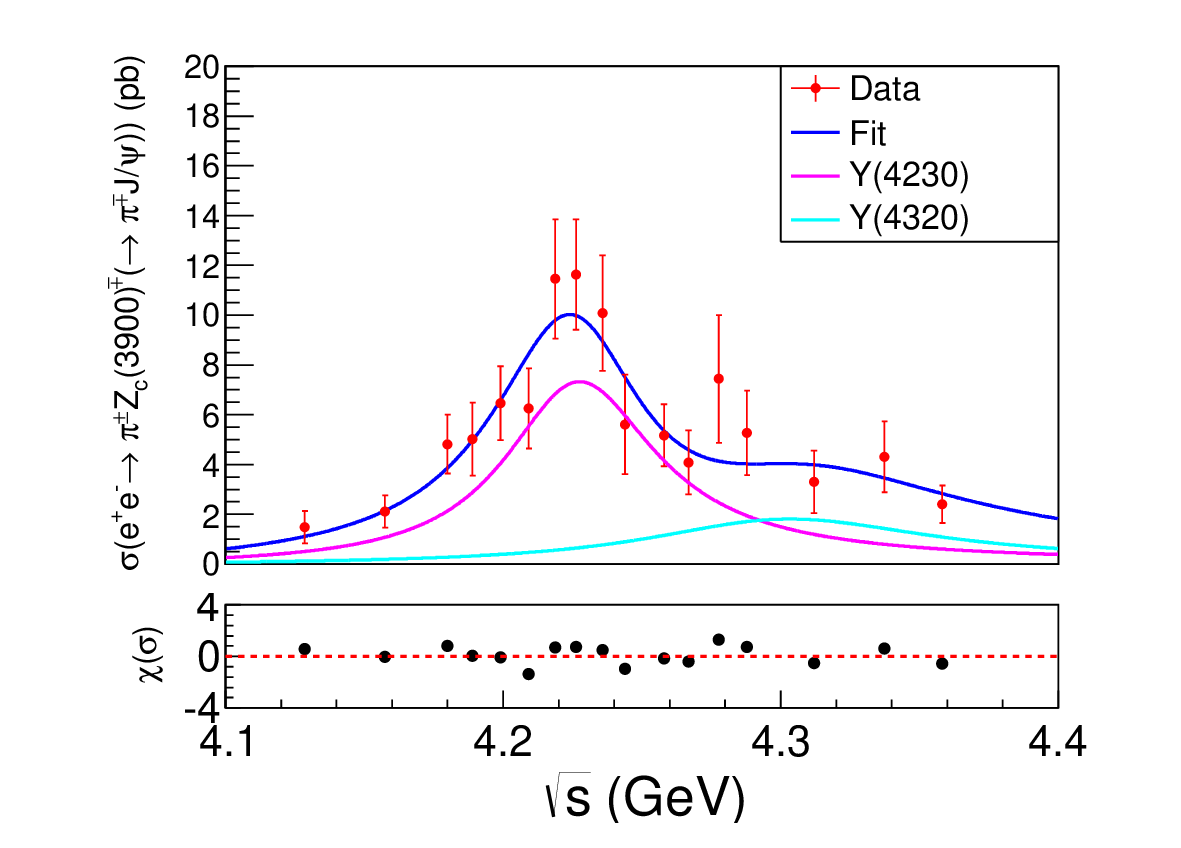}
  \includegraphics[width=0.45\textwidth, height=0.35\textwidth]{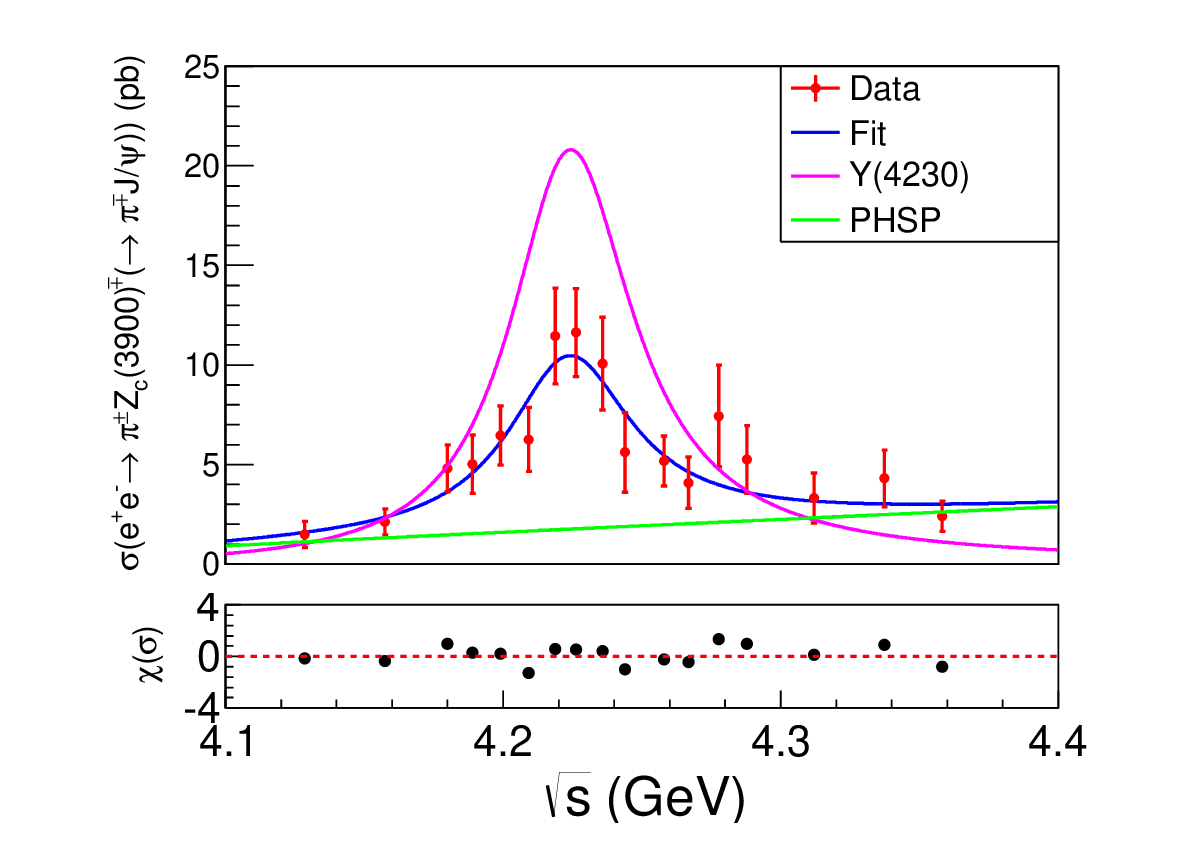}
  \put(-390, 140){(a)}
  \put(-160, 140){(b)}

  \includegraphics[width=0.45\textwidth, height=0.35\textwidth]{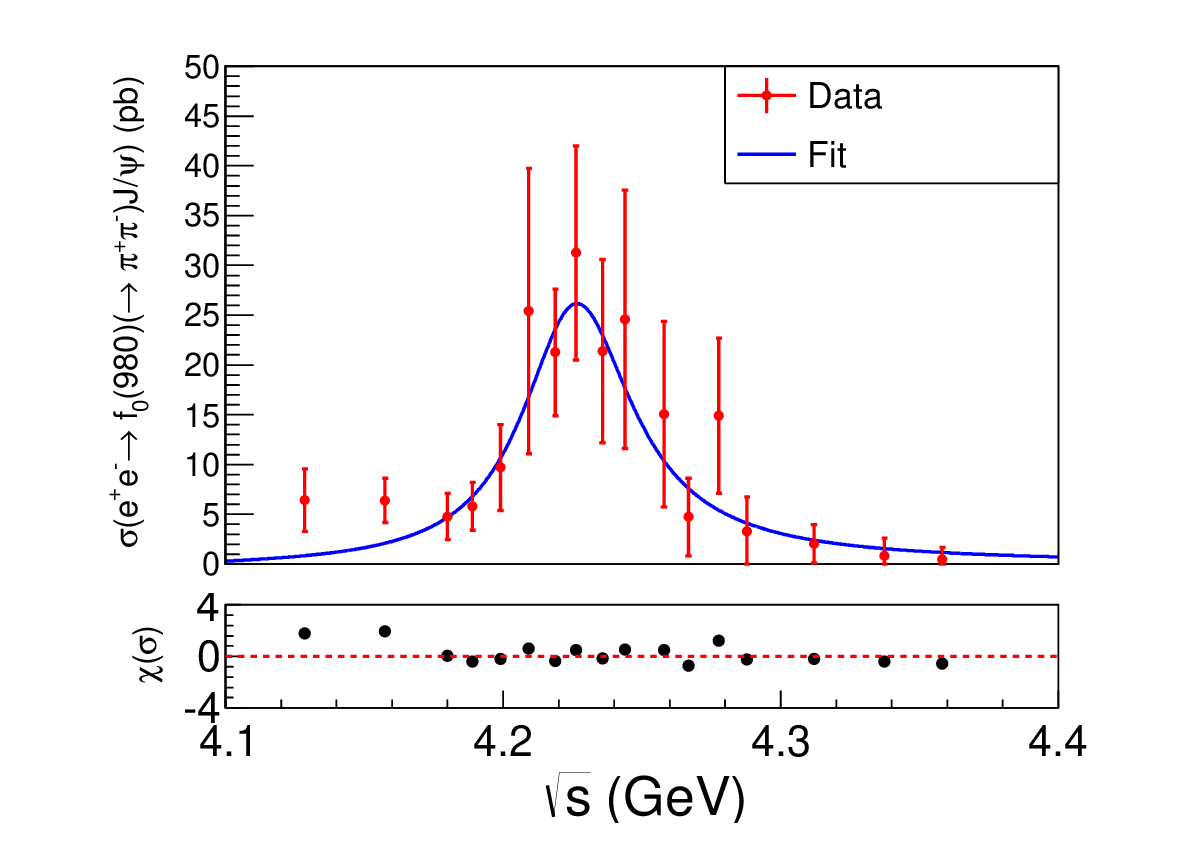}
  \includegraphics[width=0.45\textwidth, height=0.35\textwidth]{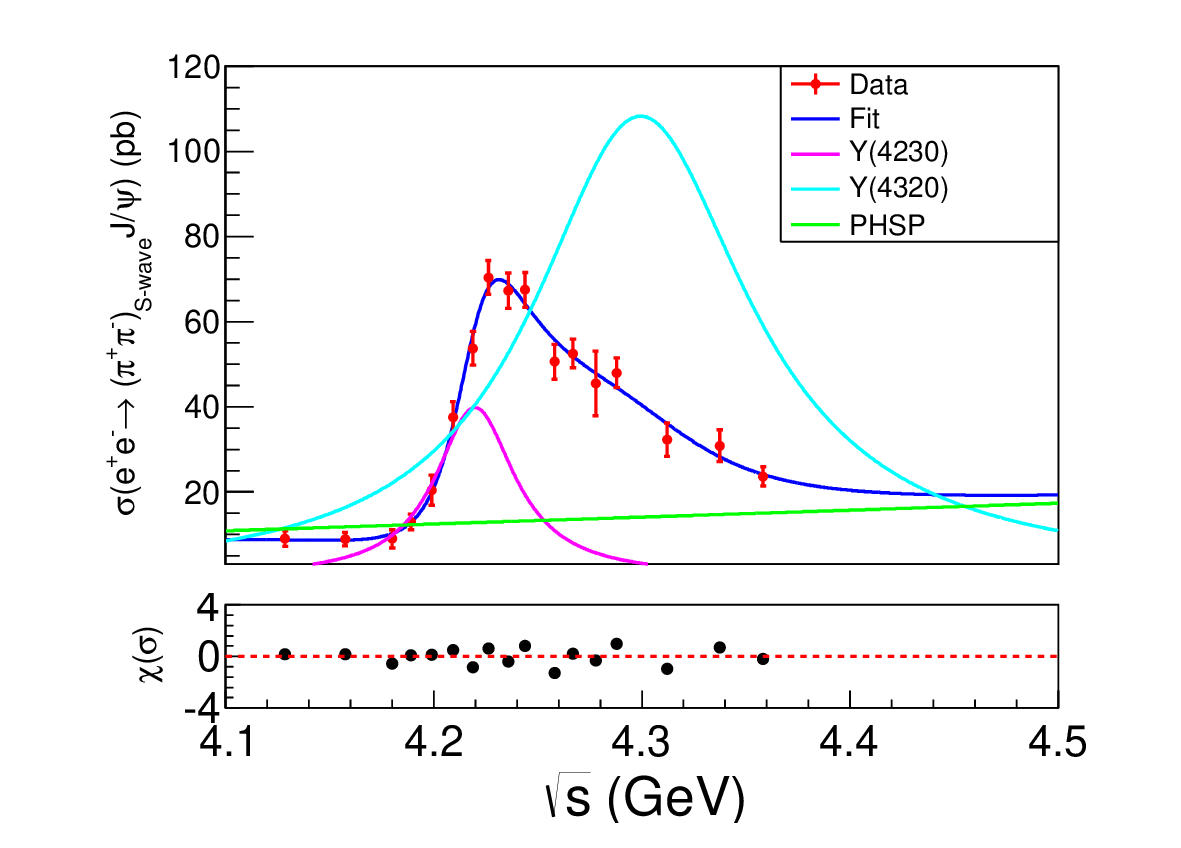}
  \put(-390, 140){(c)}
  \put(-160, 140){(d)}

\caption{\footnotesize{(a)-(b): The cross sections of $e^{+}e^{-} \rightarrow \pi^{\pm}Z_{c}(3900)^{\mp}(\rightarrow\pi^{\mp}J/\psi)$; (a) Fitted with coherent sum of two BW functions, and $\chi^{2}/ndf = 8.3/12$; (b) Fitted with coherent sum of one BW function and a two-body phase function, and $\chi^{2}/ndf = 9.1/12$; (c) The cross sections of $e^{+}e^{-} \rightarrow f_{0}(980)(\rightarrow\pi^{+}\pi^{-}) J/\psi$ and fitted with a single BW function, the $\chi^{2}/ndf = 11.1/14$; (d) The cross sections of ${e^{+}e^{-} \rightarrow \PPS J/\psi}$, and the $\chi^{2}/ndf = 6.7/11$. The $\PPS J/\psi$ is the whole of $f_{0}(500)J/\psi$, $f_{0}(980)J/\psi$, and $f_{0}(1370)J/\psi$. In these fits, the parameters of the BW function used to describe the structure around 4.3 GeV are fixed to those of the $Y(4320)$ given in Ref.~\cite{BESIII:2022qal}.}}
\label{FitCs}
\end{figure*}

 A least-square fit is performed to the cross sections of $e^{+}e^{-} \rightarrow \pi^{+}Z_{c}(3900)^{-}+c.c.\rightarrow\pi^{+}\pi^{-}J/\psi$, ${f_{0}(980)(\rightarrow\pi^{+}\pi^{-})J/\psi}$, and $(\pi^{+}\pi^{-})_{\rm{S\mbox{-}wave}} J/\psi$. For these distributions, the BW functions [Eq.~(\ref{BWCs})] are used to describe the possible resonant structures and the two-body PHSP function describes the nonresonant components. The fit results are shown in Fig.~\ref{FitCs},

\begin{equation}
\label{BWCs}
    BW_{j}(\sqrt{s})
    = \frac{M_{j}}{\sqrt{s}}
     \frac{\sqrt{12\pi \Gamma^{\rm ee}_{j}\Gamma^{\rm tot}_{j}\mathcal{B}(R_{j})}}{s-M^{2}_{j}+iM_{j}\Gamma^{\rm tot}_{j}}\times \sqrt{\frac{PS(\sqrt{s})}{PS(M_{j})}},
\end{equation}
where $M_{j}$, $\Gamma^{tot}_{j}$, and $\Gamma_{ee}$ are the mass, full width, and electric width of resonance $R_{j}$, respectively. The $\mathcal{B}(R_{j}$) is the branching fraction for $R_{j}$ decaying into $\pi^{\pm}Z_{c}(3900)^{\mp}$, $f_{0}(980)J/\psi$, and $\PPS J/\psi$. $PS(\sqrt{s}) = \frac{1}{(2\pi)^{5} (2\sqrt{s})}((s-(m_{1}+m_{2})^{2})((s-(m_{1}-m_{2})^{2}))^{1/2}$ is the standard two-body decay phase-space factor~\cite{PDG}. The $m_{1}$ and $m_{2}$ are the invariant masses of the final particles.

The cross sections of these processes are parametrized with a coherent sum of two BW functions (model I), one BW and a two-body PHSP function (model II), only one BW (model III), or a coherent sum of two BW functions and a two-body PHSP function (model IV). The cross section line shape is described by
\begin{equation}
\label{ModelX}
    \sigma_{fit}(\sqrt{s}) = \left|\sum_{j=0}^{n}{R_{j}(\sqrt{s})e^{i\phi_{j}}}\right|^{2},
\end{equation}
where $R_{j}$ represents the amplitude to describe a given resonant structure and $\phi_{j}$ is the corresponding phase. The phase $\phi_{0}$ is set to zero and the other phases are given relative to the $R_{0}$. The $R_{j}$ can be the BW or two-body PHSP function. The two-body PHSP function is described by $R(\sqrt{s}) = p_{0} \sqrt{PS(\sqrt{s})/\sqrt{s}}$, where $p_{0}$ is a free parameter.

For the $\EE \rightarrow \pi^{\pm}Z_{c}(3900)^{\mp}(\rightarrow\pi^{\mp}J/\psi)$ process, it is not possible to distinguish which model is the best at the current accuracy (Table~\ref{FitCsYres}). When describing the structure at 4.3 GeV, we fix its parameters to those of the $Y(4320)$ given in Ref.~\cite{BESIII:2022qal}. The interferences between different structures are also taken into account. The mass and width of the $Y(4220)$, and the significance of the $Y(4320)$ for these three processes in different models, are listed in Table~\ref{FitCsYres}.

\begin{table*}
\caption{The mass ($M$) and width ($\Gamma$) of the $Y(4220)$ and the significance of the $Y(4320)$ given by different processes with different models. The $\PZC$ and $f_{0}(980)J/\psi$ stand for $\EE\rightarrow\pi^{\pm}Z_{c}(3900)^{\mp}(\rightarrow\pi^{\mp}J/\psi)$ and $f_{0}(980)(\rightarrow\pi^{+}\pi^{-})J/\psi$, respectively. The $Y(4220)^{\rm ave}$ is the average of the processes $\PZC$ (in model I) and $f_{0}(980)J/\psi$ (in model III). The first uncertainty is statistical, and the second systematic.}
\centering
\renewcommand\arraystretch{1.2}
\renewcommand\tabcolsep{8.0pt}
\scalebox{1.0}{
\begin{tabular}{llll}
\hline
  \multirow{2}*{Process}          &           \multicolumn{2}{c}{$Y(4220)$}                          &      $Y(4320)$    \\
\cmidrule(rl){2-3}\cmidrule(rl){4-4}
                    &   $M$ (MeV/$c^{2}$)    &   $\Gamma$ (MeV)     &  Significance  \\
\cmidrule(rl){1-3}\cmidrule(rl){4-4}
  $\pi^{\pm}Z_{c}(3900)^{\mp}$~(model I) & $4225.7\pm6.8\pm6.8$     &   $66.5\pm16.1\pm24.1$  &  2.1$\sigma$     \\
  $\pi^{\pm}Z_{c}(3900)^{\mp}$~(model II)& $4223.1\pm6.4\pm0.6$     &   $53.8\pm19.1\pm0.3$   &  2.0$\sigma$     \\
  $f_{0}(980)J/\psi$~(model III)         & $4225.6\pm4.5\pm0.6$     &   $48.4\pm9.8\pm0.2$    &  0.5$\sigma$     \\
  $\PPS J/\psi$~(model IV)               & $4218.8\pm3.4\pm3.7$    &   $43.5\pm5.3\pm5.0$    &  11.7$\sigma$    \\
\cmidrule(rl){1-3}\cmidrule(rl){4-4}
  $Y(4220)^{\rm ave}$                    & $4225.7\pm4.1\pm3.4$     &  $57.5\pm9.4\pm12.1$     &   \\
\hline
\hline
\end{tabular}}
\label{FitCsYres}
\end{table*}

\par The systematic uncertainty of the fit comes mainly from the $\sqrt{s}$ calibration, the fit model, the parameters of $Y(4320)$, and PHSP factors. The total systematic uncertainty is given by assuming that these uncertainties are independent. The average mass and width of $Y(4220)$ are given by the processes $\PZC$ (in model I), $f_{0}(980)J/\psi$ (in model III), and ($M,~\Gamma$) = ($4225.8\pm4.1\pm3.4$~MeV/$c^{2}$, $57.5\pm9.4\pm12.1$~MeV).

\par To estimate the difference in the decay rates of $Y(4220)$ and $Y(4320)$ to $\pi^{\pm}Z_{\rm}(3900)^{\mp}$, $f_{0}(980)J/\psi$, and $(\pi^{+}\pi^{-})S_{\rm wave}J/\psi$, the ratios of $\Gamma_{ee}\mathcal{B}$ between $Y(4220)$ and $Y(4320)$ for different solutions based on the $Y(4220)$ with model IV are given, and the results are listed in Table~\ref{Pratio}. Here, $\Gamma_{ee}$ is the electronic width of the resonance, and $\mathcal{B}$ are the branching fractions for $Y(4220)/Y(4320)\rightarrow\pi^{\pm}Z_{c}(3900)^{\mp}(\rightarrow\pi^{\mp}J/\psi)$, $f_{0}(980)(\rightarrow\pi^{+}\pi^{-})J/\psi$, and $(\pi^{+}\pi^{-})_{\rm{S\mbox{-}wave}} J/\psi$. Because of the multiple solutions and large statistical uncertainties, it is hard to conclude whether the decay patterns of the $Y(4220)$ and $Y(4320)$ are different.

\begin{table*}
\caption{The ratios of $\Gamma_{ee}\mathcal{B}$ between $Y(4220)$ and $Y(4320)$ for different solutions. Here we only show the $(\Gamma_{ee}\mathcal{B})_{n}$ value of $Y(4220)$, and $n=1, 2, 3$ correspond to three different sets of solutions. ``$R_{n}$" are the ratios of $(\Gamma_{ee}\mathcal{B})_{Y(4320)}/(\Gamma_{ee}\mathcal{B})_{Y(4220)}$.}
\centering
\renewcommand\arraystretch{1.3}
\renewcommand\tabcolsep{4.0pt}
\scalebox{1.0}{
\begin{tabular}{lllllll}
\hline
\hline
   Process  &  $(\Gamma_{ee}\mathcal{B})_{1}$  &  $R_{1}$   &  $(\Gamma_{ee}\mathcal{B})_{2}$    &   $R_{2}$  &  $(\Gamma_{ee}\mathcal{B})_{3}$  & $R_{3}$     \\
\hline
  $\pi^{\pm}Z_{\rm}(3900)^{\mp}$    &  $1.2\pm0.2$ & $0.4\pm0.5$ &  $0.2\pm0.1$  & $0.8\pm1.3$ & $0.3\pm0.1$ & $3.4\pm2.1$       \\
  $f_{0}(980)J/\psi$                &  $3.6\pm0.5$  & $1.0\pm0.4$   &  $0.5\pm0.2$ & $1.8\pm2.2$  & $0.5\pm0.2$  & $3.3\pm2.5$      \\
  $(\pi^{+}\pi^{-})S_{\rm wave}J/\psi$ &  $1.2\pm0.1$ & $0.5\pm0.3$ &  $7.2\pm0.3$  & $0.1\pm0.1$ & $1.9\pm0.2$ & $7.7\pm1.0$     \\
\hline
\hline
\end{tabular}}
\label{Pratio}
\end{table*}

\section{Summary}
\par In summary, a PWA is performed on the ${\EEPPJ}$ process to measure the mass and width of the $\ZCP$ by a simultaneous fit to data at multiple energy points. The cross sections of $\EE$ to different subprocesses at different energy points are measured. Different parametrizations are used to describe the $\PPS$, and consistent cross section results of $\EE \rightarrow {\pi^{\pm}Z_{c}(3900)^{\mp}(\rightarrow\pi^{\mp}J/\psi)}$ and $\EE \rightarrow (\pi^{+}\pi^{-})_{\rm{S\mbox{-}wave}} J/\psi$ processes are obtained. The $\ZCP$ Breit-Wigner mass and width are ($M,~\Gamma$) = ($3884.6\pm0.7\pm3.3$~MeV/$c^{2}$, $37.2\pm1.3\pm6.6$~MeV). The parameters of $\ZCP$ are consistent with the CLEO-c measurements~\cite{Xiao:2013iha} and $Z_{c}(3885)^{\pm}$ in the $(D\bar{D}^{\ast})^{\pm}$ system~\cite{BESIII:2013qmu}.

\par From the fit results, the $\PPS J/\psi$ amplitude dominates the process $\EEPPJ$. The cross sections with $Y(4220)$ and $Y(4320)$ decaying in different final states do appear to have different distributions. However, due to statistical uncertainties and multiple solutions, it is difficult to conclude that the decay modes of $Y(4220)$ and $Y(4320)$ are different. In particular, the process ${\EE \rightarrow f_{0}(980)J/\psi}$ can be well described by only one BW function with parameters consistent with those of the $Y(4220)$. Finally, the mass and width of $Y(4220)$ given by the fit of the cross sections of these subprocesses are ($M,~\Gamma$) = ($4225.7\pm4.1\pm3.4$~MeV/$c^{2}$, $57.5\pm9.4\pm12.1$~MeV).

\acknowledgements
BESIII Collaboration thanks the staff of BEPCII and the IHEP computing center for their strong support. This work is supported in part by the National Key R\&D Program of China under Contracts No. 2020YFA0406300, No.2020YFA0406400, and No.2023YFA1606000; the National Natural Science Foundation of China (NSFC) under Contracts No. 11635010, No.11735014, No.11835012, No.11935015, No.11935016, No.11935018, No.11961141012, No.12025502, No.12035009, No.12035013, No.12061131003, No.12192260, No.12192261, No.12192262, No.12192263, No.12192264, No.12192265, No.12221005, No.12225509, and No.12235017; the Chinese Academy of Sciences (CAS) Large-Scale Scientific Facility Program; the CAS Center for Excellence in Particle Physics (CCEPP); Joint Large-Scale Scientific Facility Funds of the NSFC and CAS under Contract No. U1832207; the CAS Key Research Program of Frontier Sciences under Contracts No. QYZDJ-SSW-SLH003, and No. QYZDJ-SSW-SLH040; the 100 Talents Program of CAS; Hubei University of Automotive Technology under Contract No. BK202318; the Institute of Nuclear and Particle Physics (INPAC) and Shanghai Key Laboratory for Particle Physics and Cosmology; the Fundamental Research Funds of Shandong University; the European Union's Horizon 2020 research and innovation programme under Marie Sklodowska-Curie grant agreement under Contract No. 894790; the German Research Foundation DFG under Contracts No. 455635585, Collaborative Research Center CRC 1044, FOR5327, GRK 2149; Istituto Nazionale di Fisica Nucleare, Italy; the Ministry of Development of Turkey under Contract No. DPT2006K-120470; the National Research Foundation of Korea under Contract No. NRF-2022R1A2C1092335; the National Science and Technology fund of Mongolia; the National Science Research and Innovation Fund (NSRF) via the Program Management Unit for Human Resources and Institutional Development, Research and Innovation of Thailand under Contract No. B16F640076; the Polish National Science Centre under Contract No. 2019/35/O/ST2/02907; the Swedish Research Council; the U. S. Department of Energy under Contract No. DE-FG02-05ER41374.

\section{Data availability}
The data that support the findings of this article are openly available~\cite{DataAvailability}, embargo periods may apply.

\newpage

\begin{center}
\section*{Appendix}
\end{center}

\setcounter{table}{0}
\renewcommand{\thetable}{T\arabic{table}}
\setcounter{figure}{0}
\renewcommand{\thefigure}{F\arabic{figure}}

\par
\begin{figure*}
\centering
  \includegraphics[width=0.45\textwidth, height=0.3\textwidth]{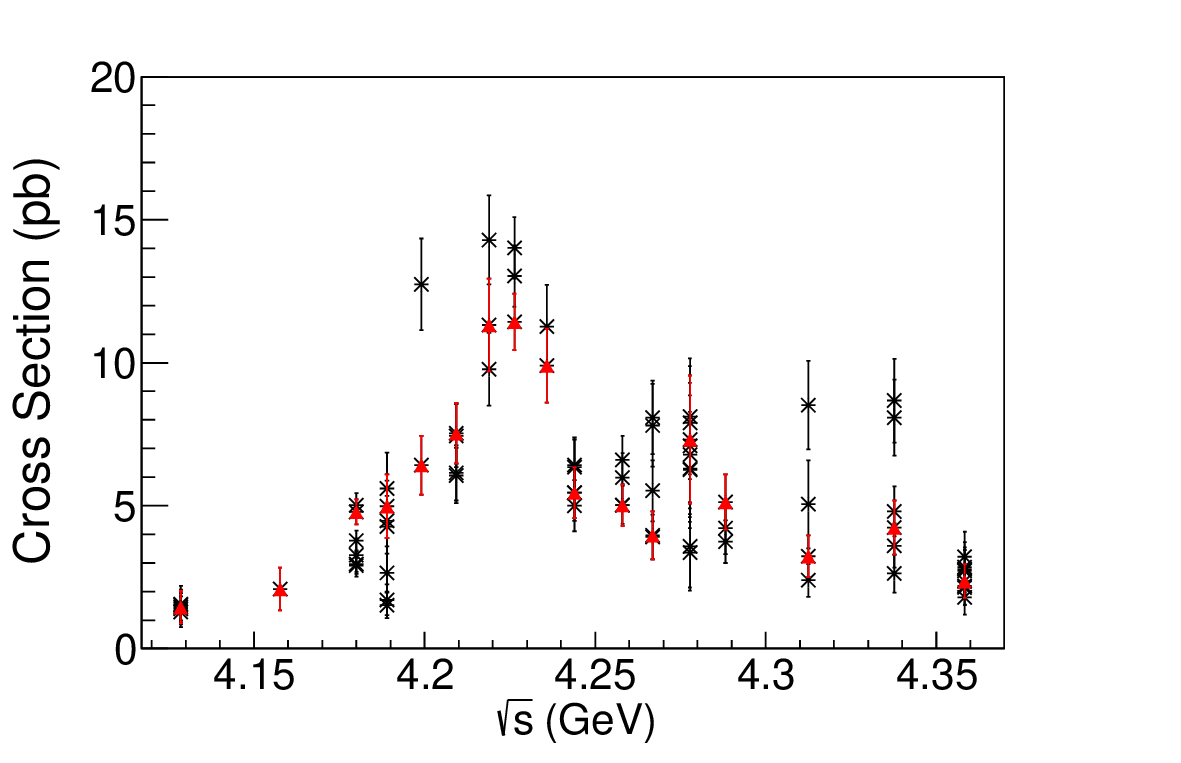}\hspace{1pt}
  \includegraphics[width=0.45\textwidth, height=0.3\textwidth]{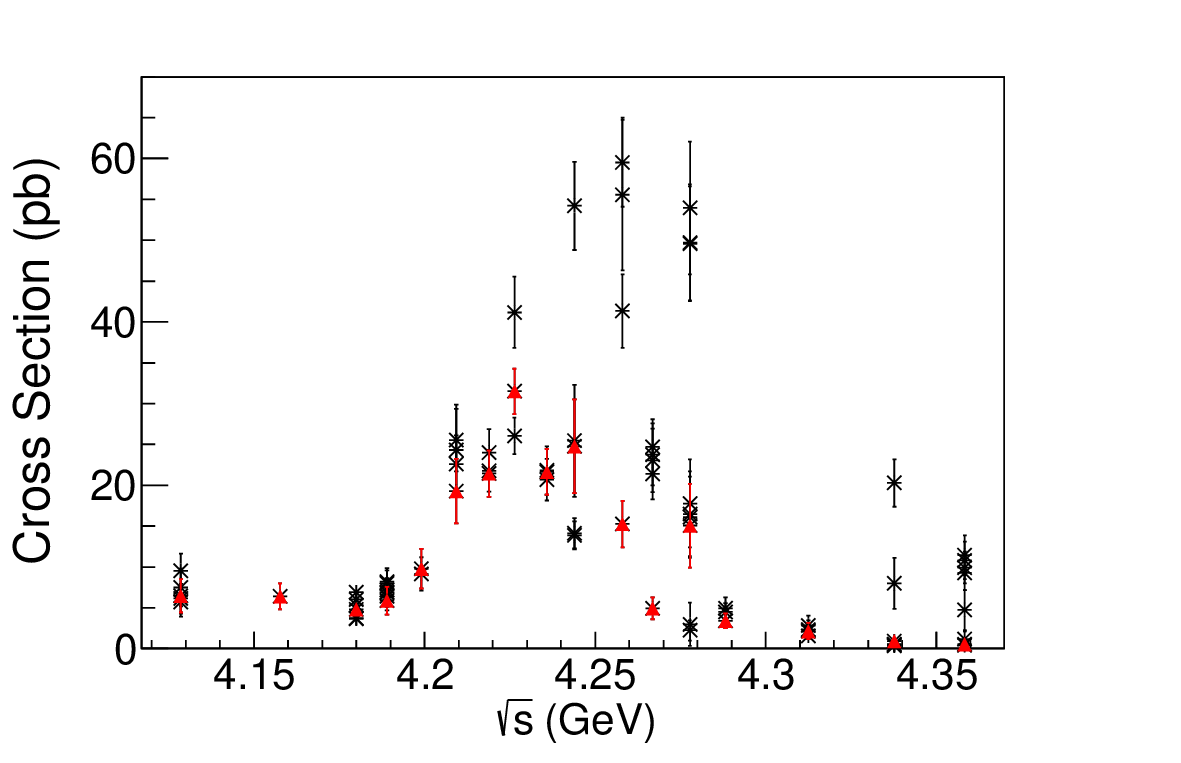}\hspace{1pt}
  \includegraphics[width=0.45\textwidth, height=0.3\textwidth]{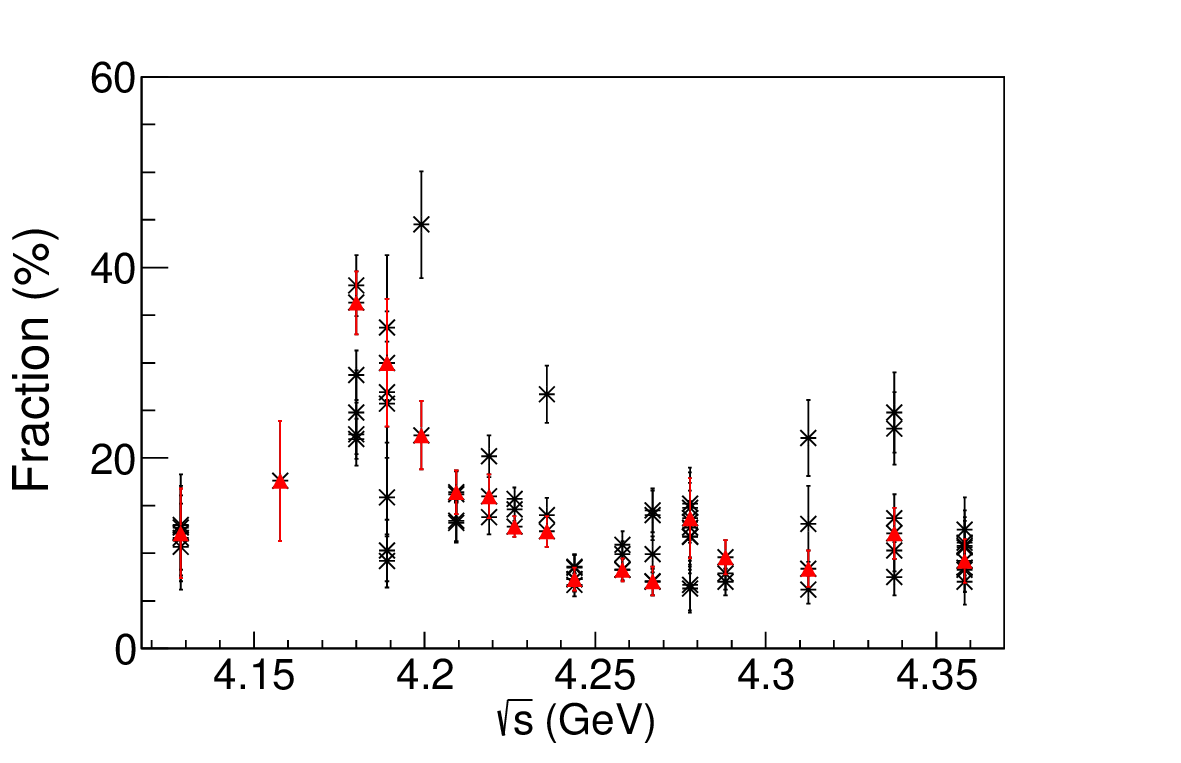}\hspace{1pt}
  \includegraphics[width=0.45\textwidth, height=0.3\textwidth]{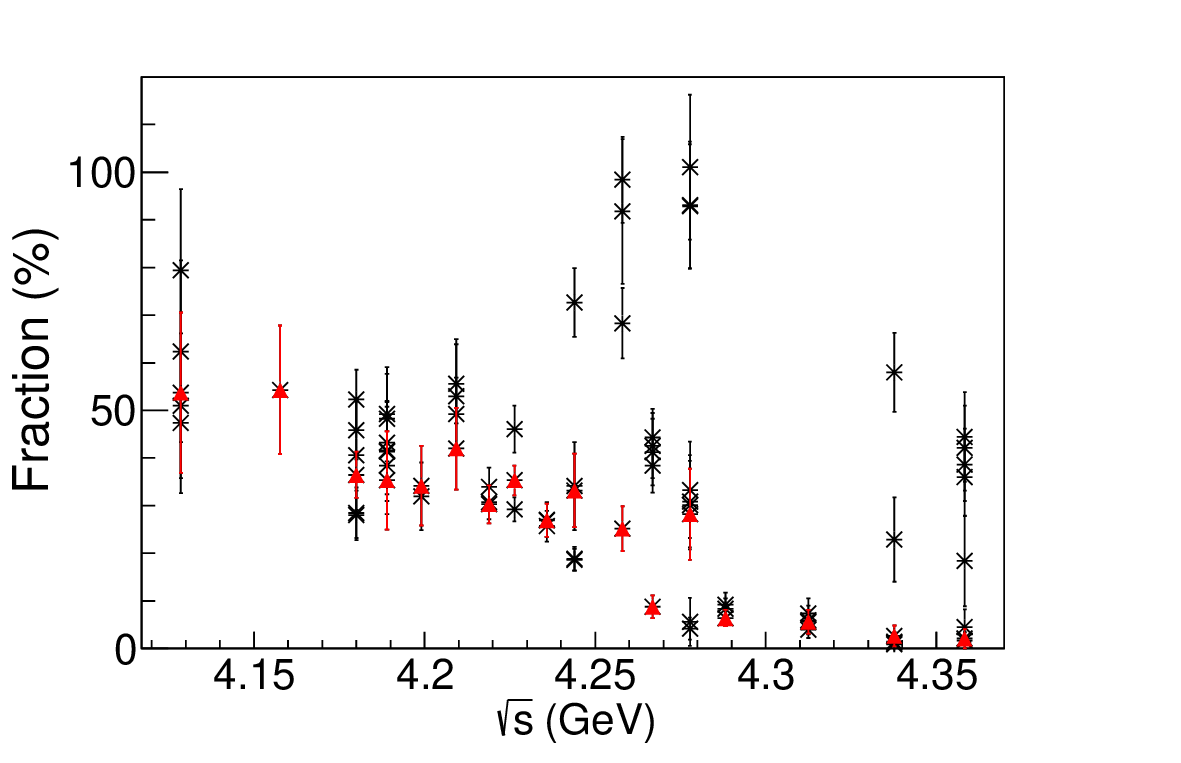}\hspace{1pt}
  \put(-360, 150){(a)}
  \put(-130, 150){(b)}
  \put(-360, -6){(c)}
  \put(-130, -6){(d)}
\caption{\footnotesize{The cross sections and fractions of $\EE\rightarrow\pi^{+}Z_{c}(3900)^{-}+c.c.\rightarrow\pi^{+}\pi^{-}J/\psi$ and $f_{0}(980)(\rightarrow\pi^{+}\pi^{-})J/\psi$ given by 10,000 random initialization fits at each energy point.} Figures (a) and (c) are the cross sections and fractions distributions of $\pi^{\pm}Z_{c}(3900)^{\mp}+c.c.$, respectively. Figures (b) and (d) are the cross section and fraction distributions of $f_{0}(980)J/\psi$, respectively. The red marks correspond to the solution with the minimum FCN value at each energy point. The corrections in Table~\ref{MWCorrection} are not considered here.}
\label{Mult_Solution_CS_Fractions}
\end{figure*}

 Figure~\ref{Mult_Solution_CS_Fractions} shows the cross sections and fractions of $\pi^{\pm}Z_{c}(3900)^{\mp}$ and $f_{0}(980)J/\psi$ corresponding to different FCN values that appeared in 10,000 random initialization fits at each energy point, respectively. Some solutions with the same FCN but fewer cumulative times, such as less than 100 times, are not shown. The cumulative number of solutions of the same FCN and the fractions of the subprocesses of these solutions are listed in Tables~\ref{FCNallBW1} and~\ref{FCNallBW2}.

\par Tables~\ref{totfraction} and ~\ref{totCrosssection} list the fractions and cross sections of $\EE\rightarrow\pi^{\pm}Z_{c}(3900)^{\mp}(\rightarrow\pi^{\mp}J/\psi)$, $f_{0}(980)(\rightarrow\pi^{+}\pi^{-})J/\psi$, and $(\pi^{+}\pi^{-})_{\rm{S\mbox{-}wave}} J/\psi$ at different energy points by using the Breit-Wigner and K-matrix methods to parametrize the $\PPS$ in the PWA of $\EEPPJ$. Table~\ref{InterferenceTerm4230} lists the interference magnitude between each subprocess at $\sqrt{s} = 4.2263$ and 4.2580 GeV.

\par
\begin{figure*}
\centering
  \includegraphics[width=0.45\textwidth, height=0.21\textwidth]{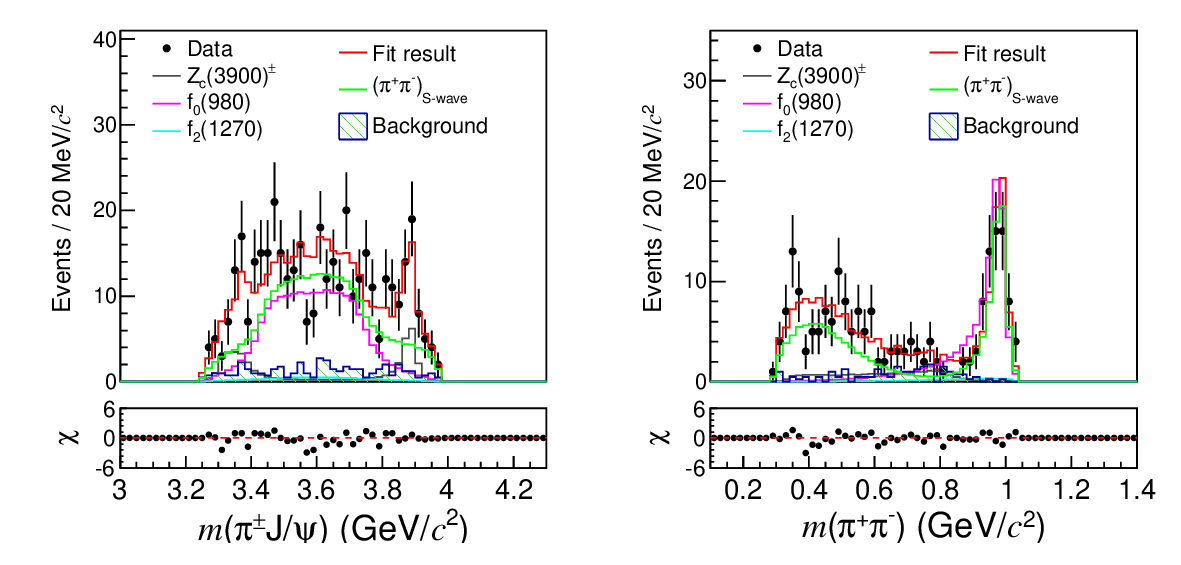}
  \includegraphics[width=0.45\textwidth, height=0.21\textwidth]{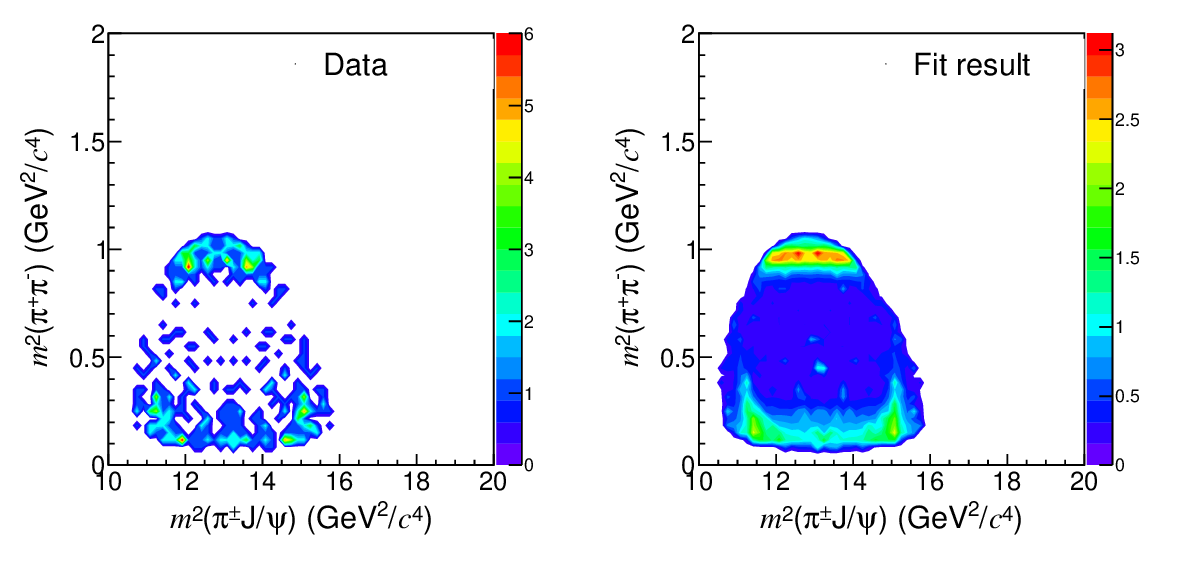}
  \includegraphics[width=0.45\textwidth, height=0.21\textwidth]{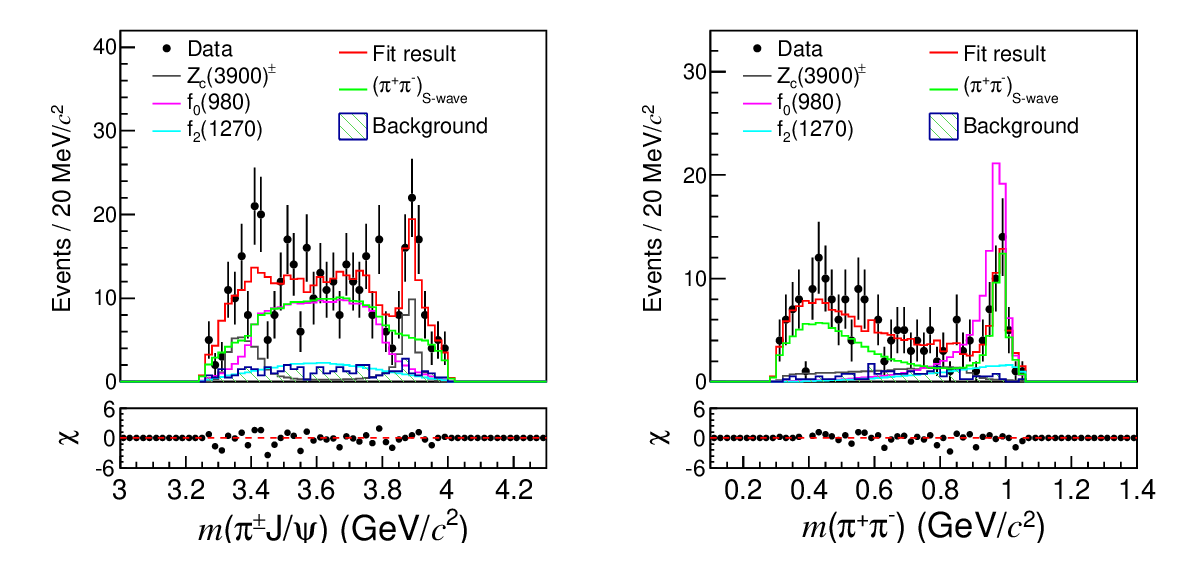}
  \includegraphics[width=0.45\textwidth, height=0.21\textwidth]{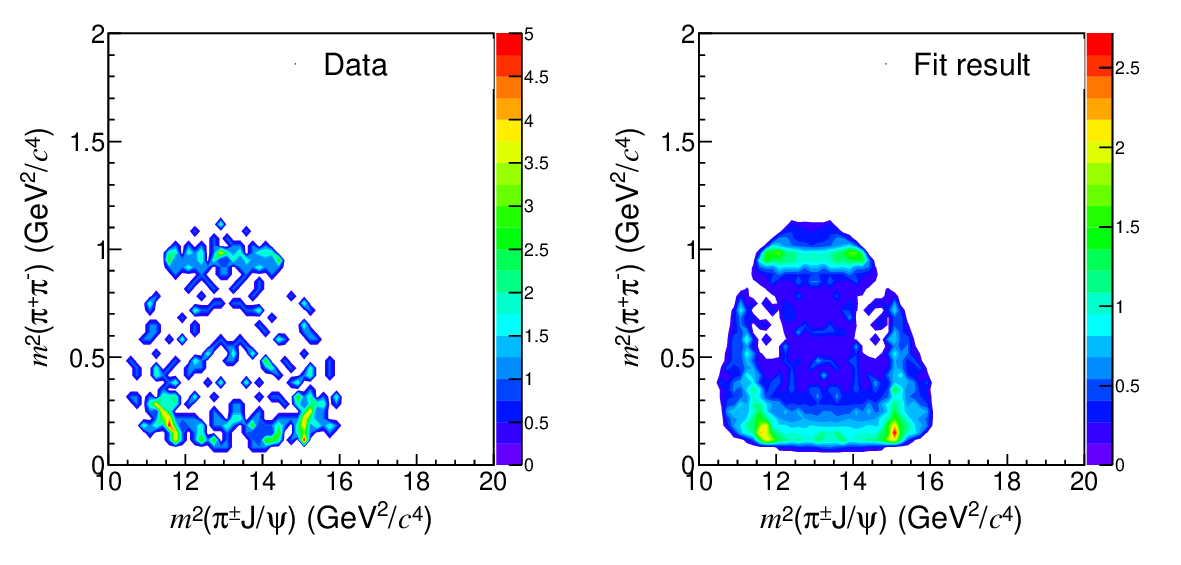}
  \includegraphics[width=0.45\textwidth, height=0.21\textwidth]{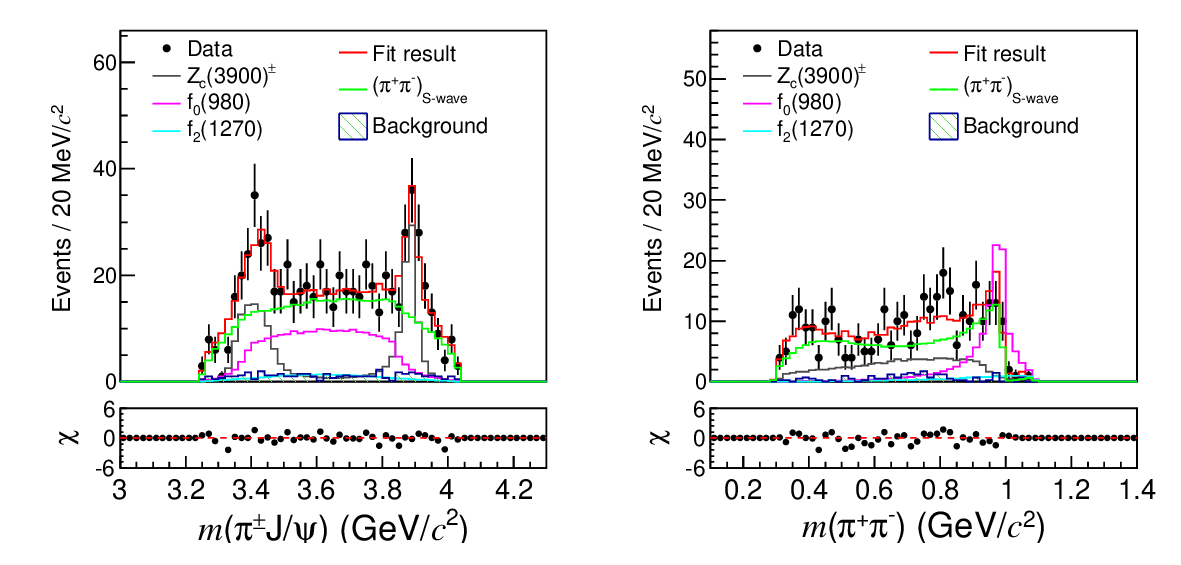}
  \includegraphics[width=0.45\textwidth, height=0.21\textwidth]{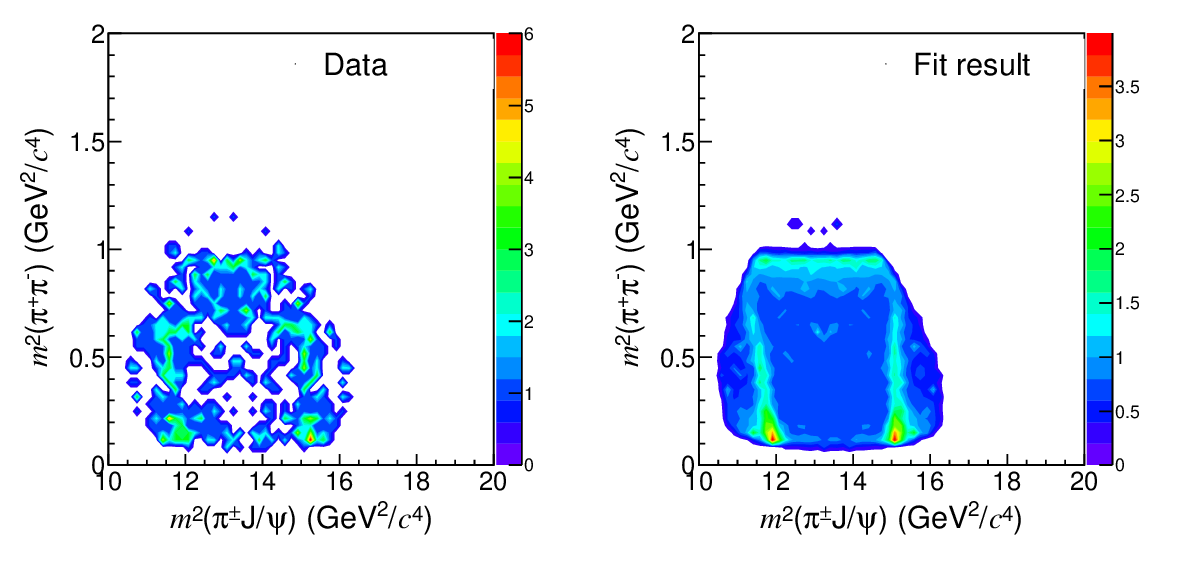}
  \includegraphics[width=0.45\textwidth, height=0.21\textwidth]{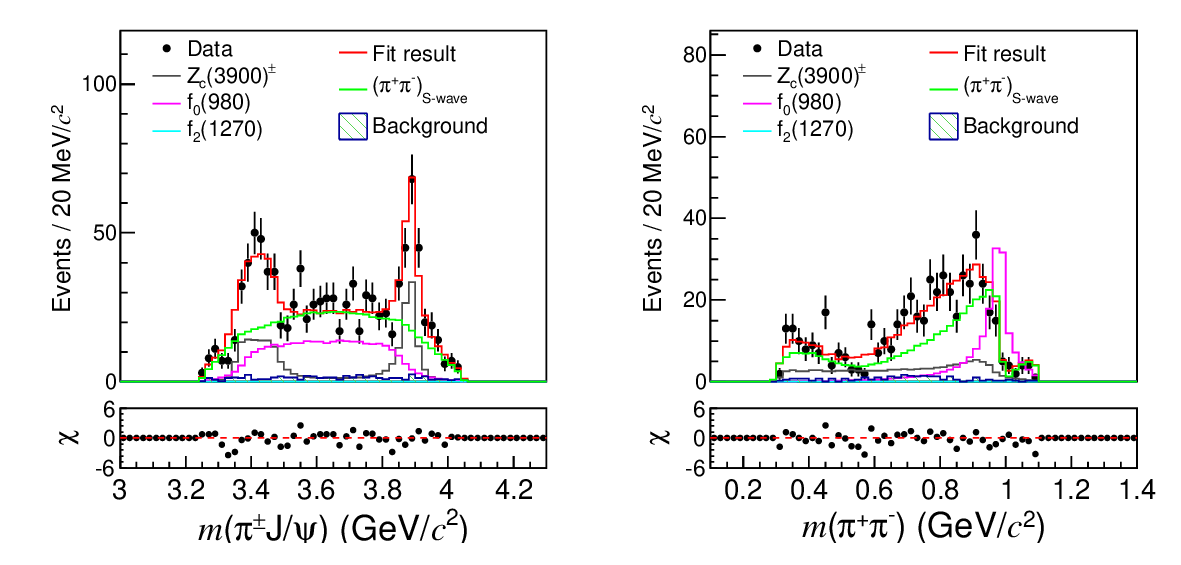}
  \includegraphics[width=0.45\textwidth, height=0.21\textwidth]{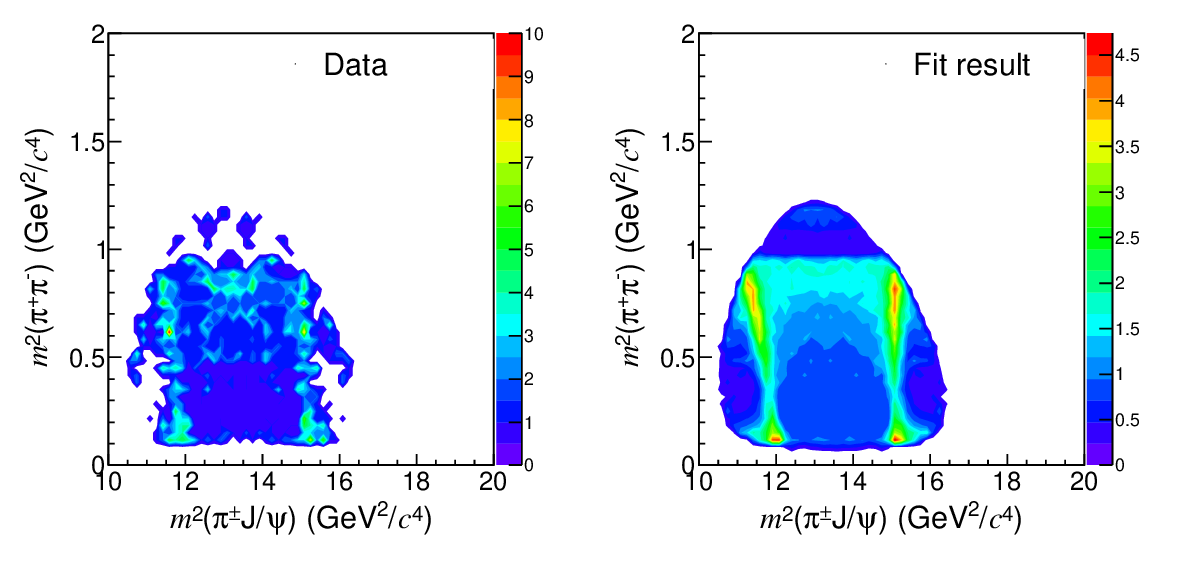}
\caption{\footnotesize{The PWA fit results of $\sqrt{s}=$4.1271 (1st row), 4.1567 (2nd row), 4.1888 (3rd row), 4.1989 (4th row) GeV samples.} The plots show the distributions of $m(\pi^{\pm}J/\psi)$ (1st column), $m(\pi^{+}\pi^{-})$ (2nd column), and Dalitz plots from data (3rd column) and the fit result (4th column). The $\chi^{2}/ndf$ for each term are listed in Table~\ref{chisqvsndf1}. }
\label{PWAresultTotal1}
\end{figure*}

\begin{figure*}
\centering
  \includegraphics[width=0.45\textwidth, height=0.21\textwidth]{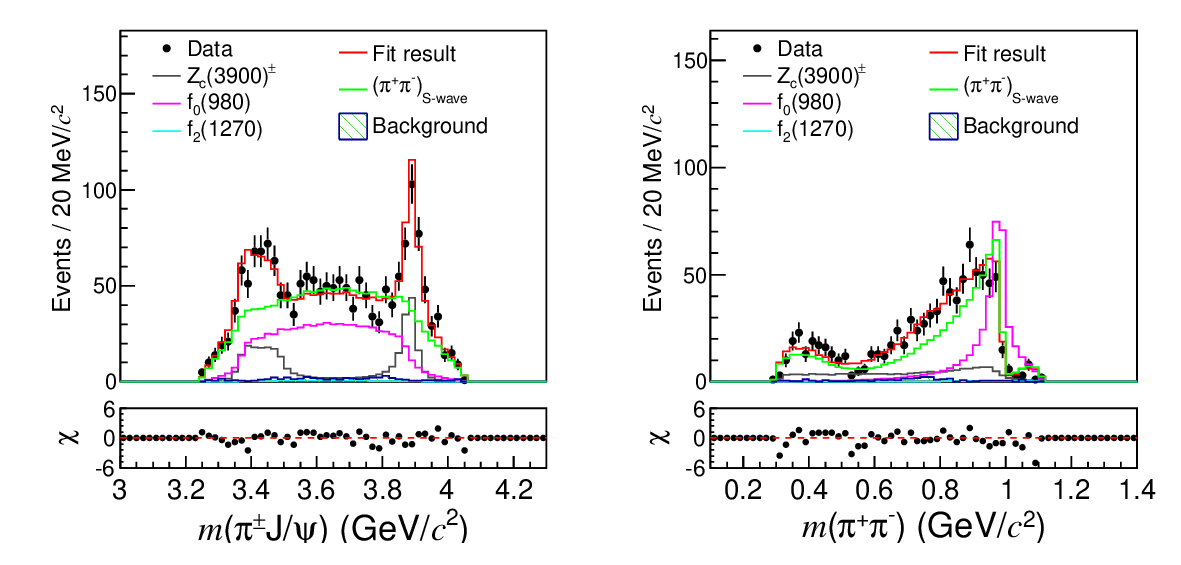}
  \includegraphics[width=0.45\textwidth, height=0.21\textwidth]{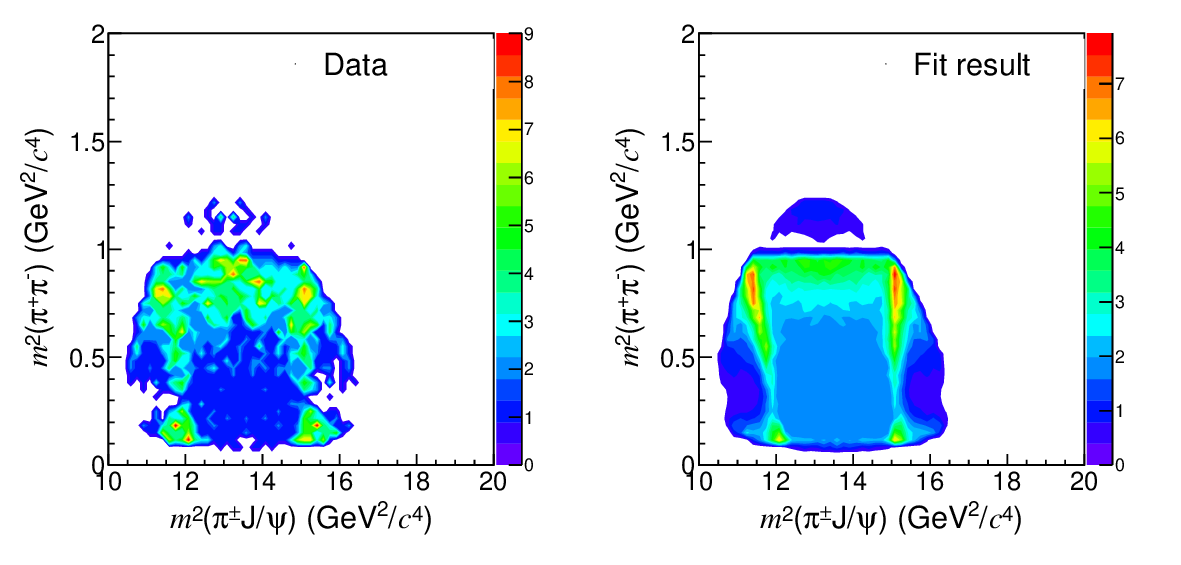}
  \includegraphics[width=0.45\textwidth, height=0.21\textwidth]{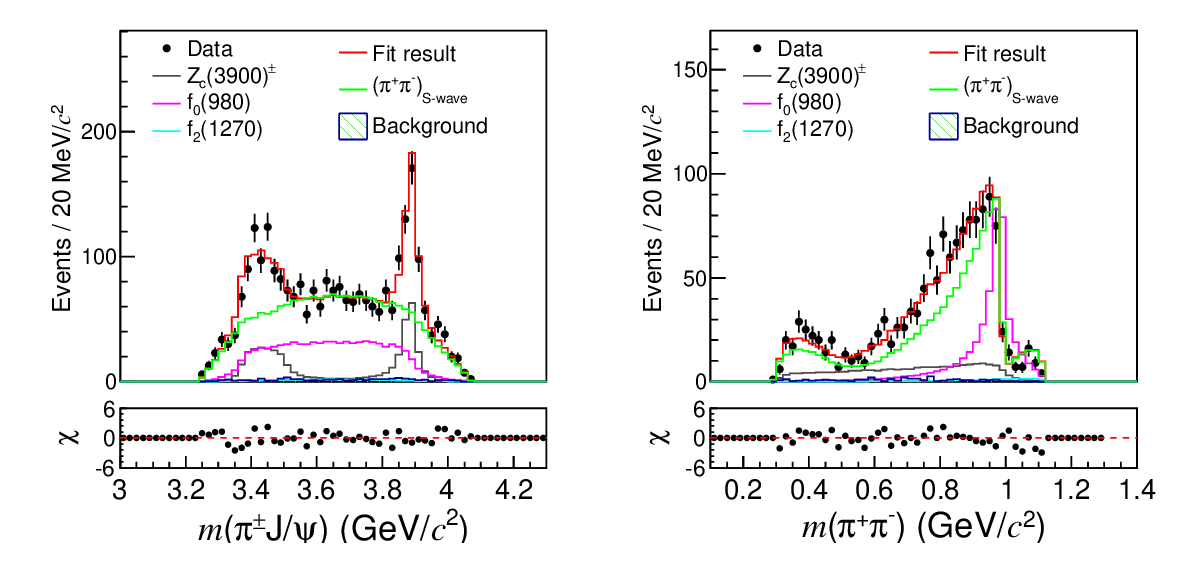}
  \includegraphics[width=0.45\textwidth, height=0.21\textwidth]{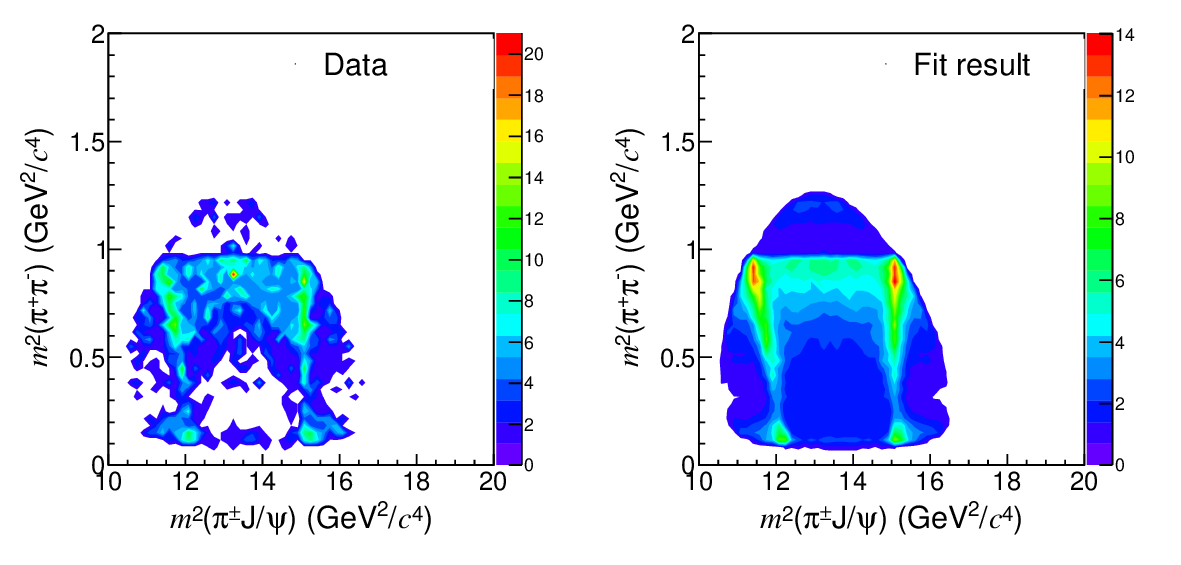}
  \includegraphics[width=0.45\textwidth, height=0.21\textwidth]{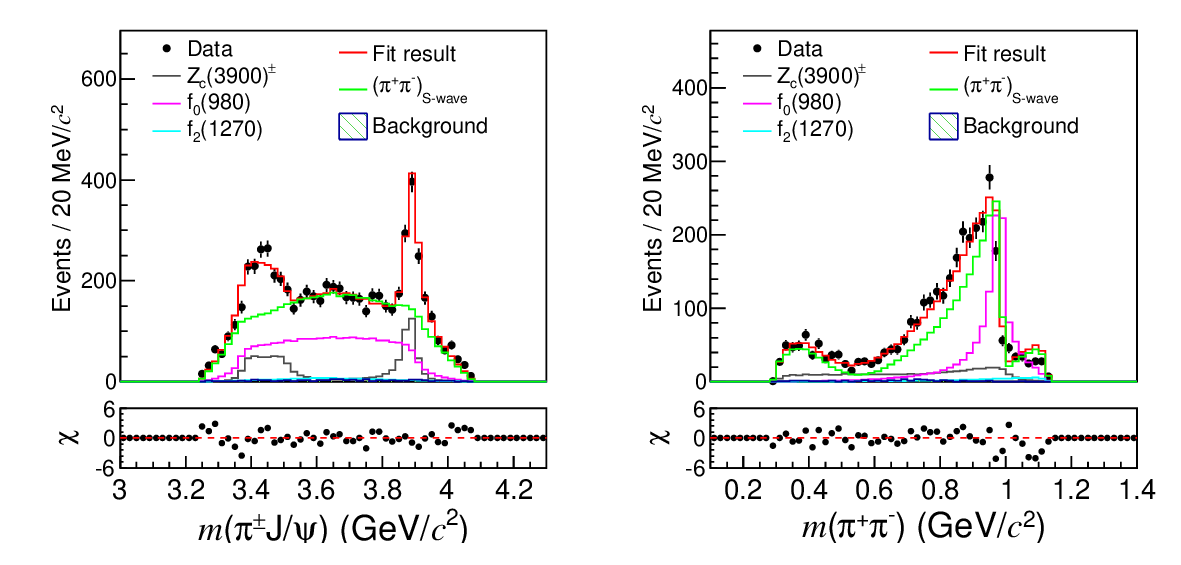}
  \includegraphics[width=0.45\textwidth, height=0.21\textwidth]{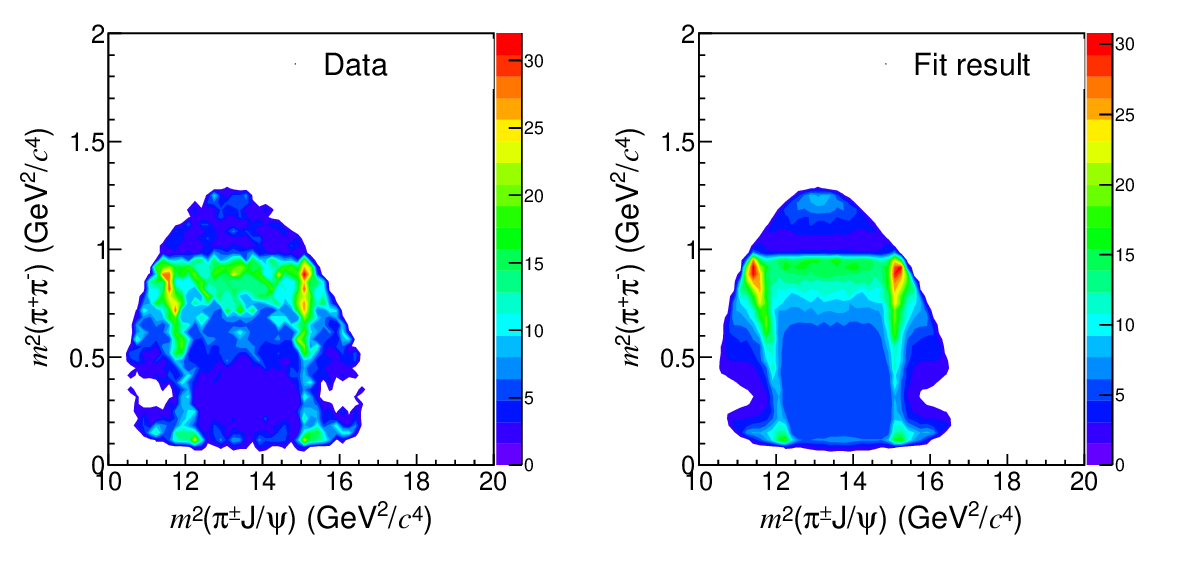}
  \includegraphics[width=0.45\textwidth, height=0.21\textwidth]{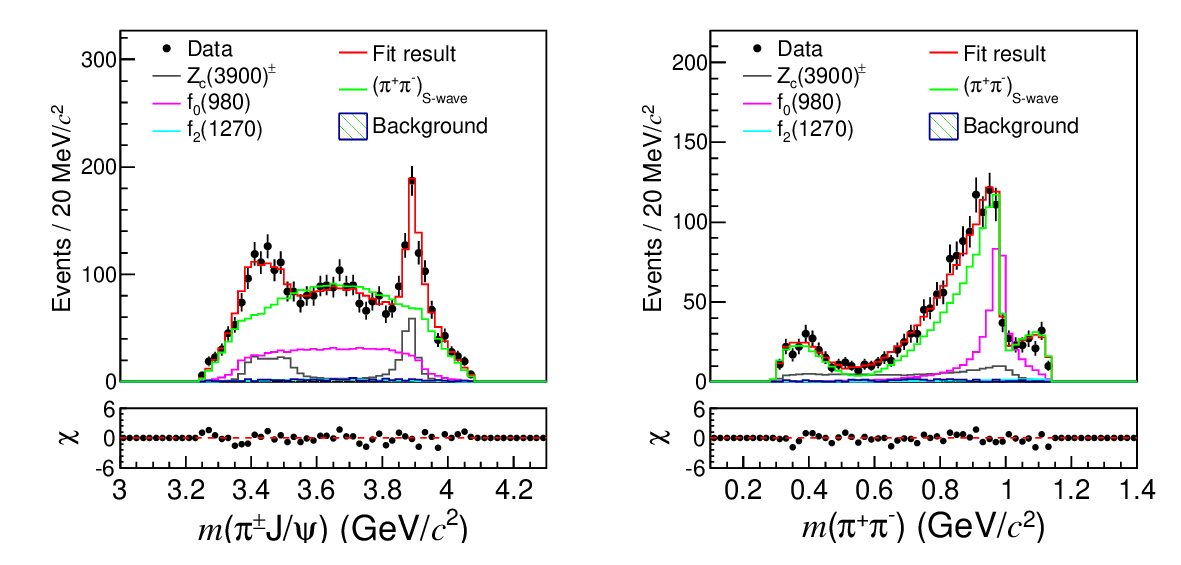}
  \includegraphics[width=0.45\textwidth, height=0.21\textwidth]{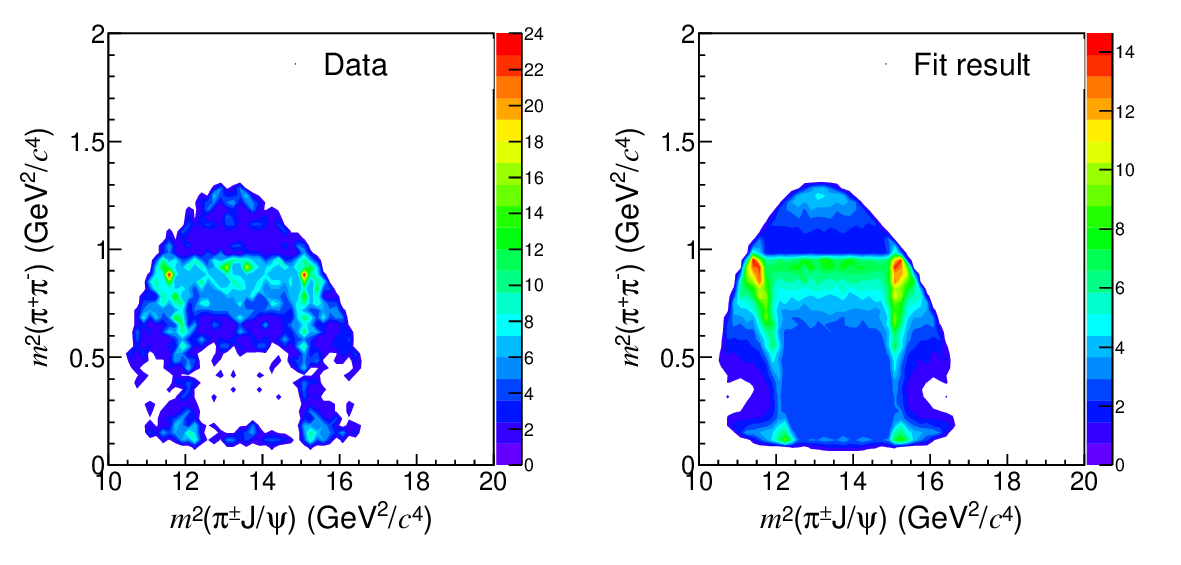}
\caption{\footnotesize{The PWA fit results of $\sqrt{s}=$4.2091 (1st row), 4.2187 (2nd row), 4.2263 (3rd row), 4.2357 (4th row) GeV samples.} The plots show the distributions of $m(\pi^{\pm}J/\psi)$ (1st column), $m(\pi^{+}\pi^{-})$ (2nd column), and Dalitz plots from data (3rd column) and the fit result (4th column). The $\chi^{2}/ndf$ for each term are listed in Table~\ref{chisqvsndf1}.}
\label{PWAresultTotal2}
\end{figure*}

\begin{figure*}
\centering
  \includegraphics[width=0.45\textwidth, height=0.21\textwidth]{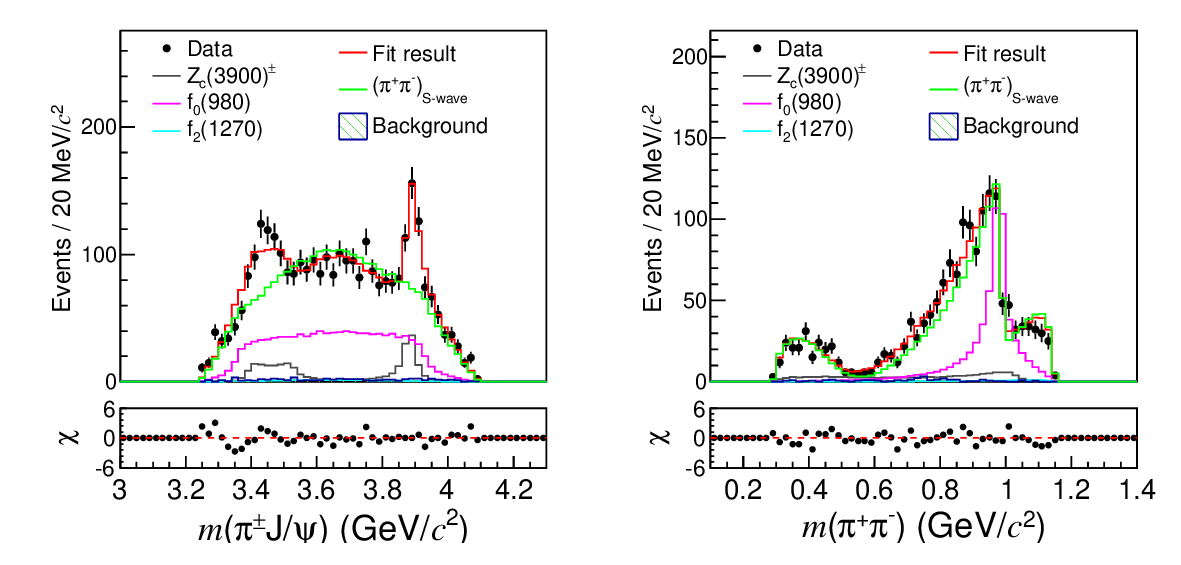}
  \includegraphics[width=0.45\textwidth, height=0.21\textwidth]{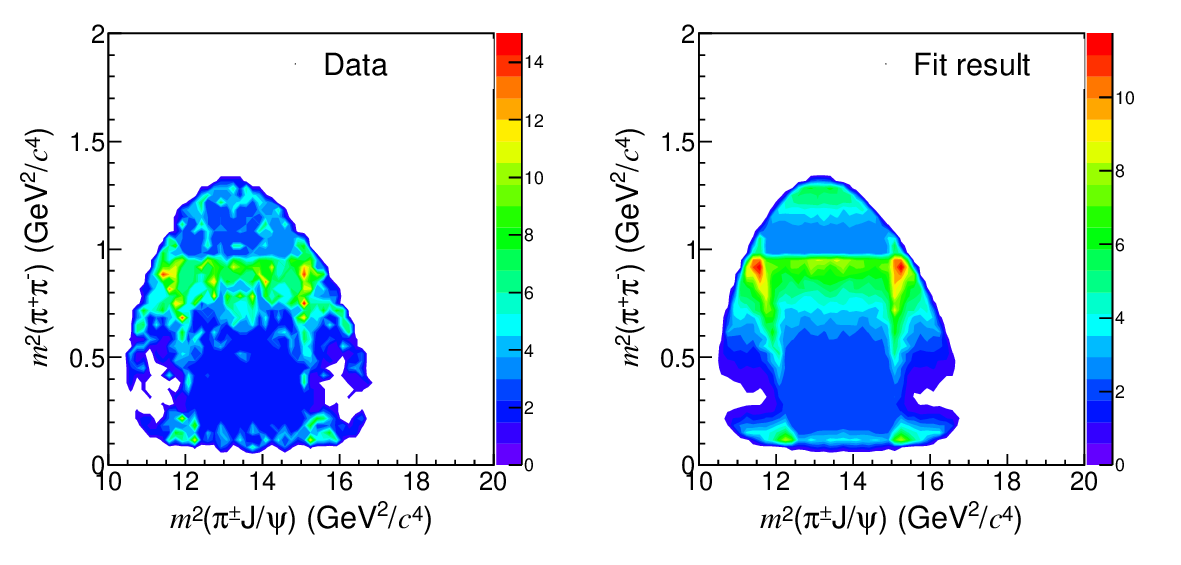}
  \includegraphics[width=0.45\textwidth, height=0.21\textwidth]{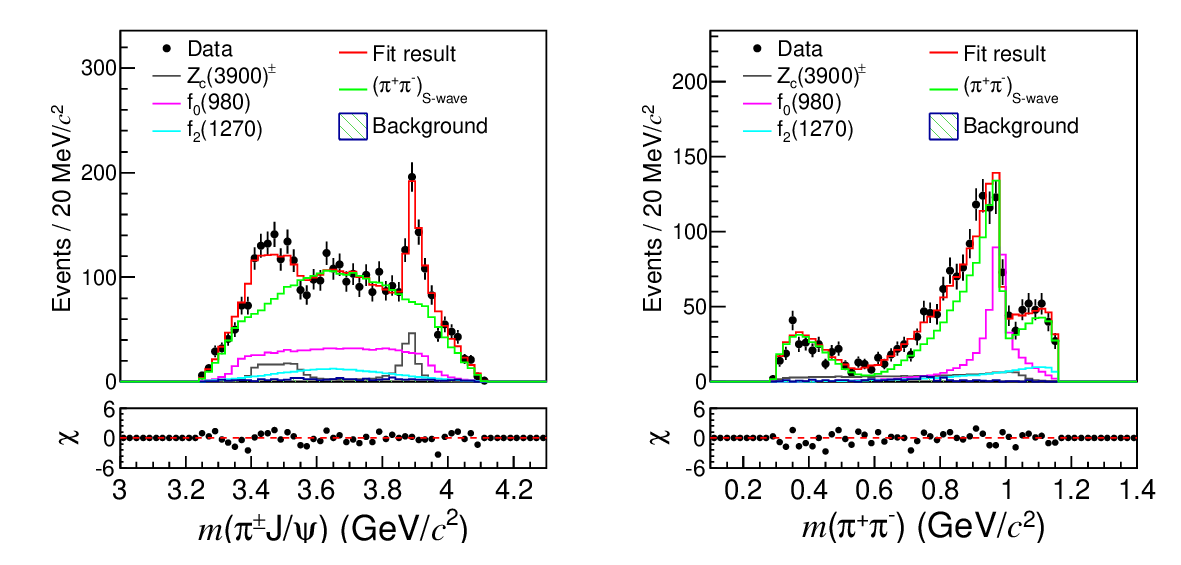}
  \includegraphics[width=0.45\textwidth, height=0.21\textwidth]{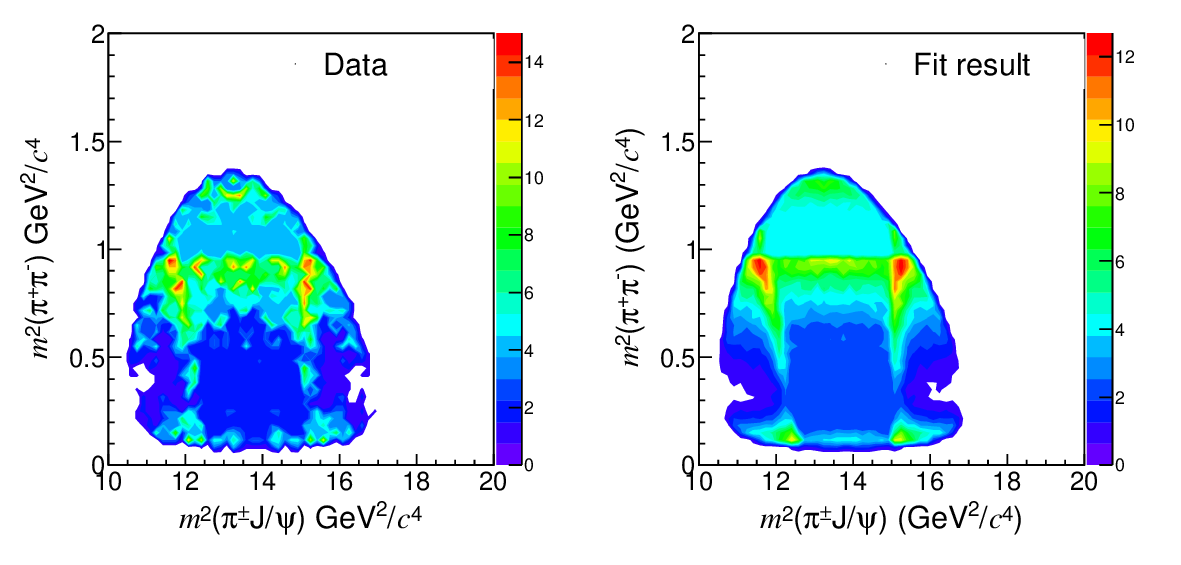}
  \includegraphics[width=0.45\textwidth, height=0.21\textwidth]{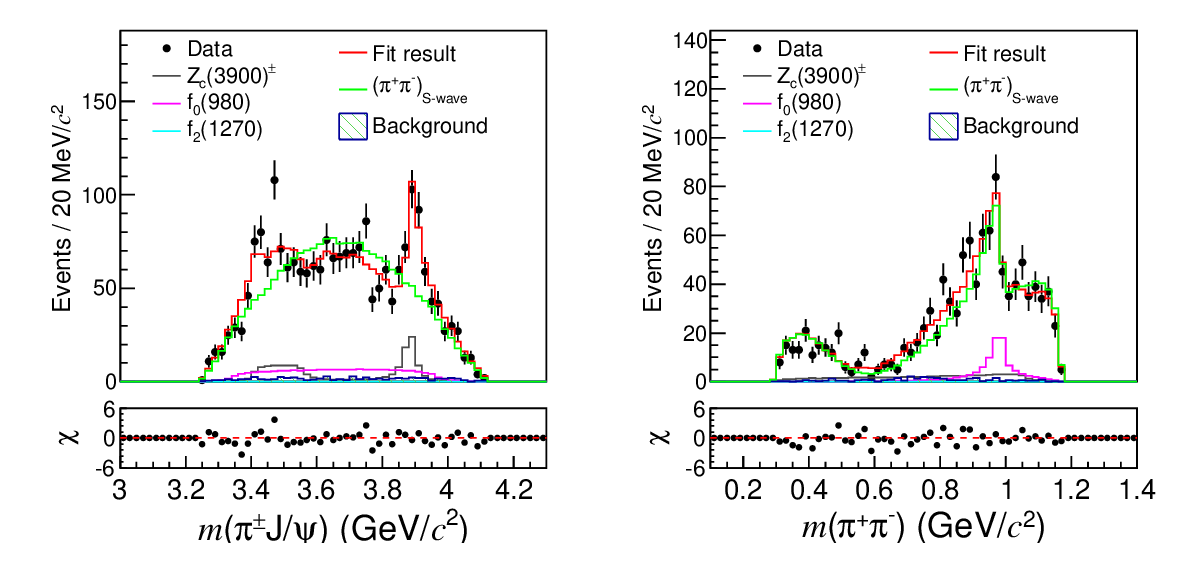}
  \includegraphics[width=0.45\textwidth, height=0.21\textwidth]{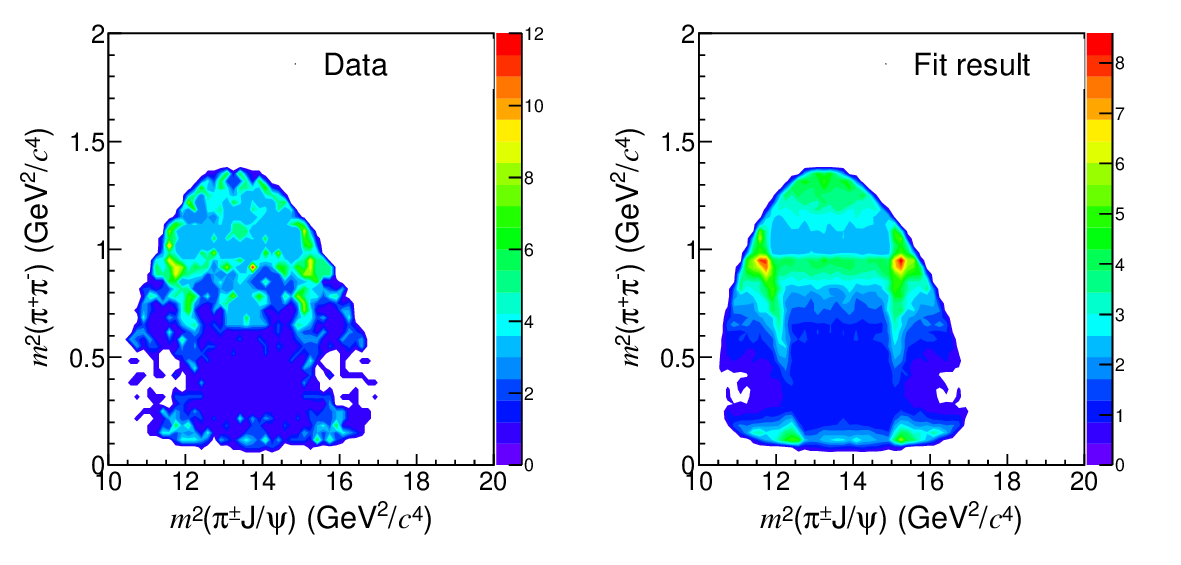}
  \includegraphics[width=0.45\textwidth, height=0.21\textwidth]{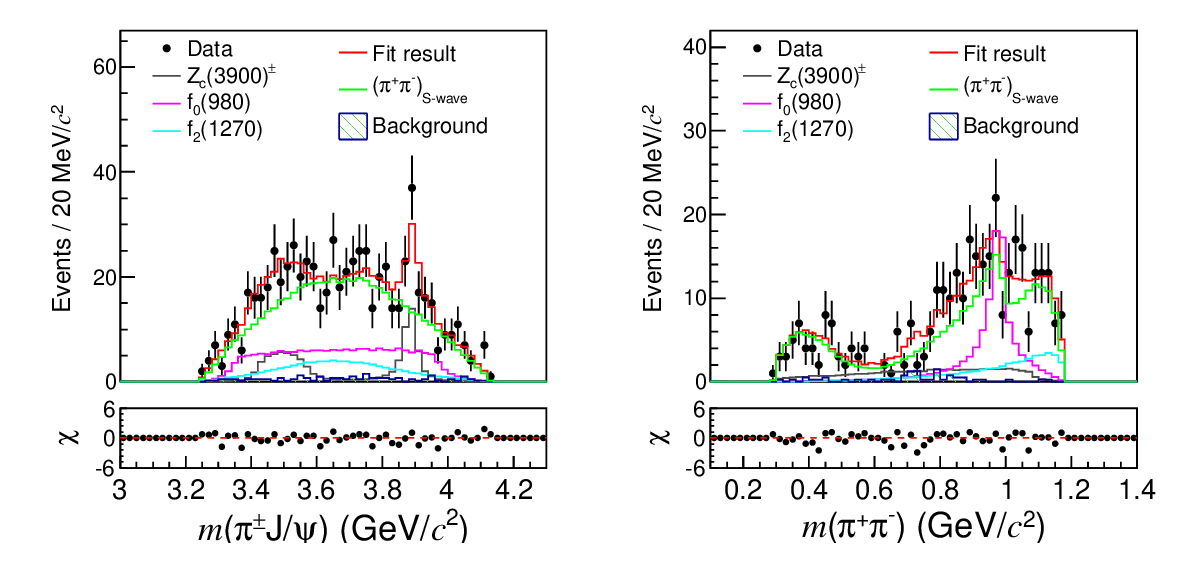}
  \includegraphics[width=0.45\textwidth, height=0.21\textwidth]{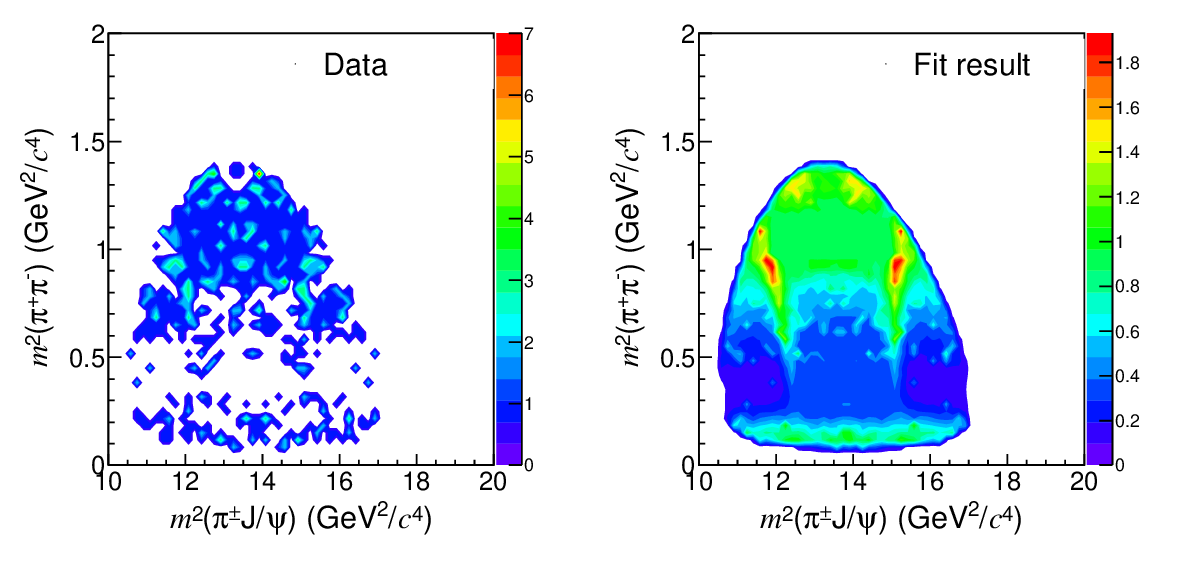}
\caption{\footnotesize{The PWA fit results of $\sqrt{s}=$4.2438 (1st row), 4.2580 (2nd row), 4.2667 (3rd row), 4.2776 (4th row) GeV samples.} The plots show the distributions of $m(\pi^{\pm}J/\psi)$ (1st column), $m(\pi^{+}\pi^{-})$ (2nd column), and Dalitz plots from data (3rd column) and the fit result (4th column). The $\chi^{2}/ndf$ for each term are listed in Table~\ref{chisqvsndf1}. }
\label{PWAresultTotal3}
\end{figure*}

\begin{figure*}
\centering
  \includegraphics[width=0.45\textwidth, height=0.21\textwidth]{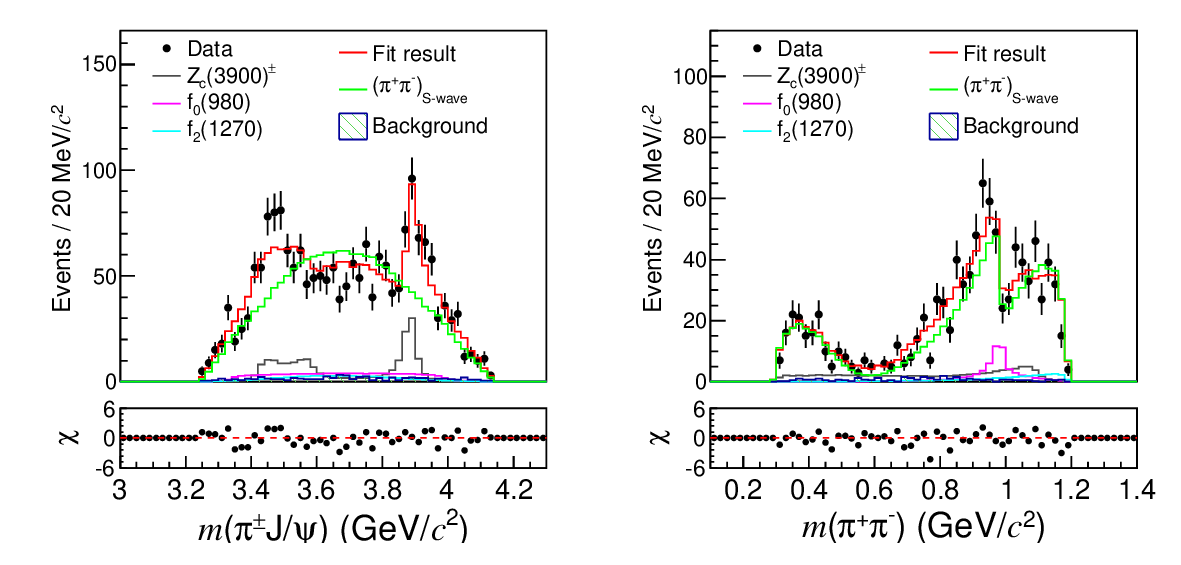}
  \includegraphics[width=0.45\textwidth, height=0.21\textwidth]{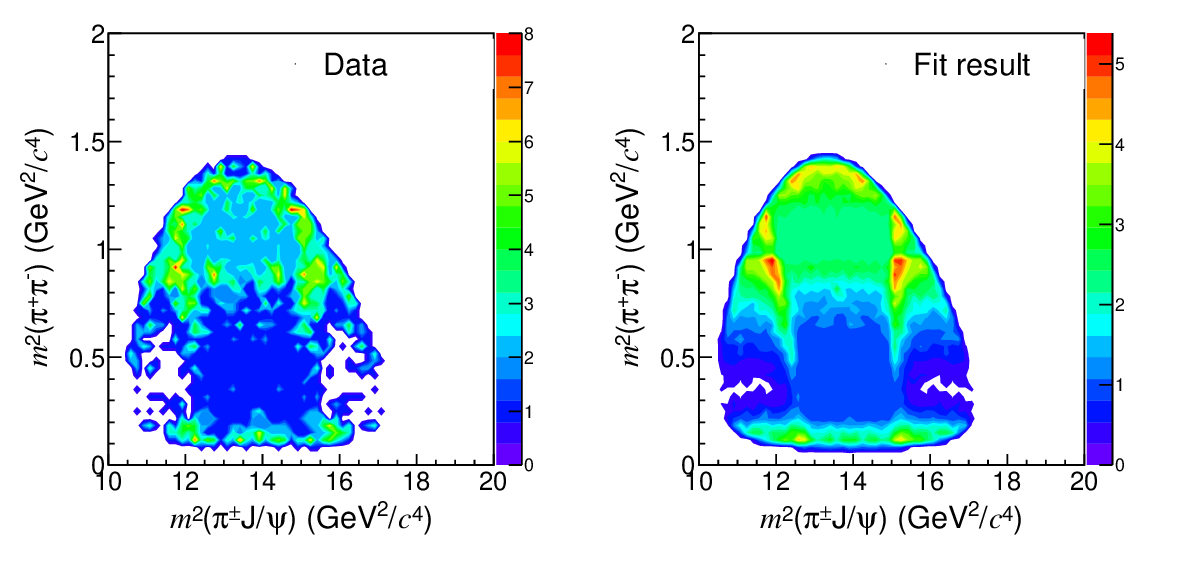}
  \includegraphics[width=0.45\textwidth, height=0.21\textwidth]{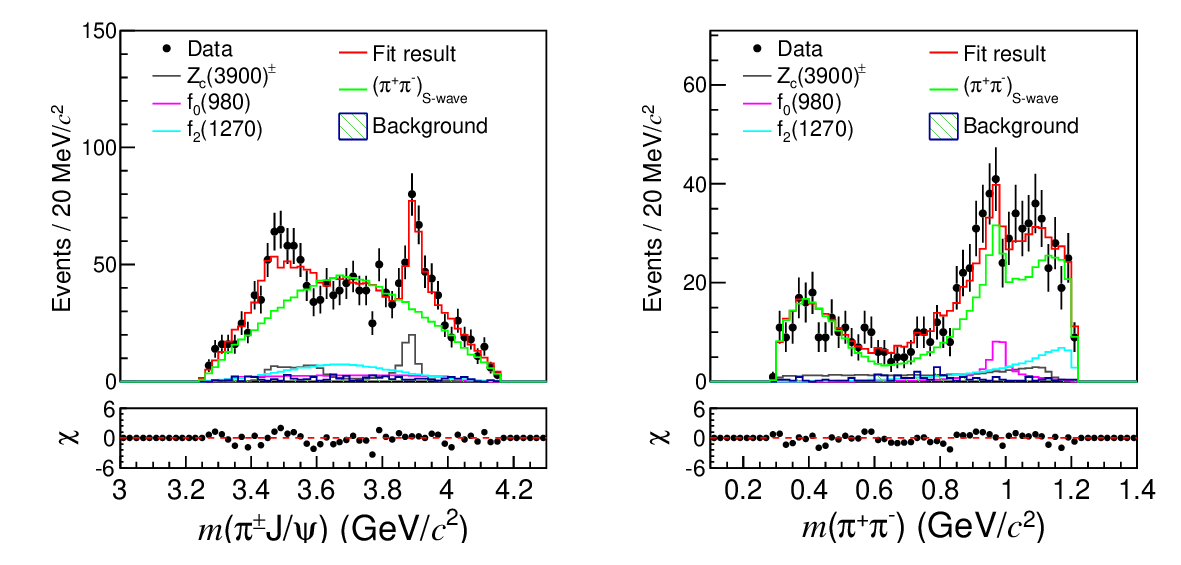}
  \includegraphics[width=0.45\textwidth, height=0.21\textwidth]{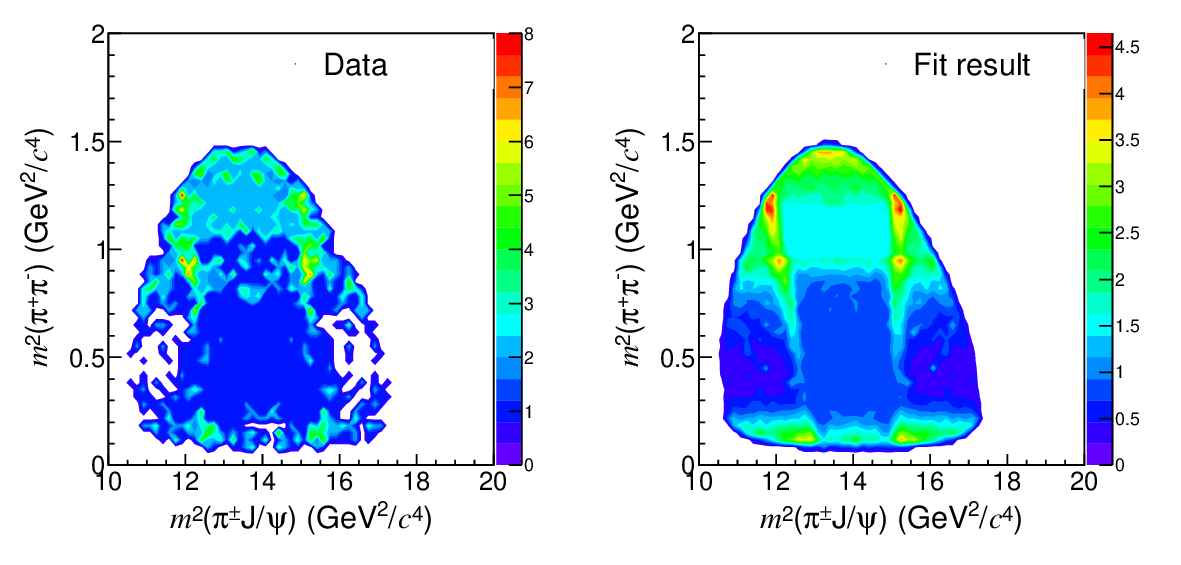}
  \includegraphics[width=0.45\textwidth, height=0.21\textwidth]{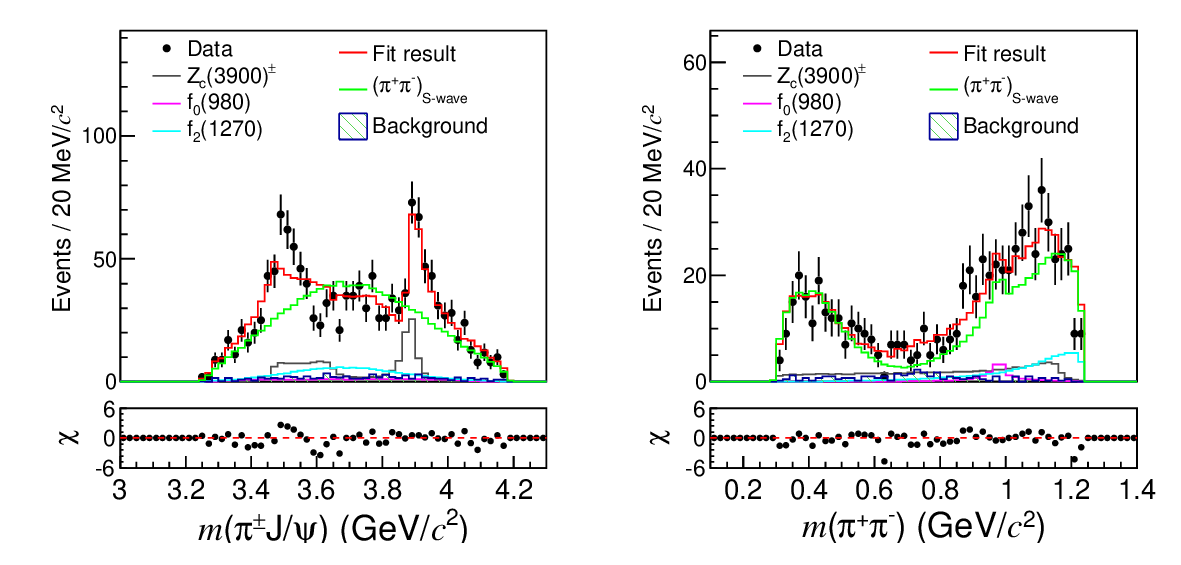}
  \includegraphics[width=0.45\textwidth, height=0.21\textwidth]{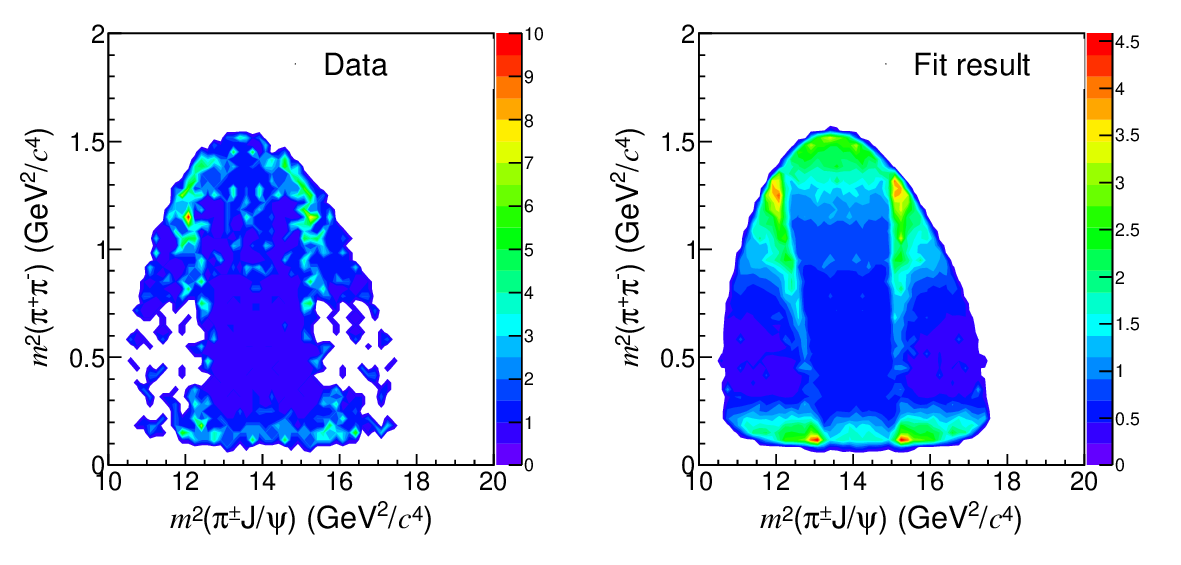}
  \includegraphics[width=0.45\textwidth, height=0.21\textwidth]{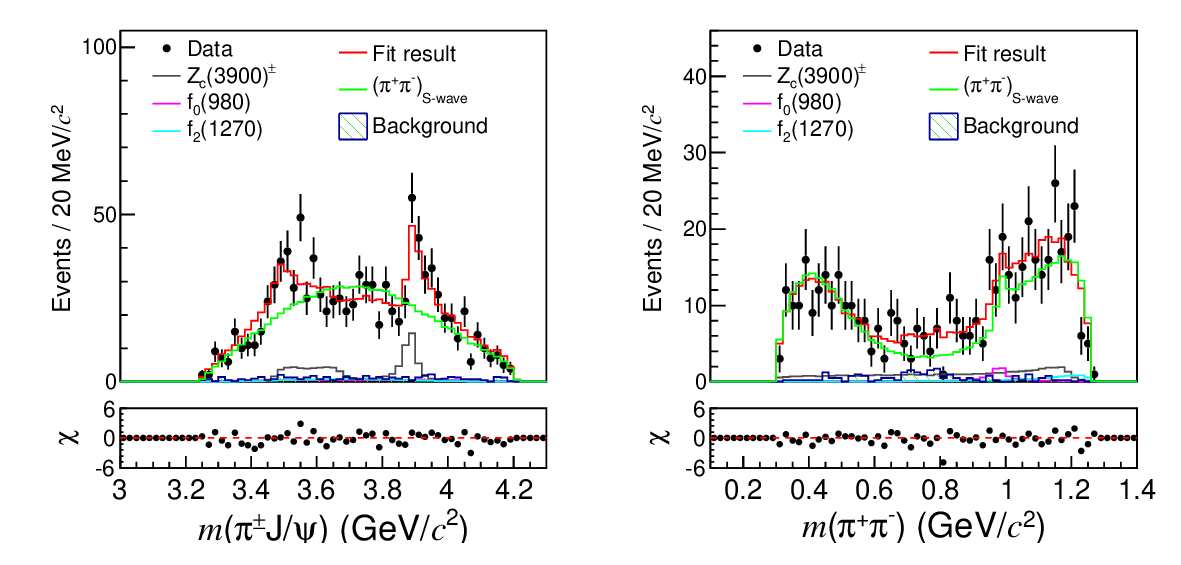}
  \includegraphics[width=0.45\textwidth, height=0.21\textwidth]{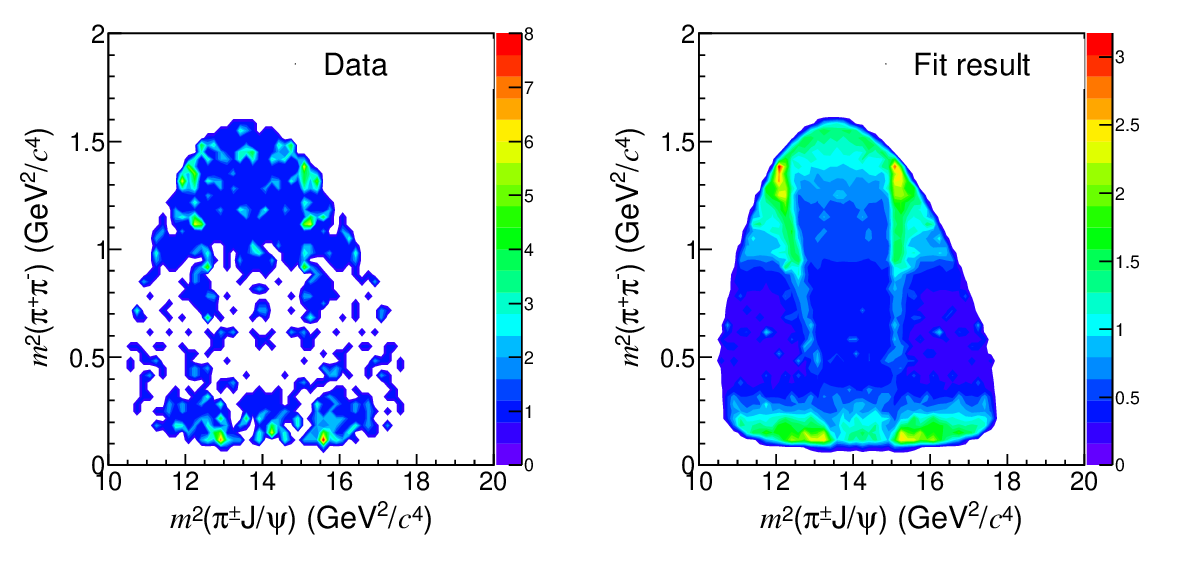}
\caption{\footnotesize{The PWA fit results of $\sqrt{s}=$4.2866 (1st row), 4.3115 (2nd row), 4.3370 (3rd row), 4.3583 (4th row) GeV samples.} The plots show the distributions of $m(\pi^{\pm}J/\psi)$ (1st column), $m(\pi^{+}\pi^{-})$ (2nd column), and Dalitz plots from data (3rd column) and the fit result (4th column). The $\chi^{2}/ndf$ for each term are listed in Table~\ref{chisqvsndf1}.}
\label{PWAresultTotal4}
\end{figure*}

 Figures~\ref{PWAresultTotal1},~\ref{PWAresultTotal2},~\ref{PWAresultTotal3}, and ~\ref{PWAresultTotal4} show the PWA fit results of the other 16 energy points and the corresponding $\chi^{2}/ndf$ are listed in Table~\ref{chisqvsndf1}. The $\chi^{2}/ndf$ for Dalitz distributions are calculated by the adaptive binning scheme~\cite{LHCb:2024yzj}.

\par
\begin{figure*}
\centering
  \includegraphics[width=0.21\textwidth, height=0.21\textwidth]{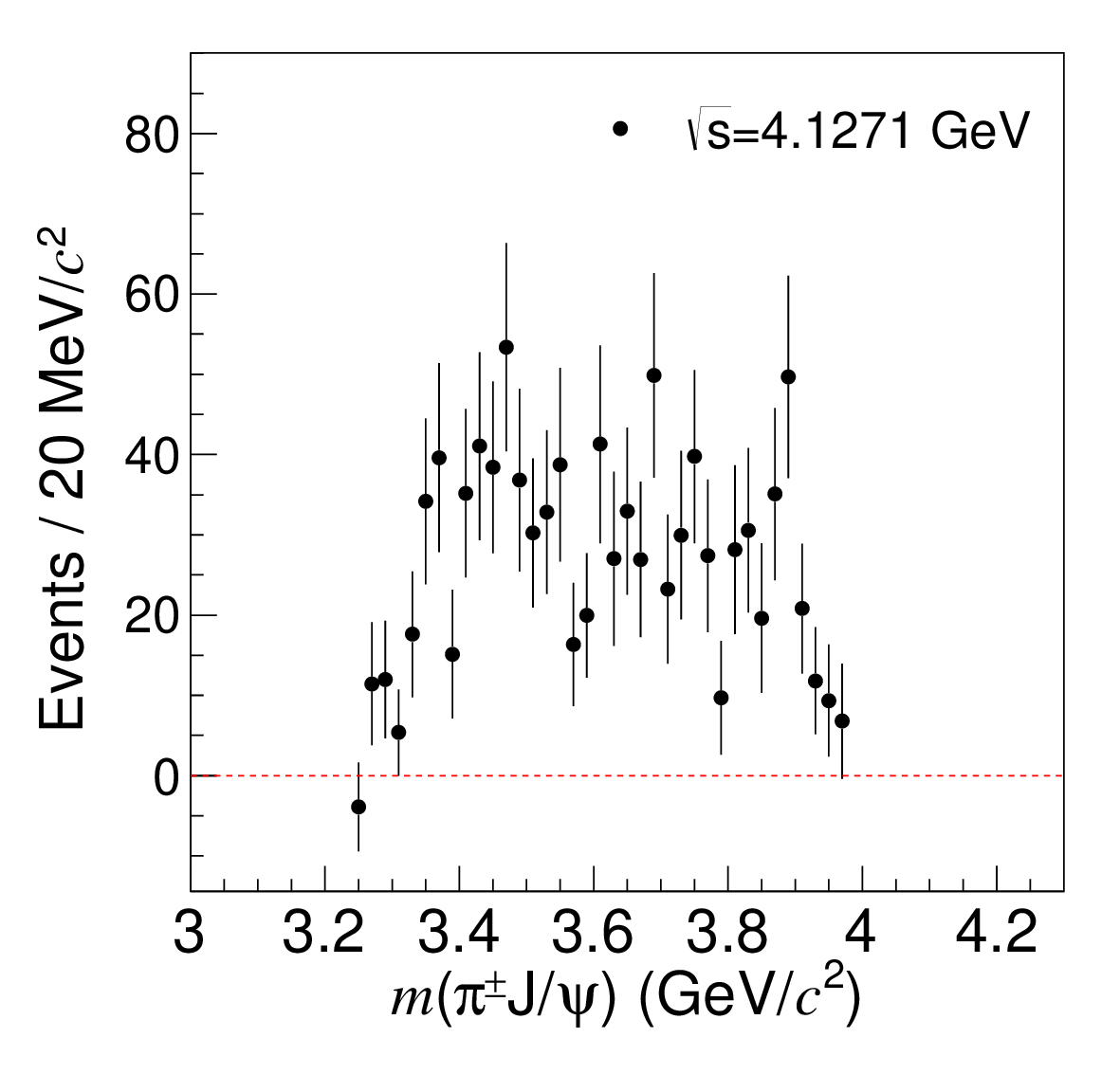}
  \includegraphics[width=0.21\textwidth, height=0.21\textwidth]{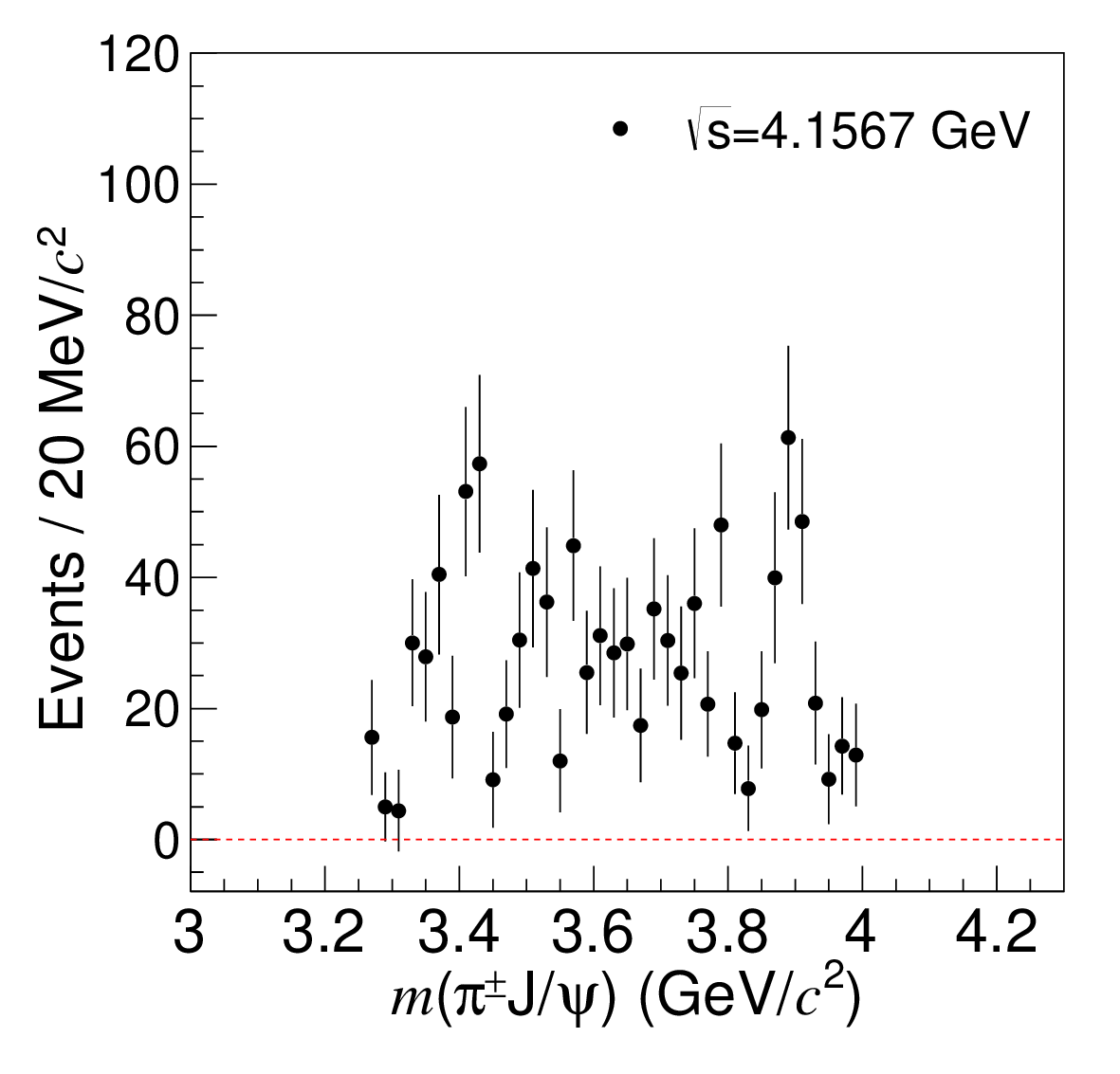}
  \includegraphics[width=0.21\textwidth, height=0.21\textwidth]{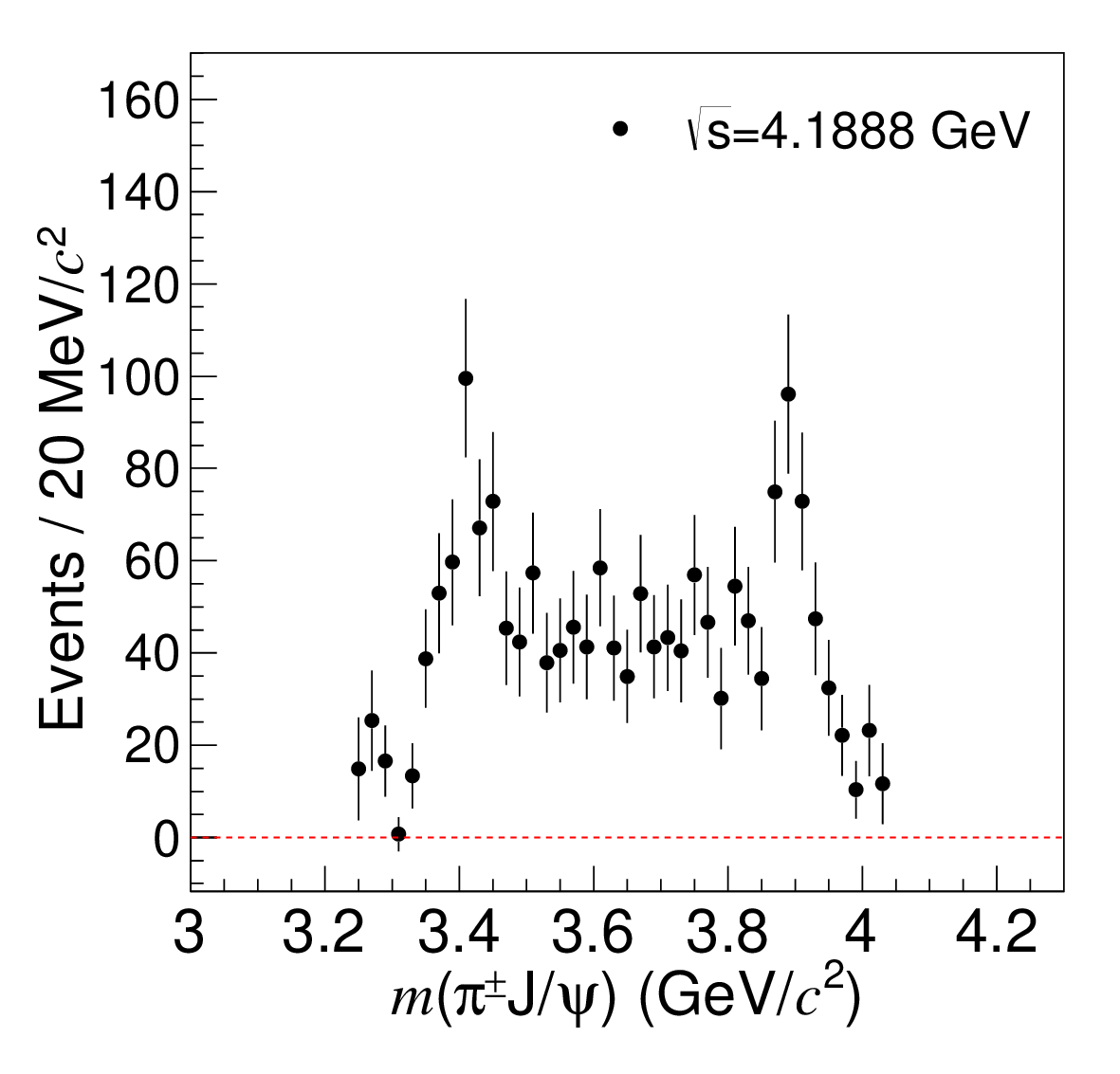}
  \includegraphics[width=0.21\textwidth, height=0.21\textwidth]{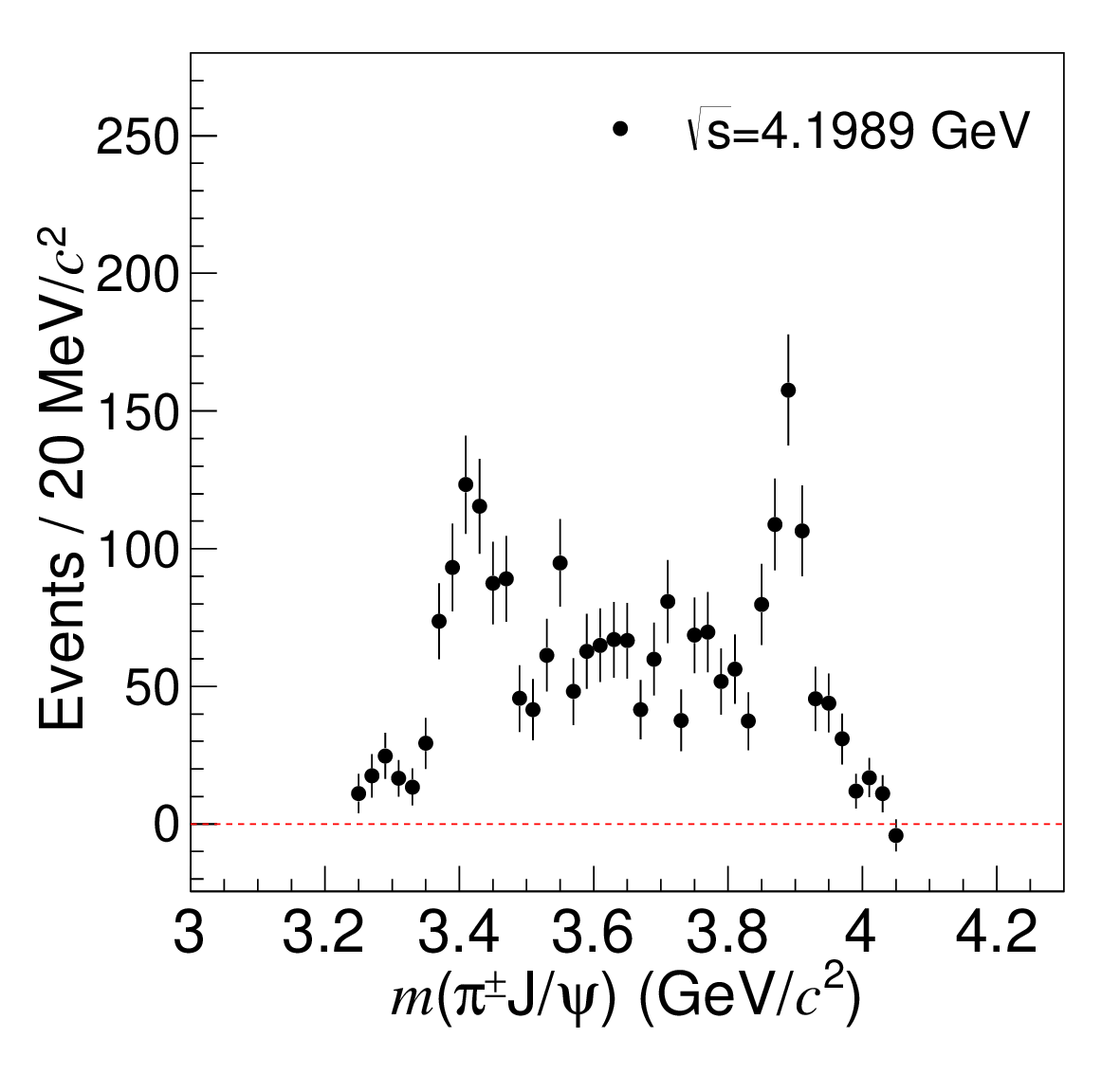}
  \includegraphics[width=0.21\textwidth, height=0.21\textwidth]{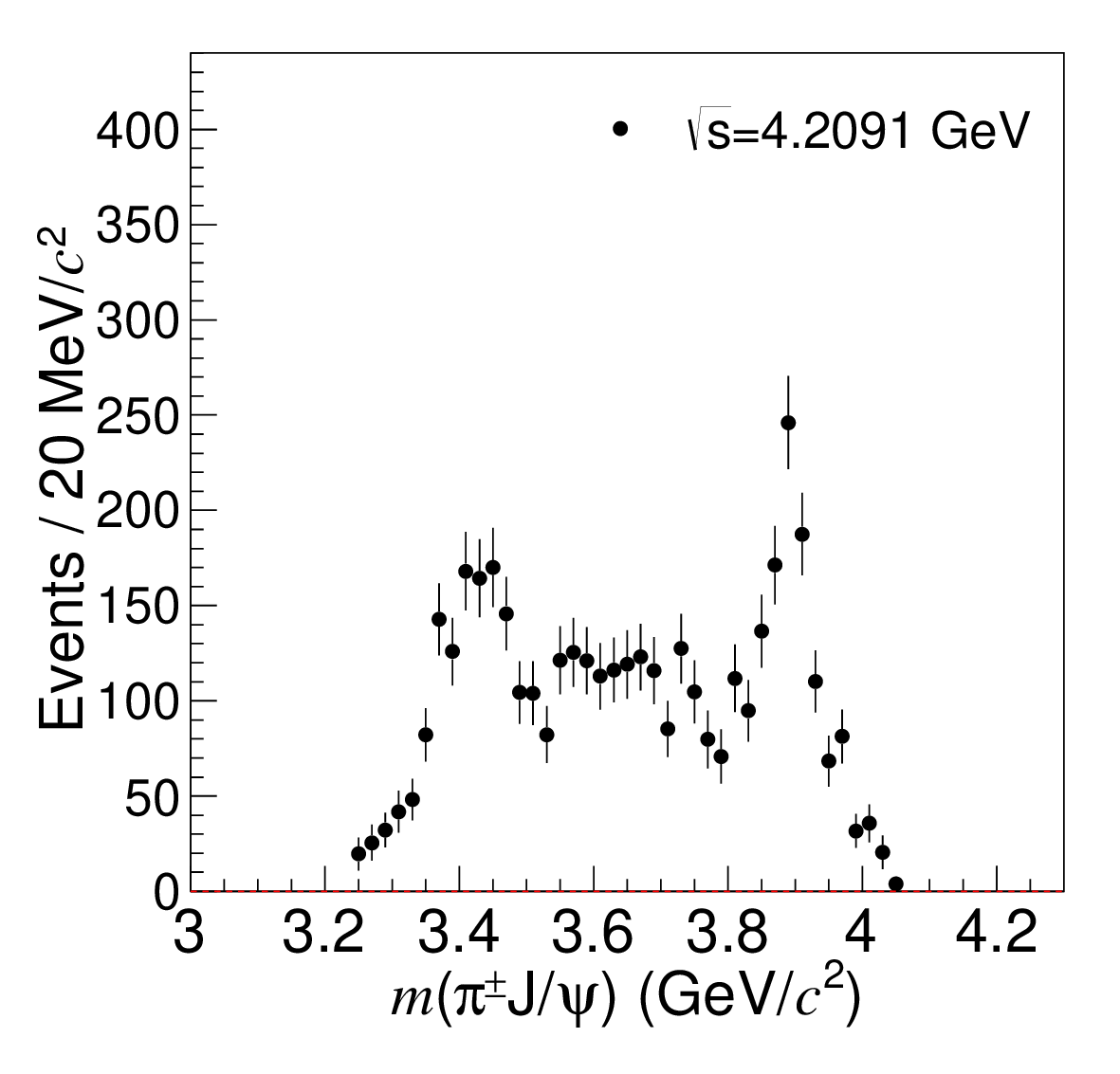}
  \includegraphics[width=0.21\textwidth, height=0.21\textwidth]{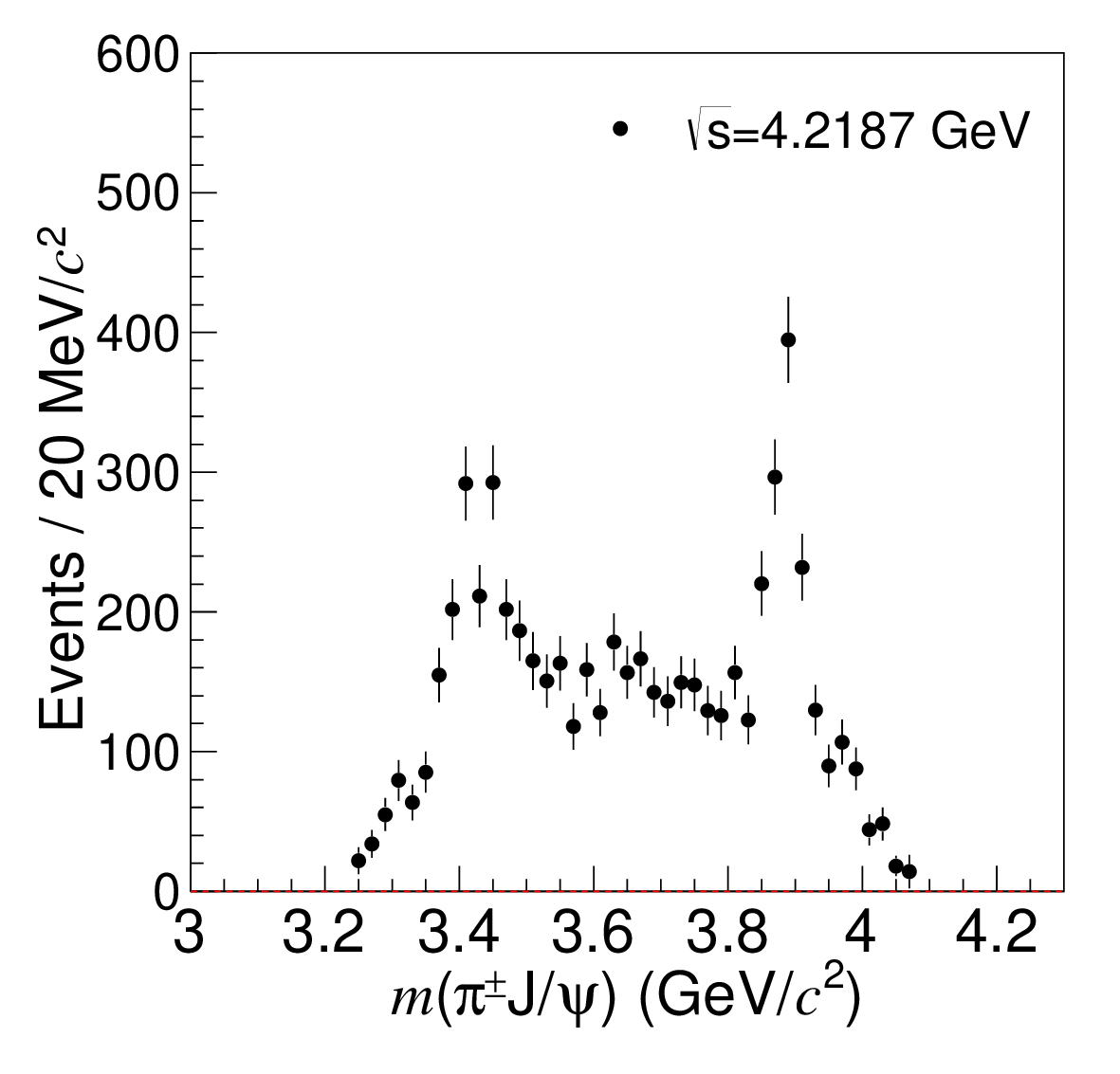}
  \includegraphics[width=0.21\textwidth, height=0.21\textwidth]{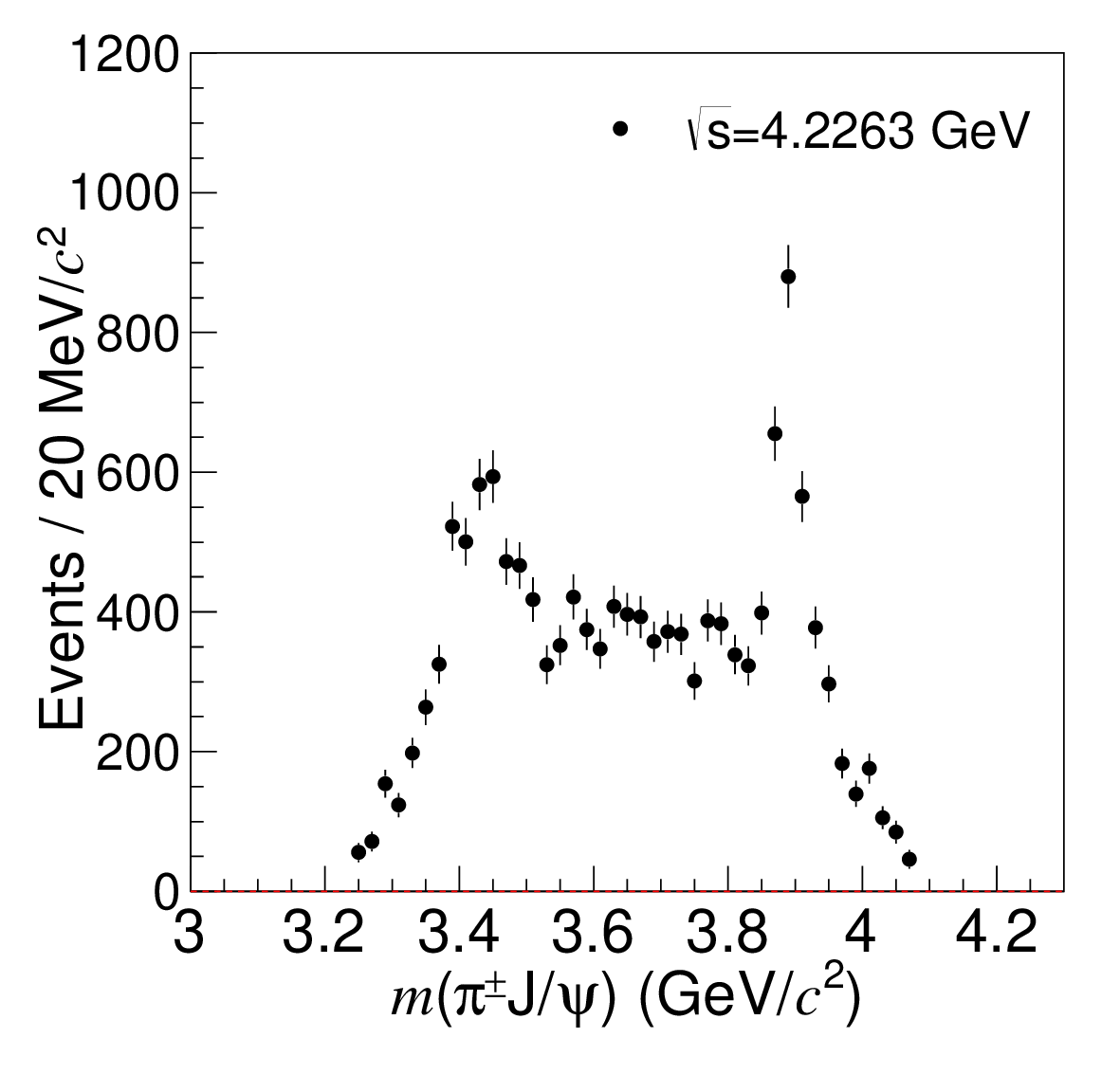}
  \includegraphics[width=0.21\textwidth, height=0.21\textwidth]{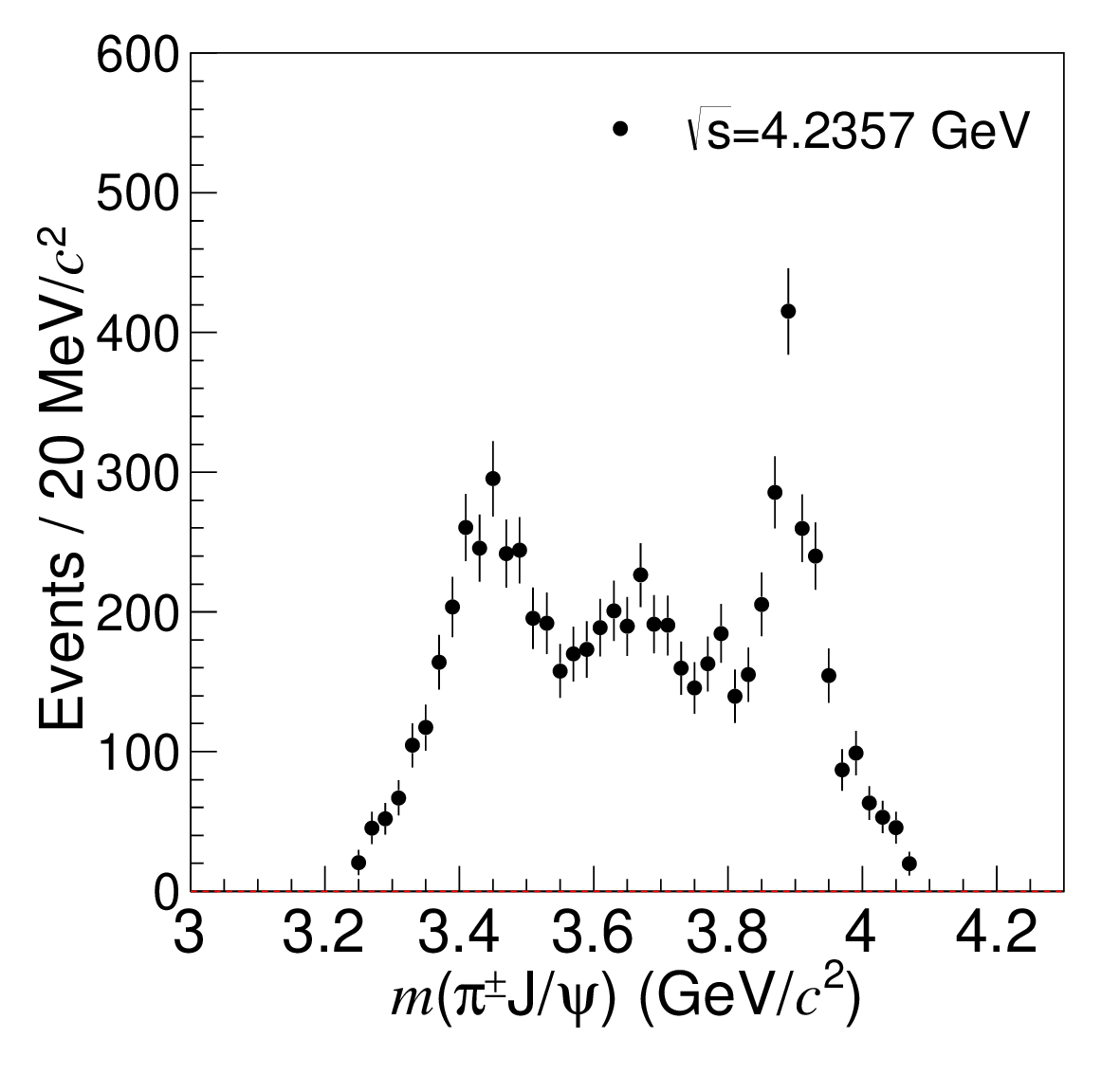}
  \includegraphics[width=0.21\textwidth, height=0.21\textwidth]{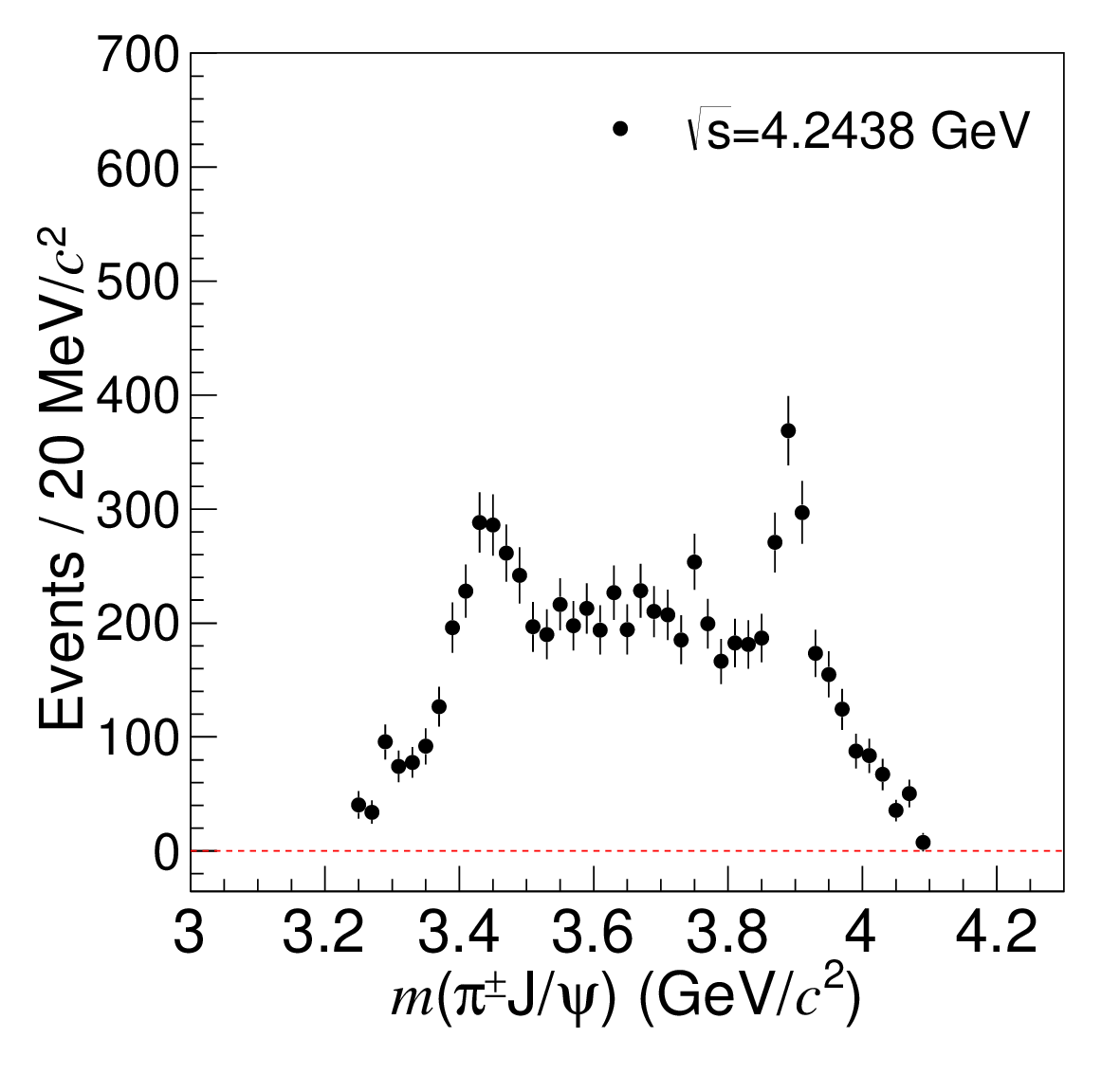}
  \includegraphics[width=0.21\textwidth, height=0.21\textwidth]{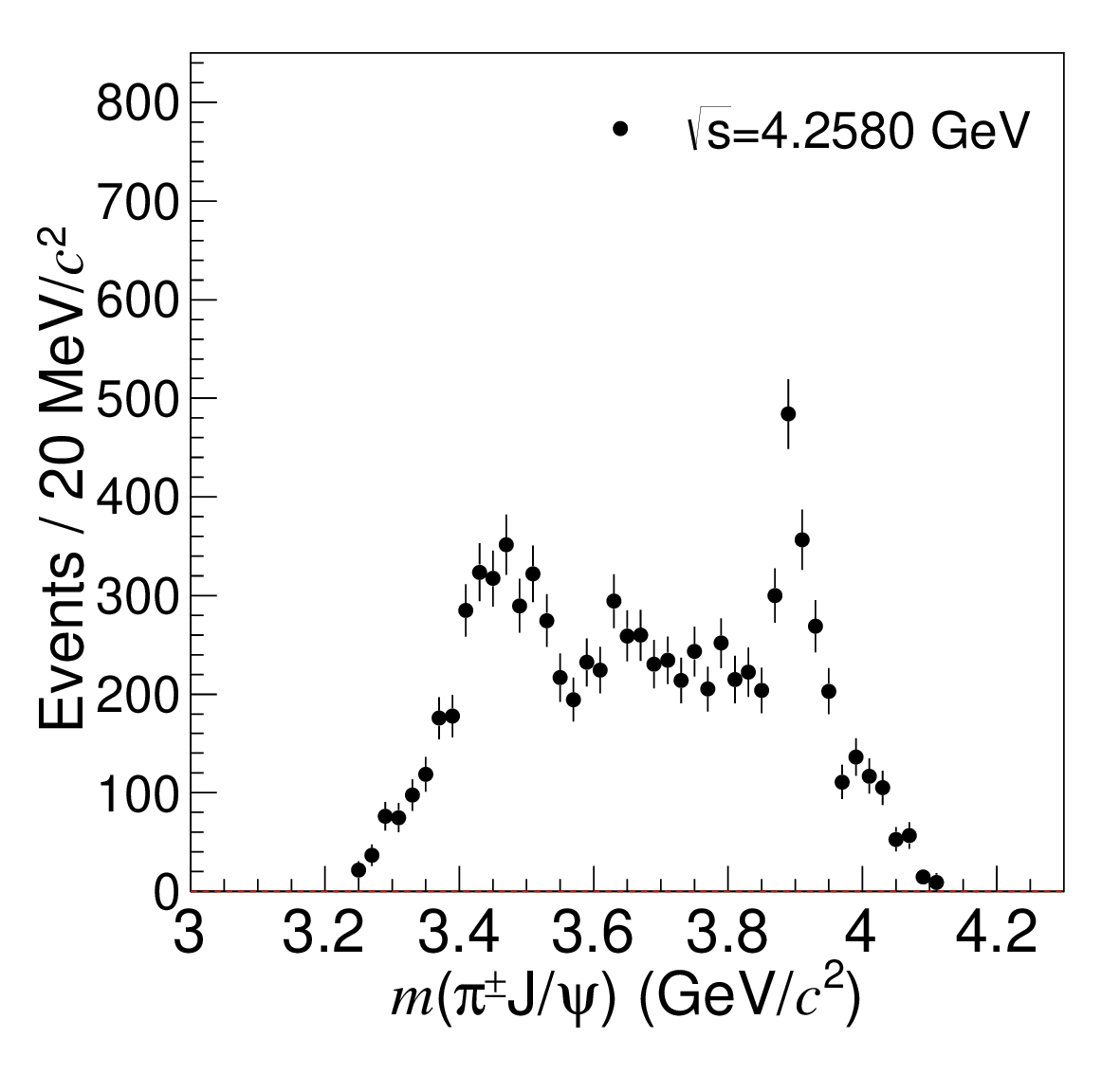}
  \includegraphics[width=0.21\textwidth, height=0.21\textwidth]{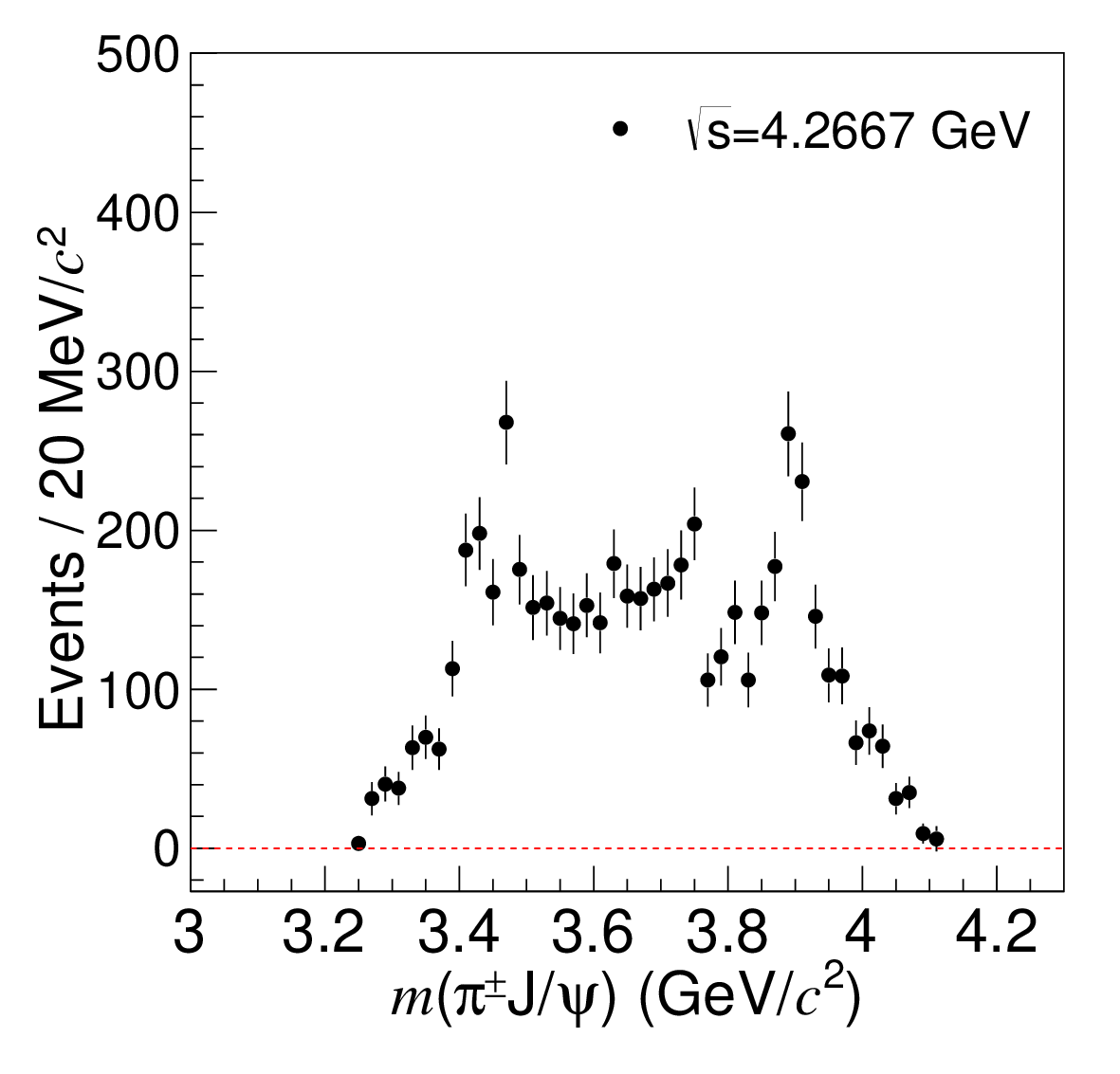}
  \includegraphics[width=0.21\textwidth, height=0.21\textwidth]{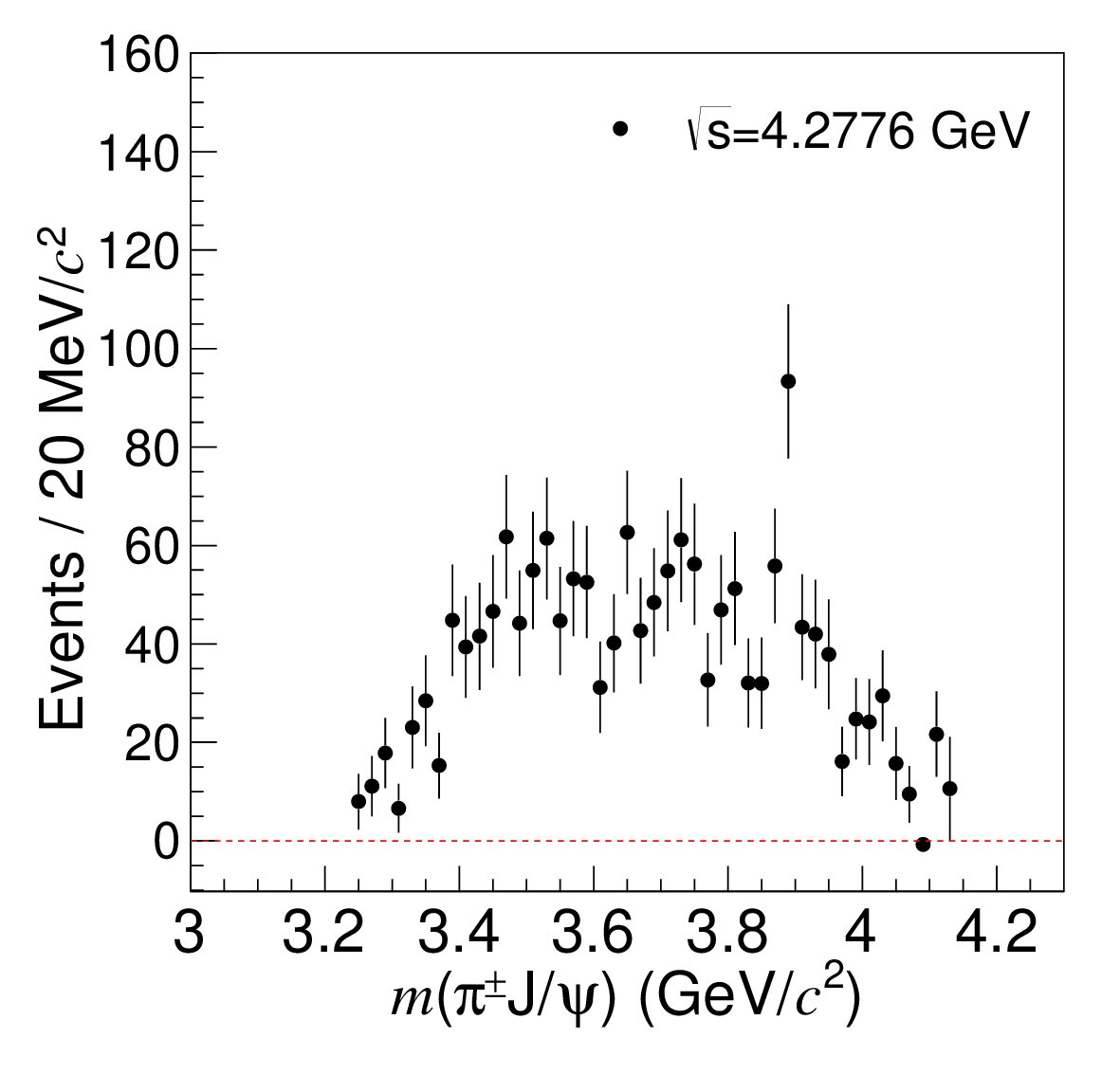}
  \includegraphics[width=0.21\textwidth, height=0.21\textwidth]{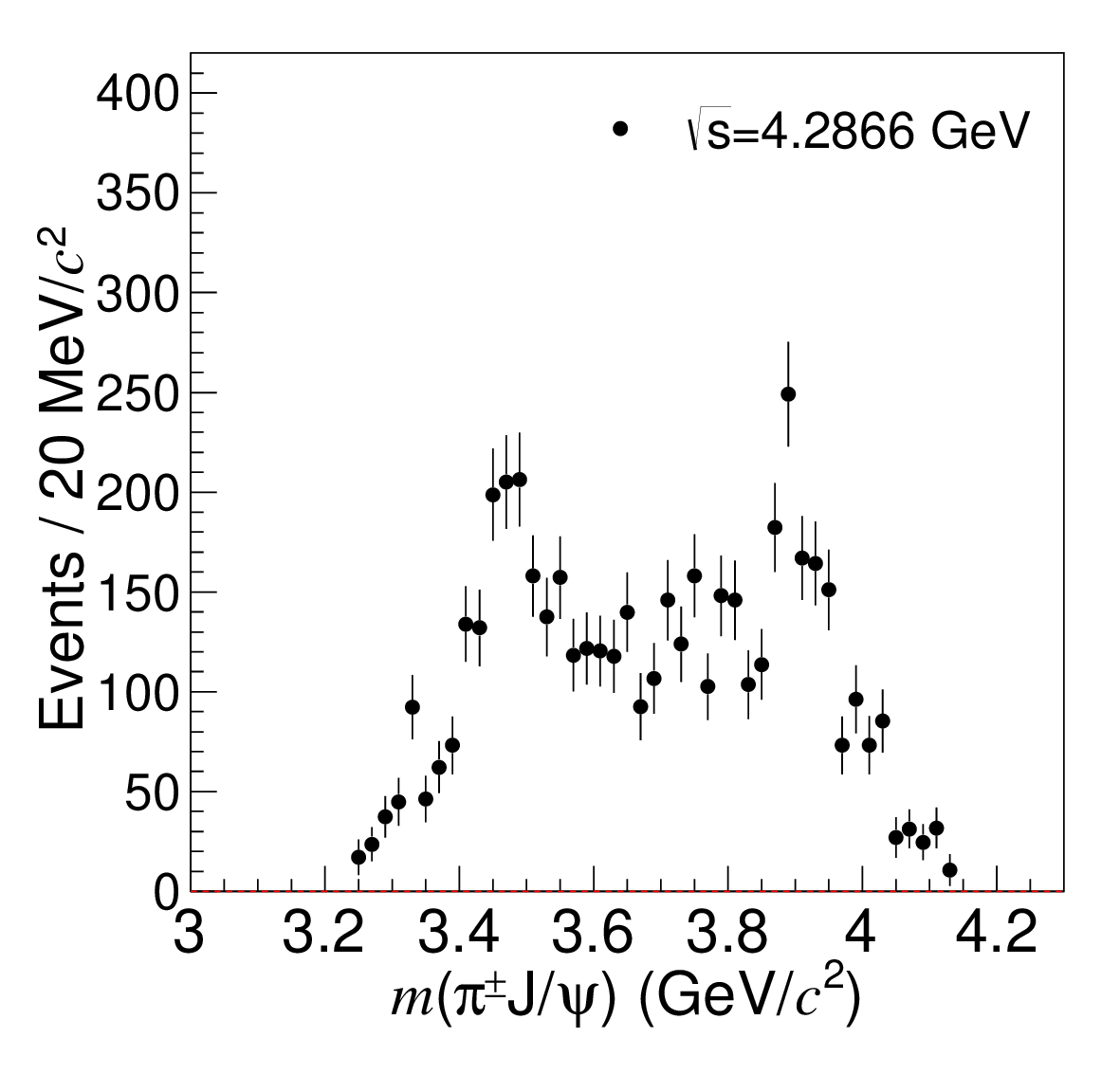}
  \includegraphics[width=0.21\textwidth, height=0.21\textwidth]{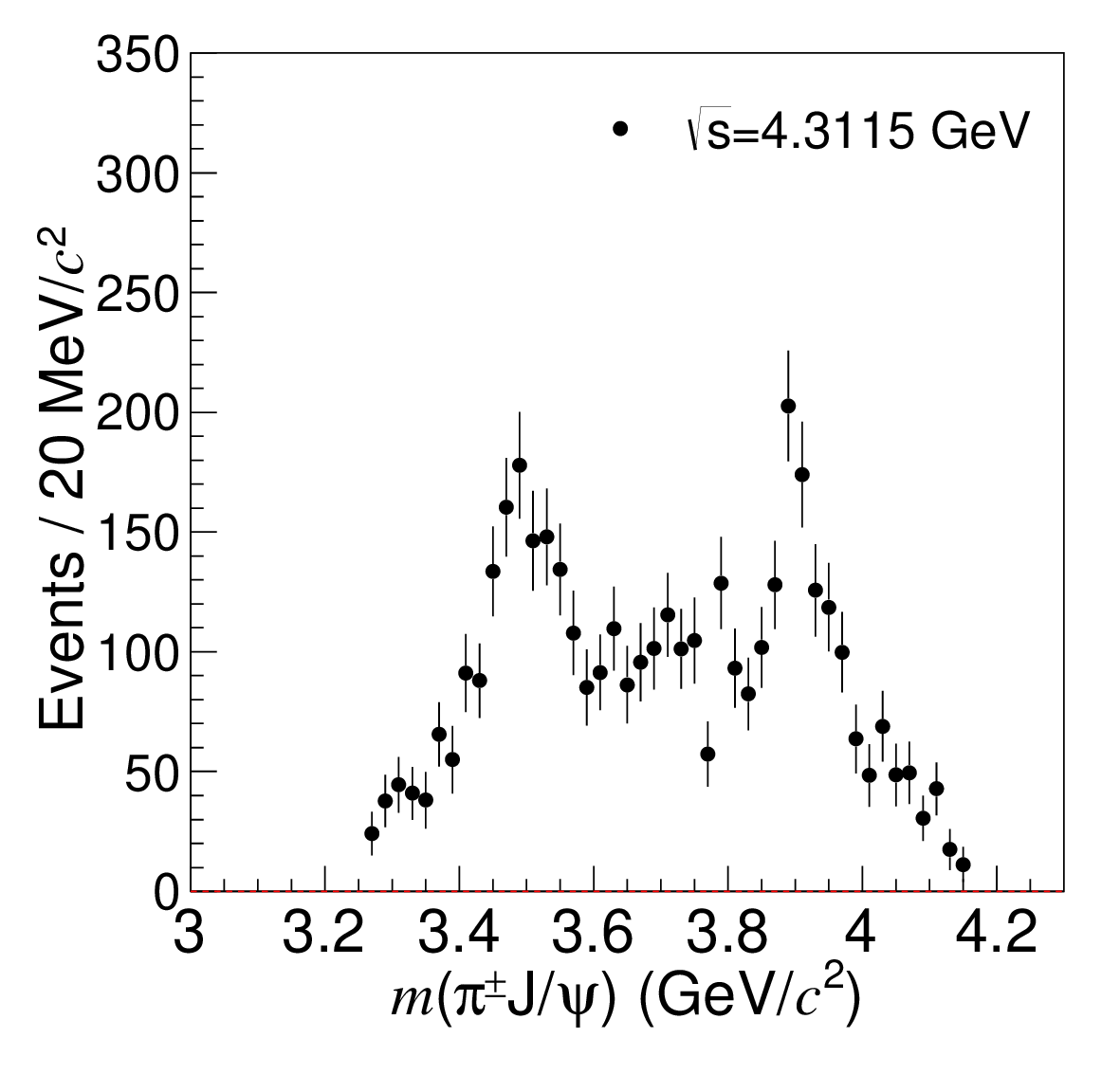}
  \includegraphics[width=0.21\textwidth, height=0.21\textwidth]{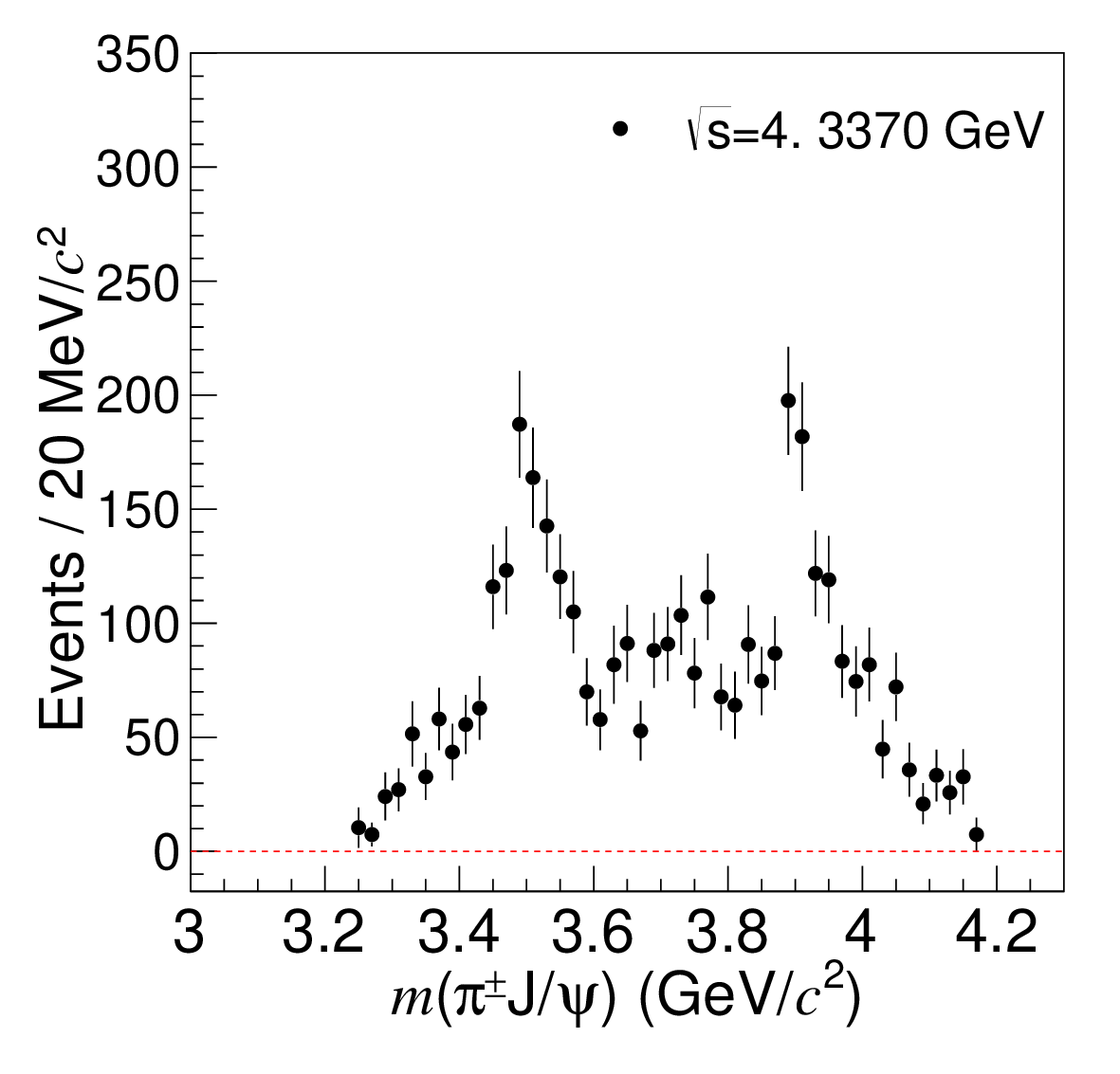}
  \includegraphics[width=0.21\textwidth, height=0.21\textwidth]{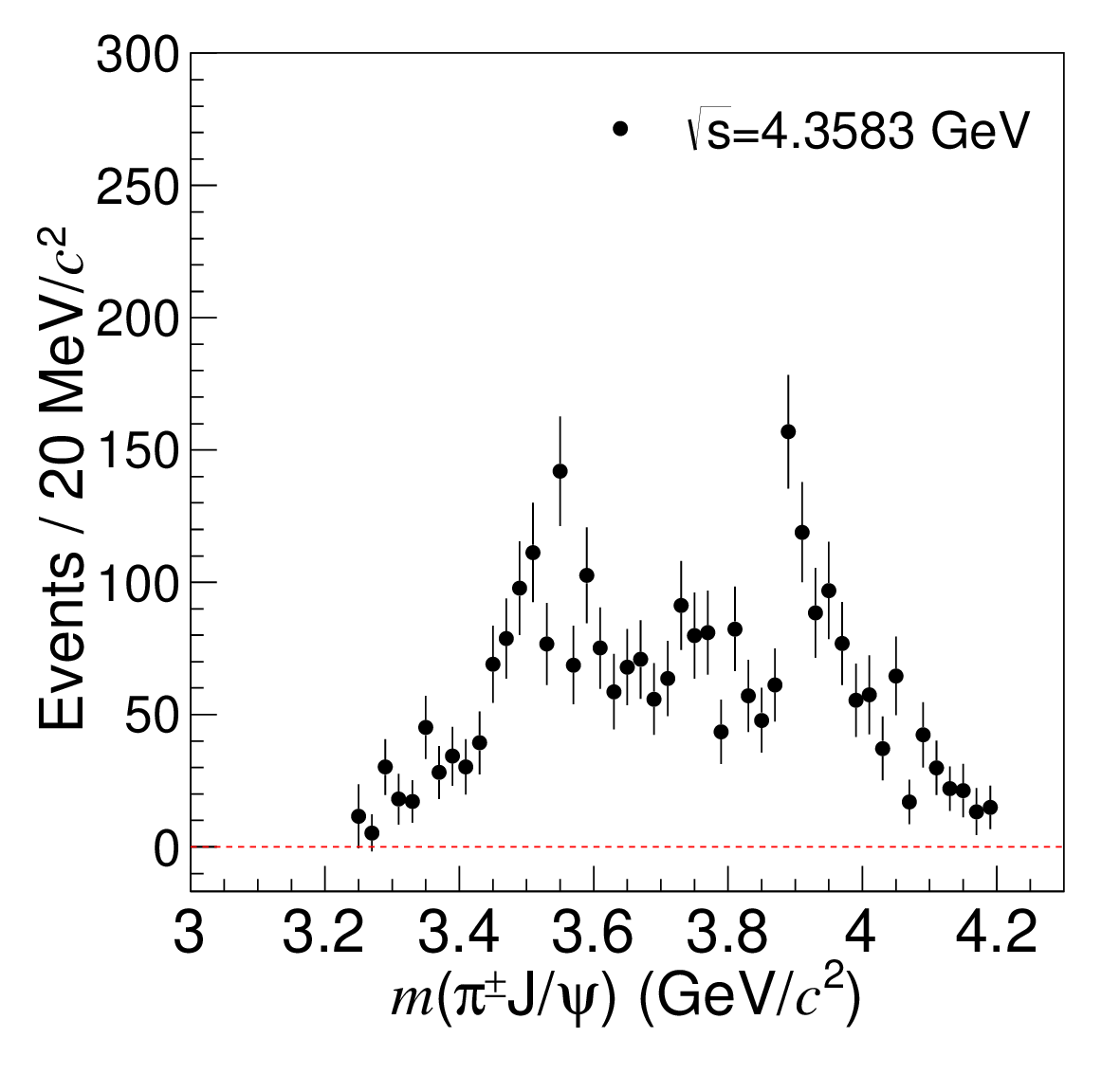}
\caption{The $m(\pi^{\pm} J/\psi)$ distribution after the efficiency correction for the other 16 energy points except the $\sqrt{s}=$4.1780 GeV sample. The sideband events are used as an estimate of the background events and subtracted from events in the signal region.}
\label{CorrectMasspiJpsi}
\end{figure*}

\begin{figure*}
\centering
  \includegraphics[width=0.21\textwidth, height=0.21\textwidth]{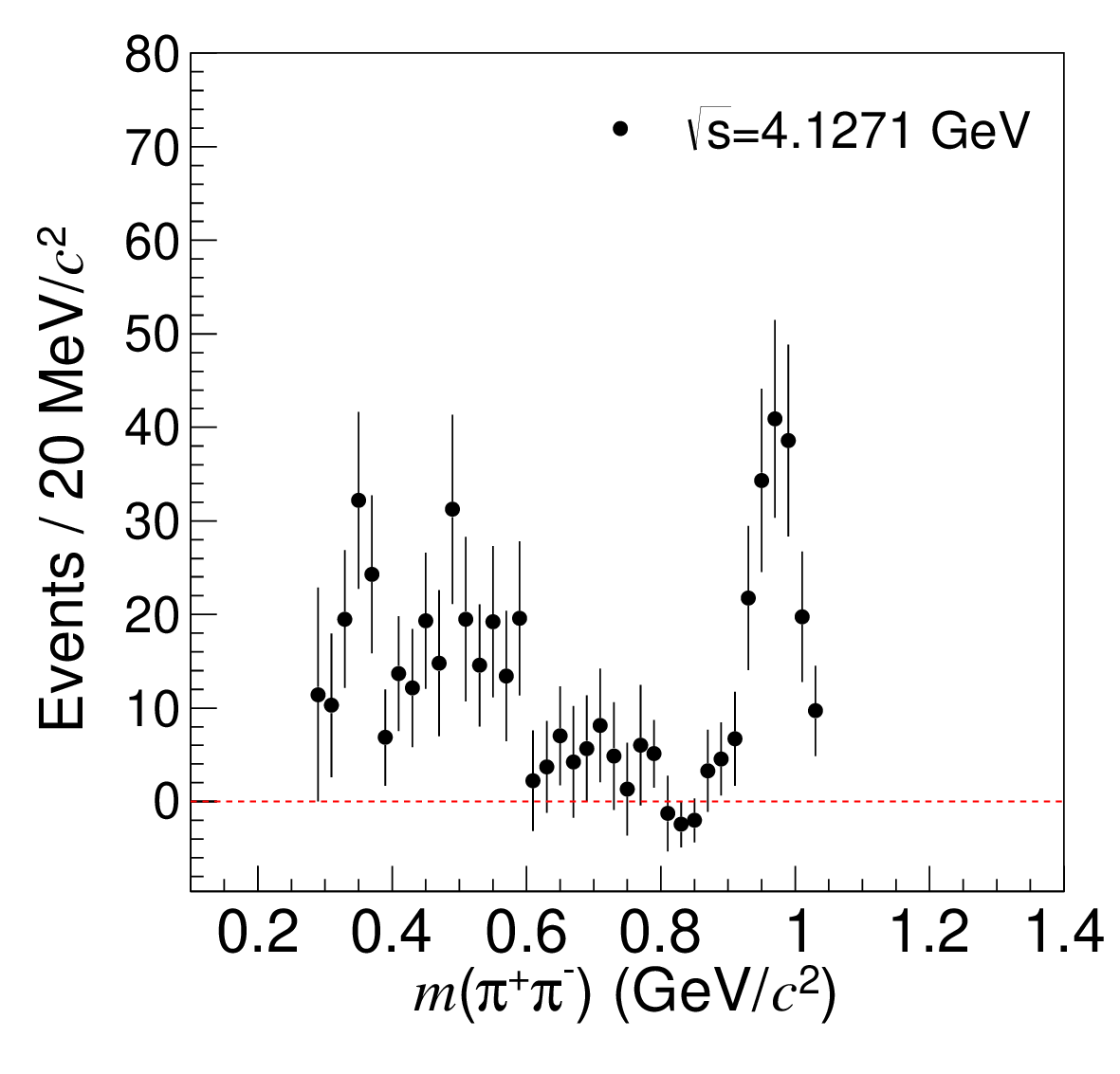}
  \includegraphics[width=0.21\textwidth, height=0.21\textwidth]{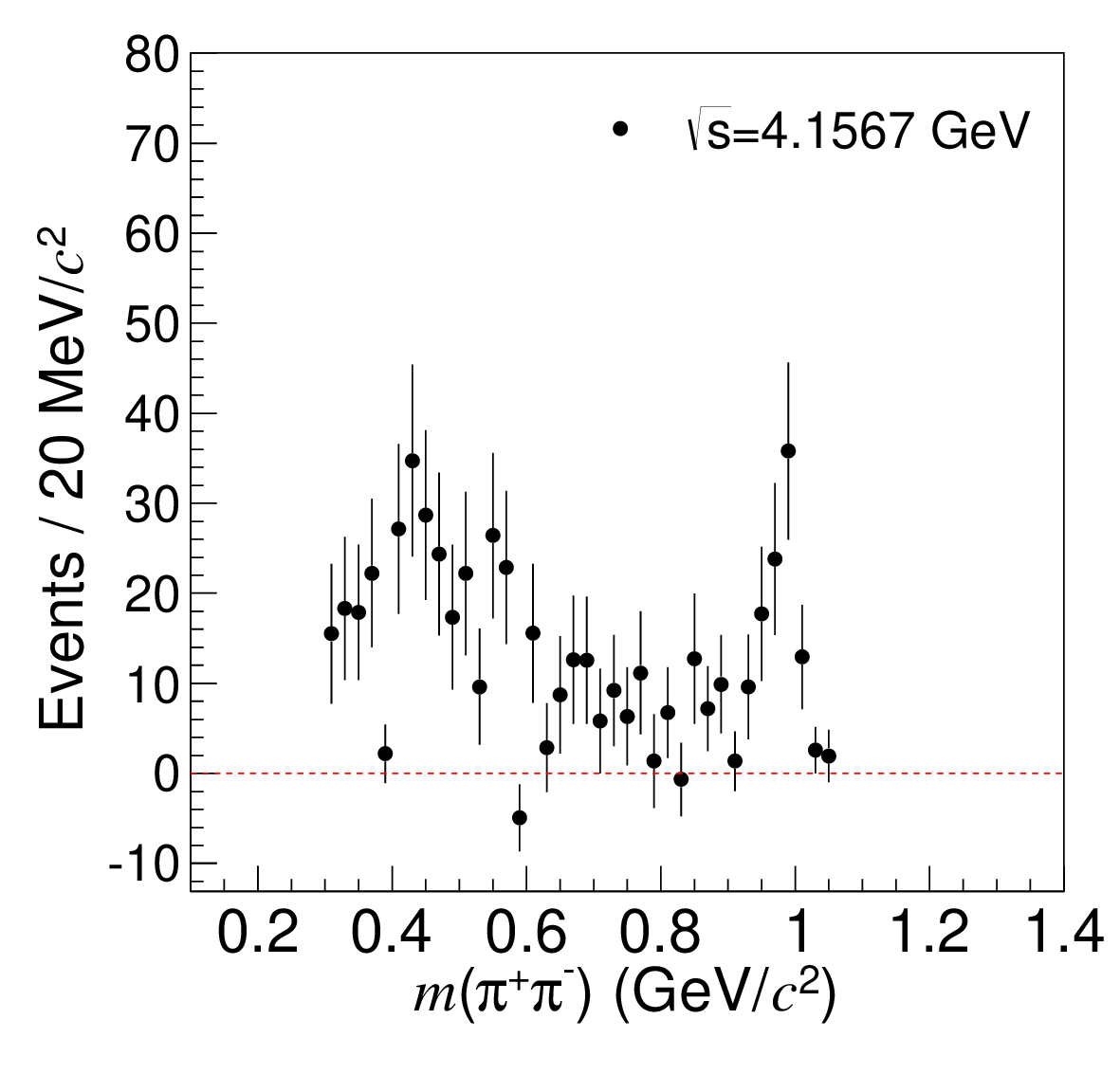}
  \includegraphics[width=0.21\textwidth, height=0.21\textwidth]{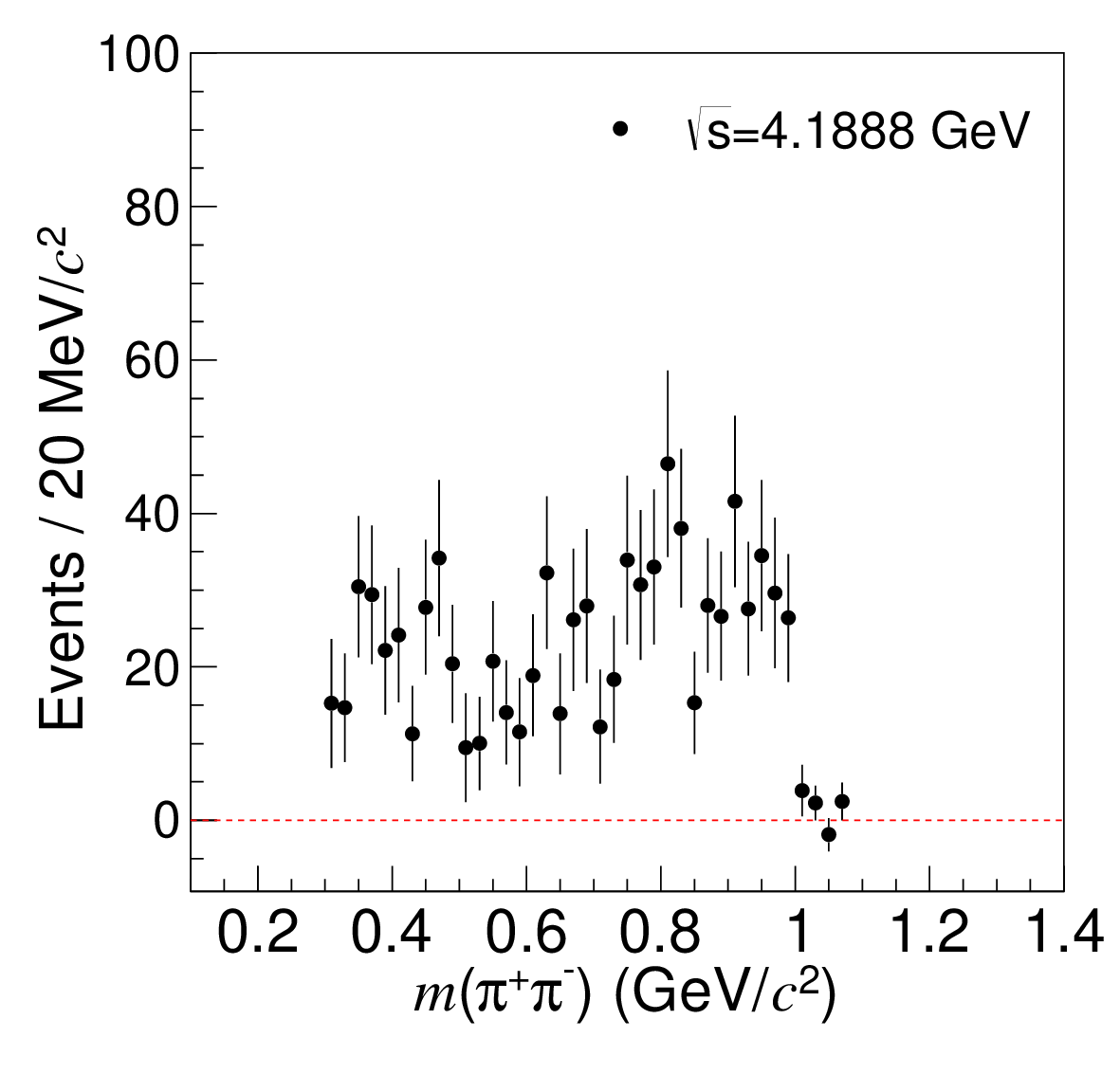}
  \includegraphics[width=0.21\textwidth, height=0.21\textwidth]{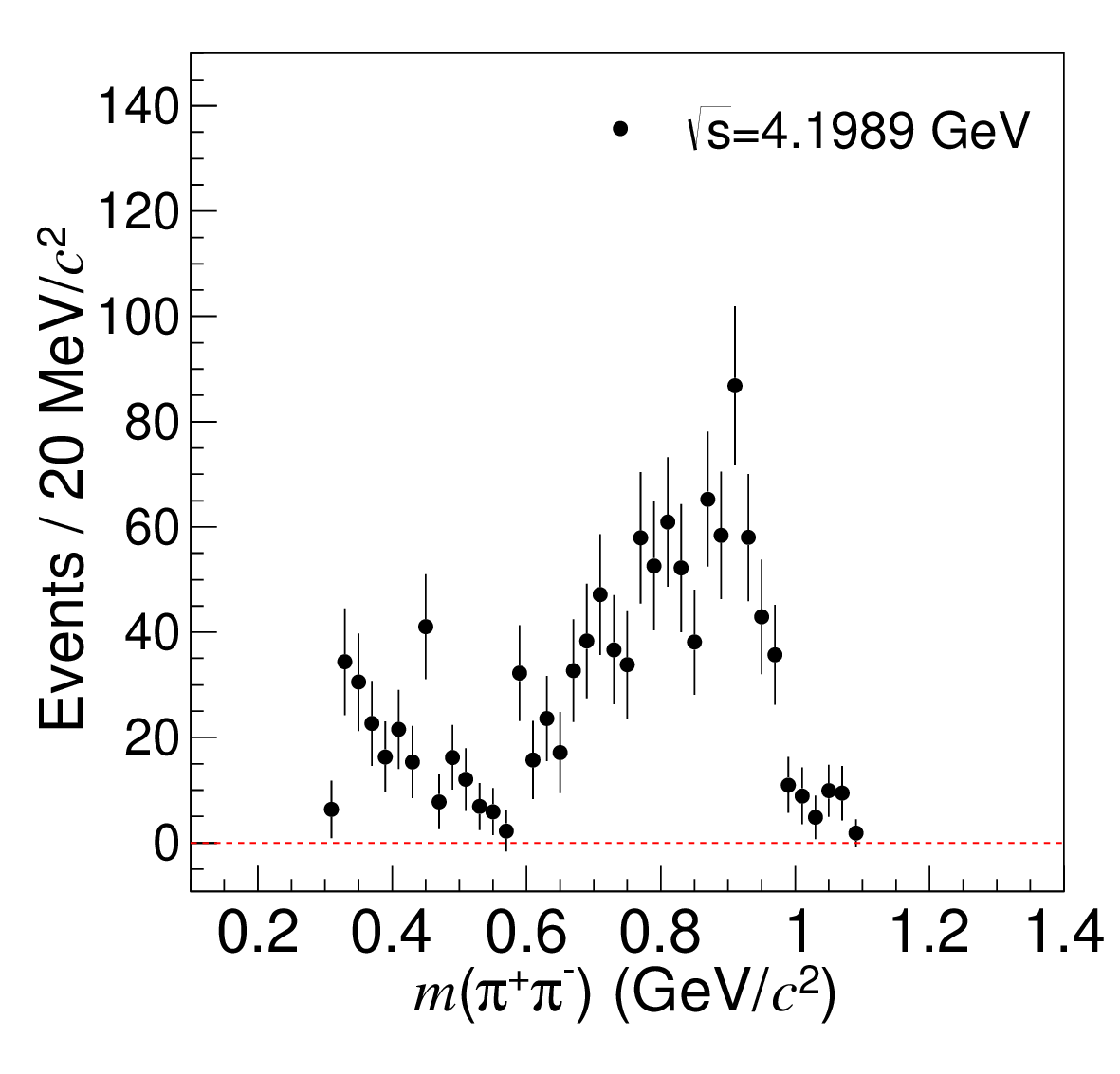}
  \includegraphics[width=0.21\textwidth, height=0.21\textwidth]{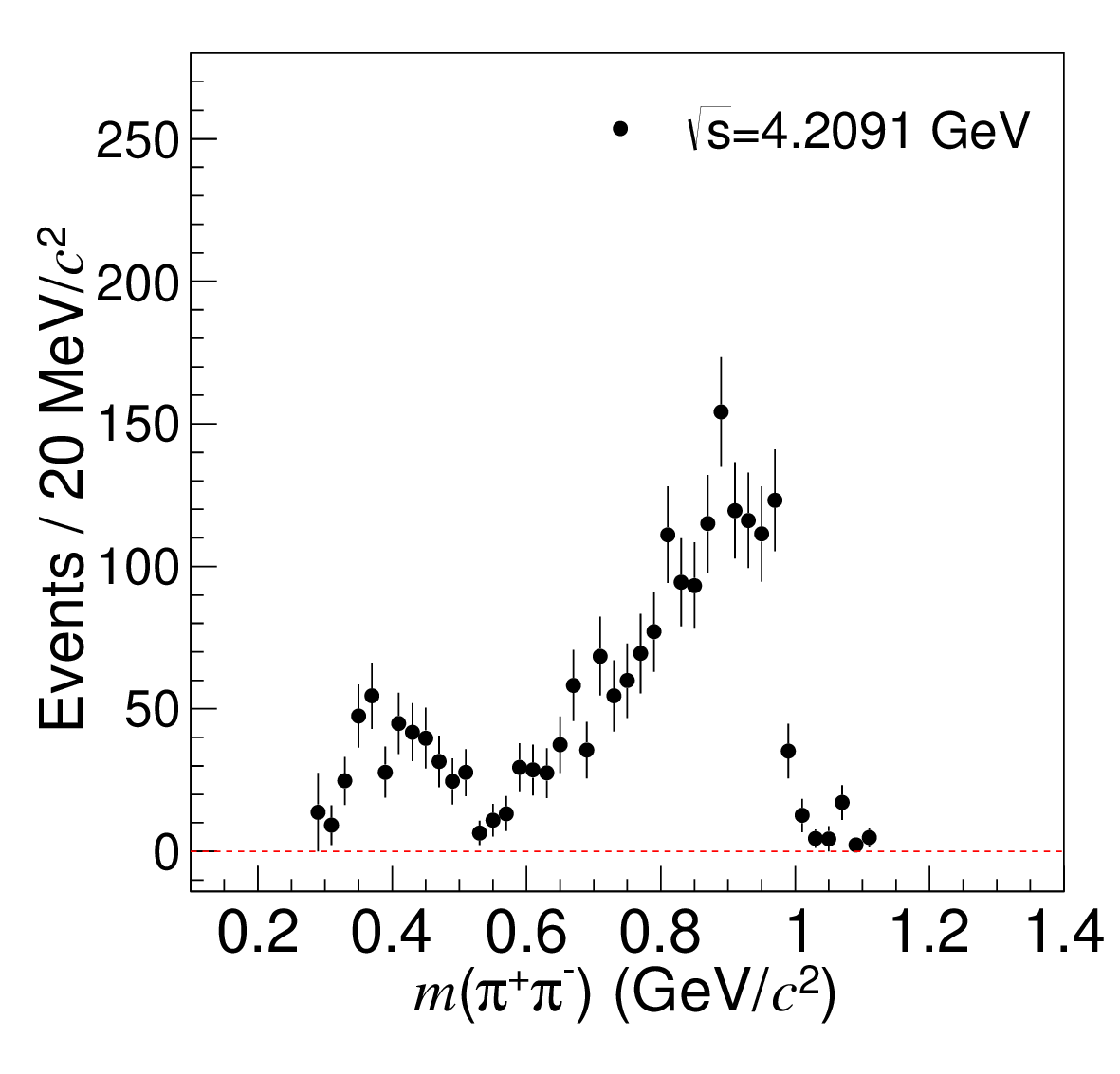}
  \includegraphics[width=0.21\textwidth, height=0.21\textwidth]{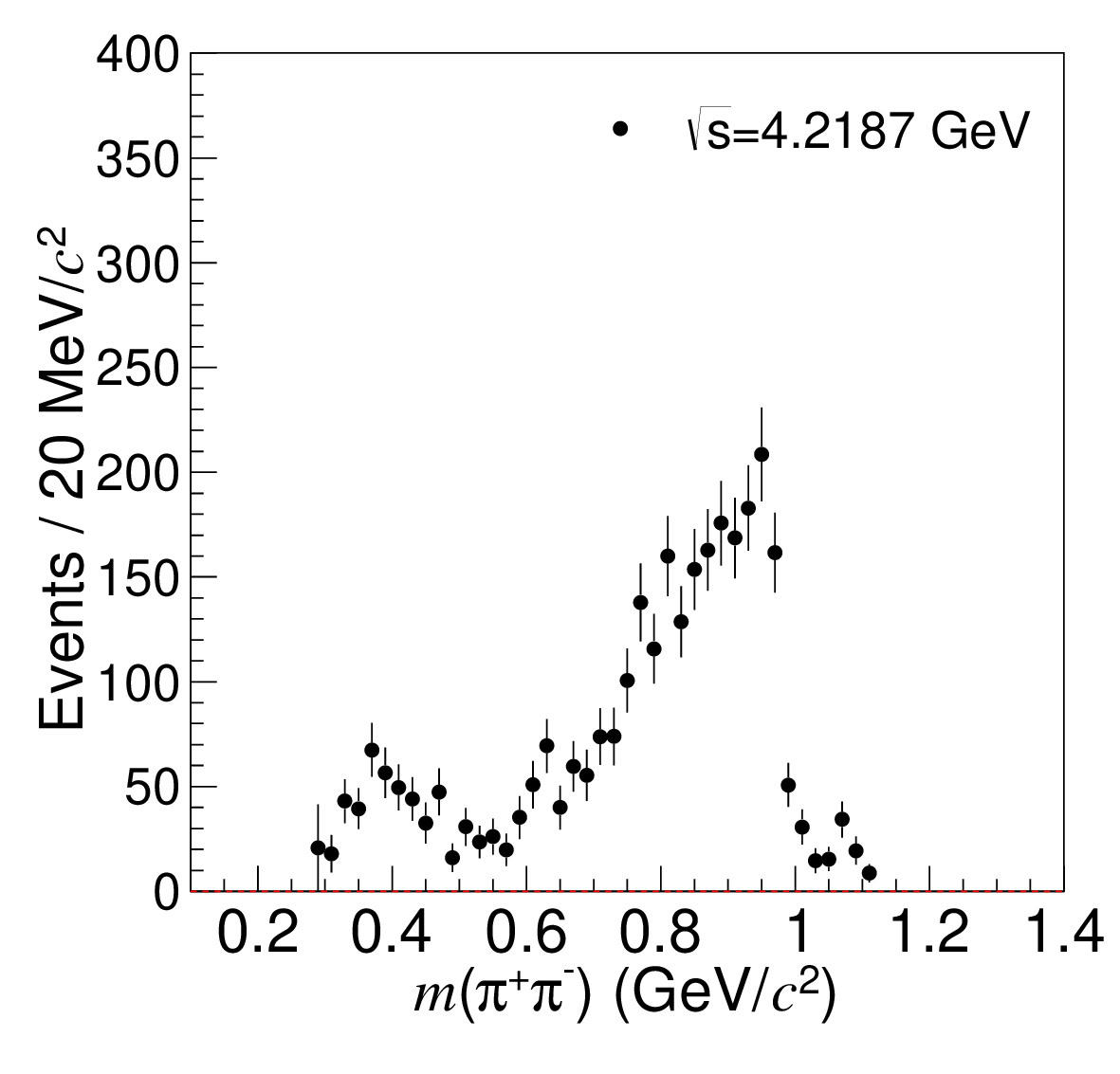}
  \includegraphics[width=0.21\textwidth, height=0.21\textwidth]{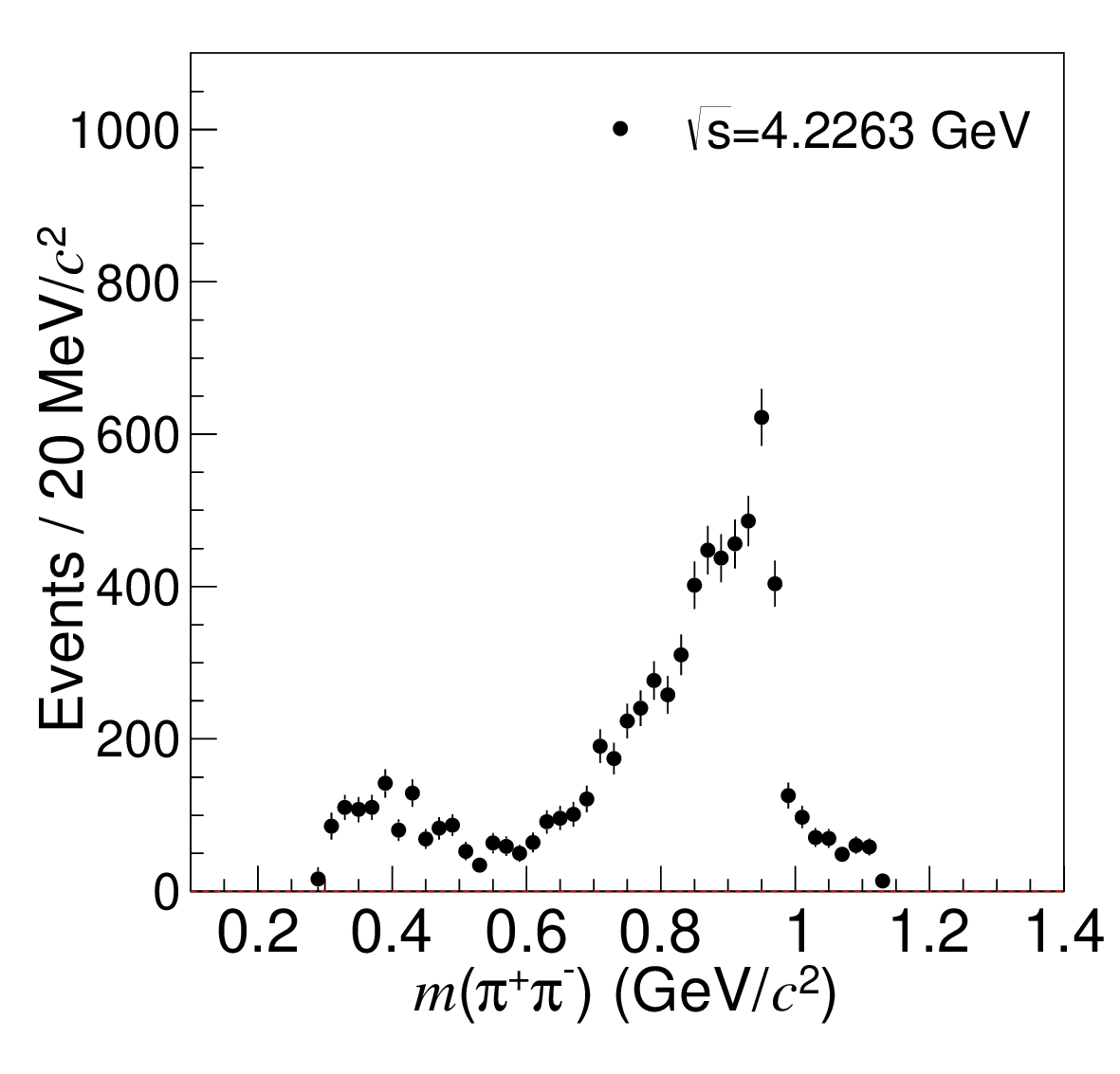}
  \includegraphics[width=0.21\textwidth, height=0.21\textwidth]{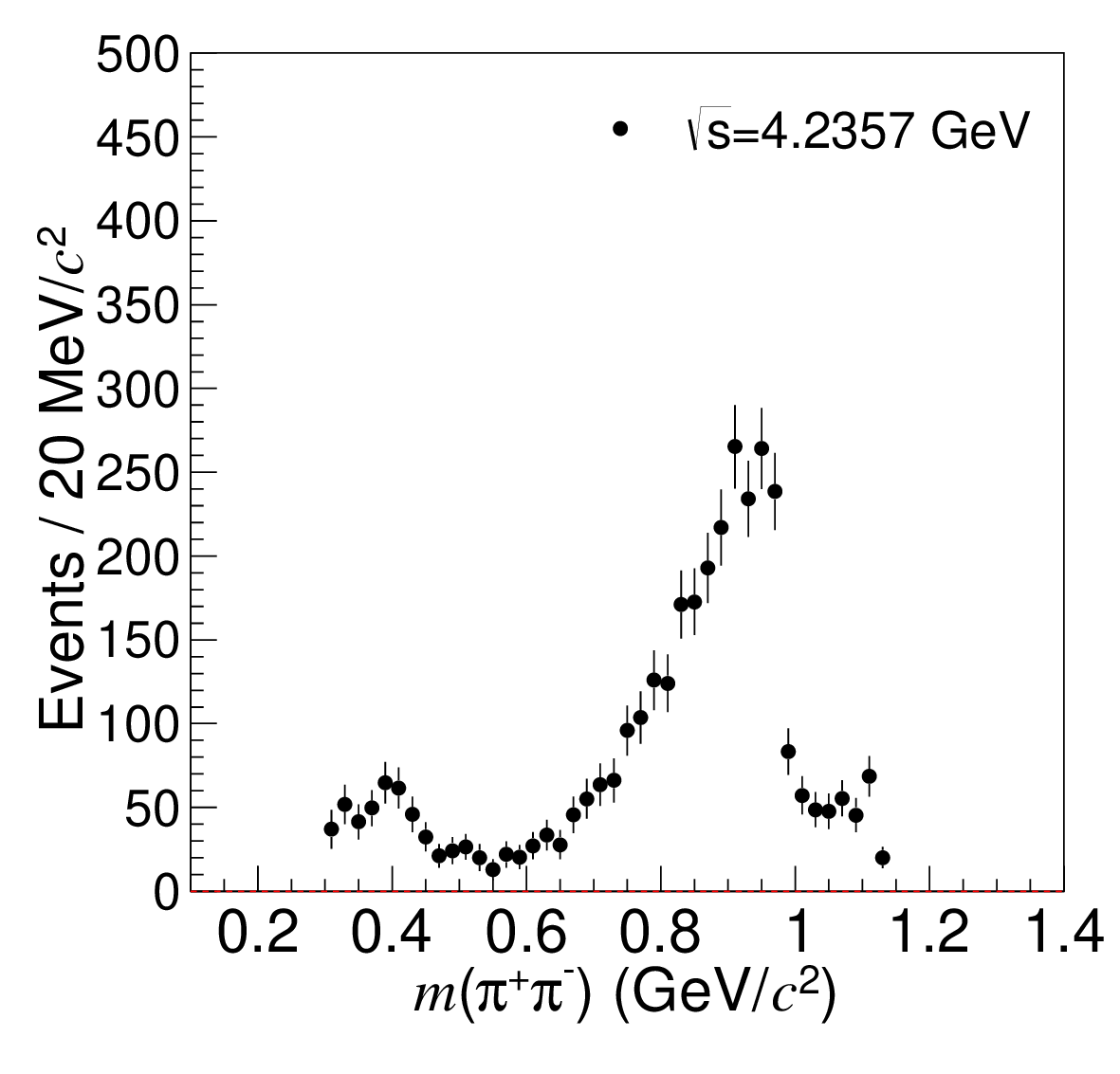}
  \includegraphics[width=0.21\textwidth, height=0.21\textwidth]{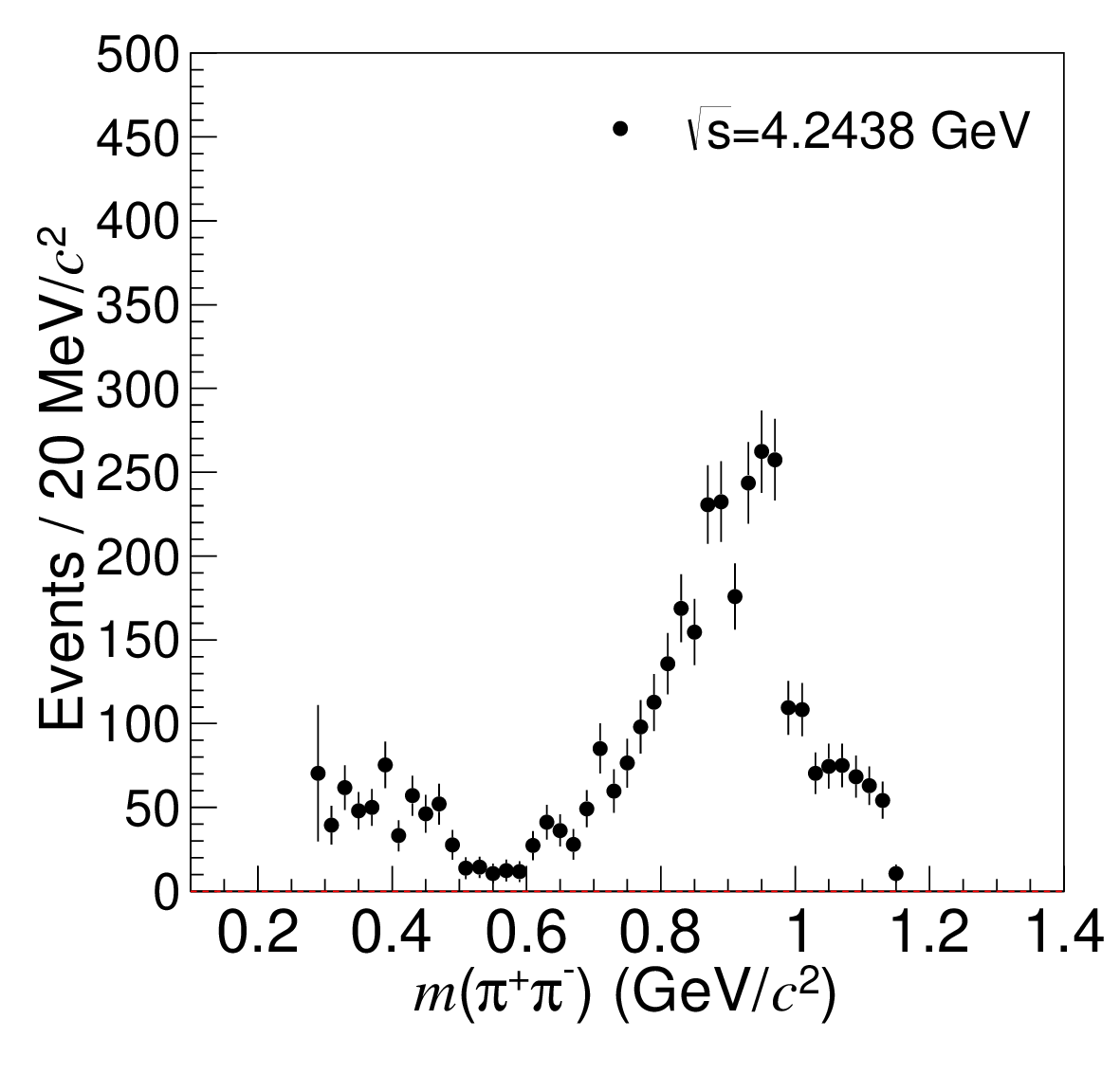}
  \includegraphics[width=0.21\textwidth, height=0.21\textwidth]{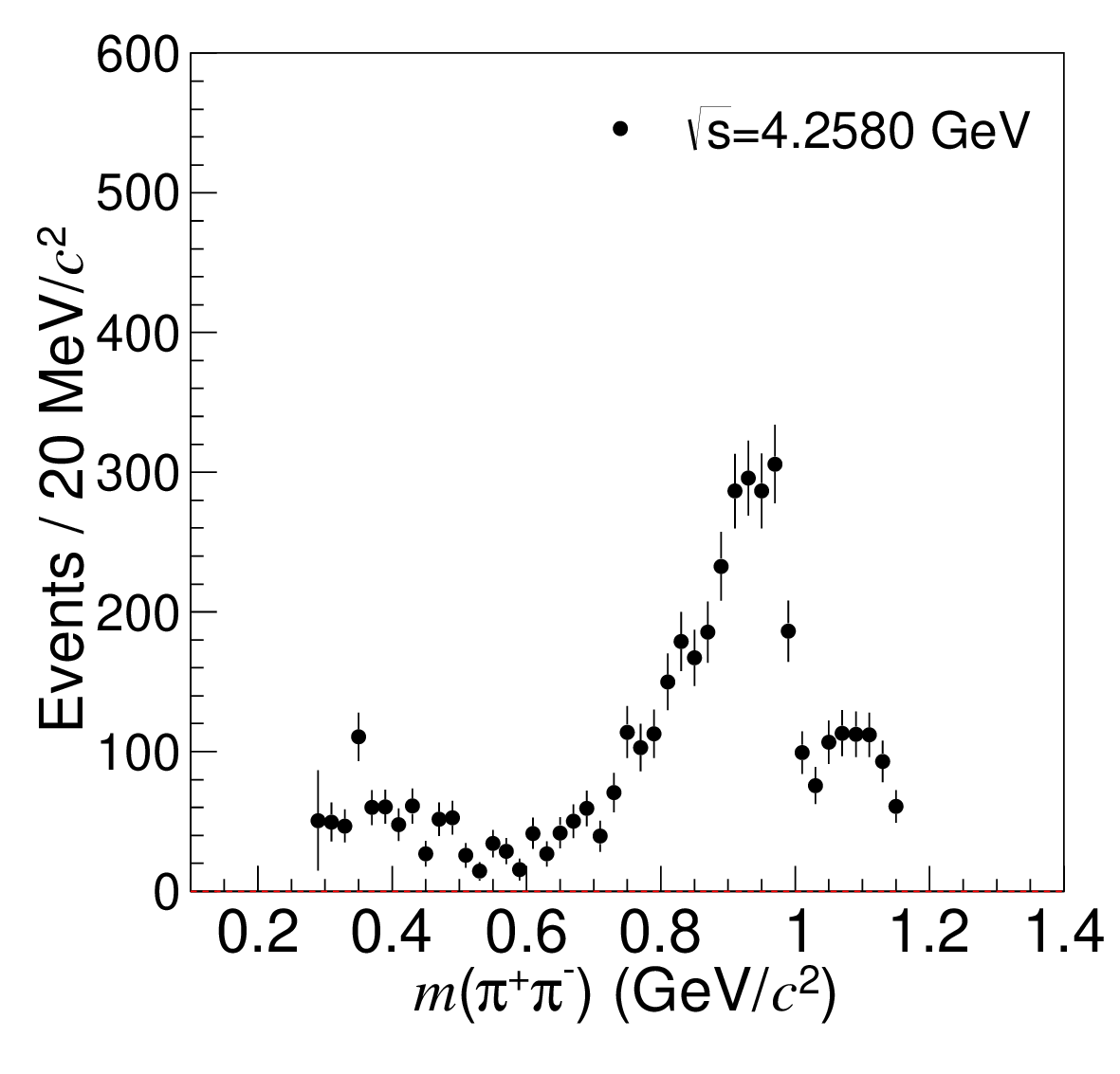}
  \includegraphics[width=0.21\textwidth, height=0.21\textwidth]{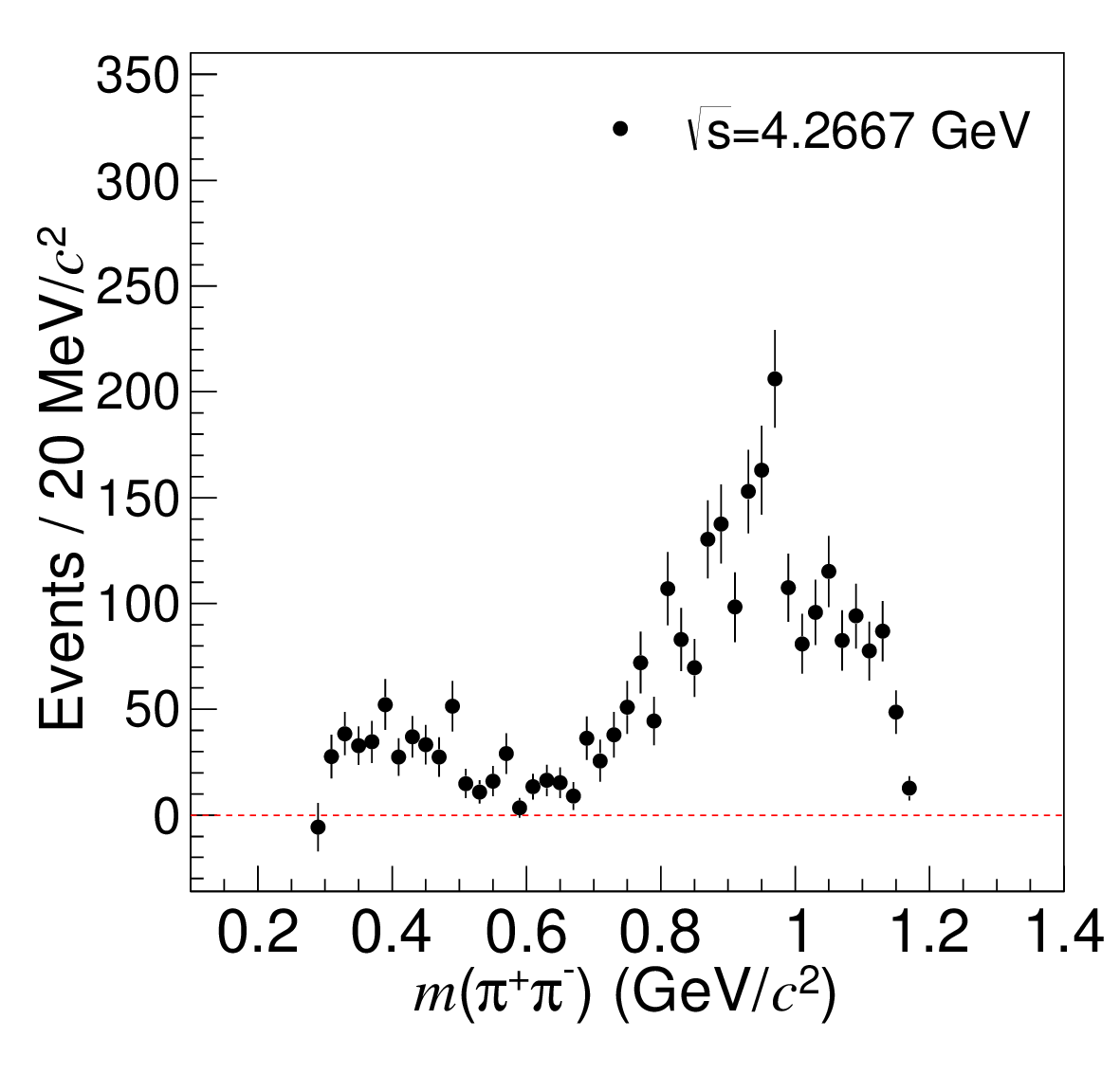}
  \includegraphics[width=0.21\textwidth, height=0.21\textwidth]{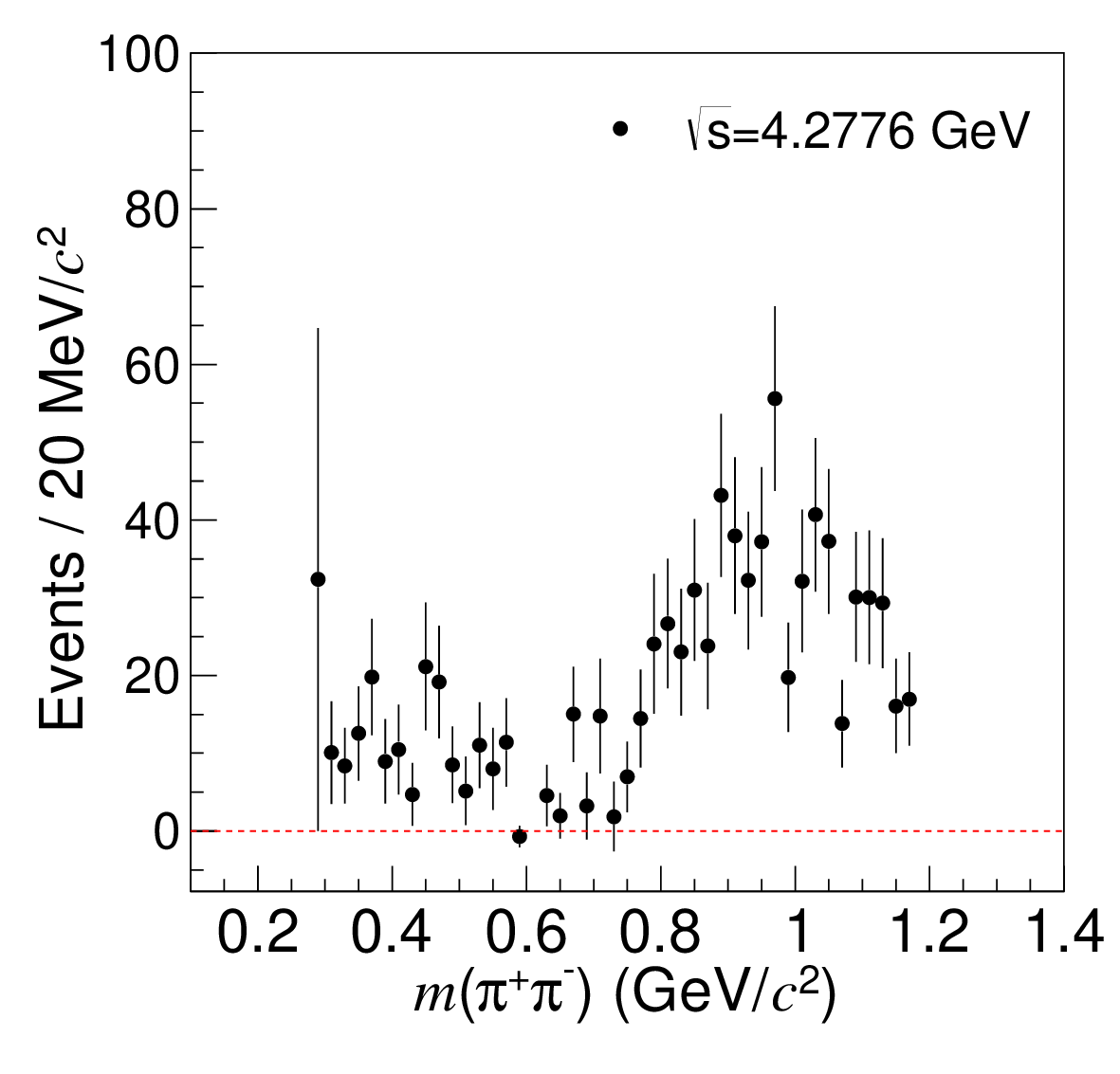}
  \includegraphics[width=0.21\textwidth, height=0.21\textwidth]{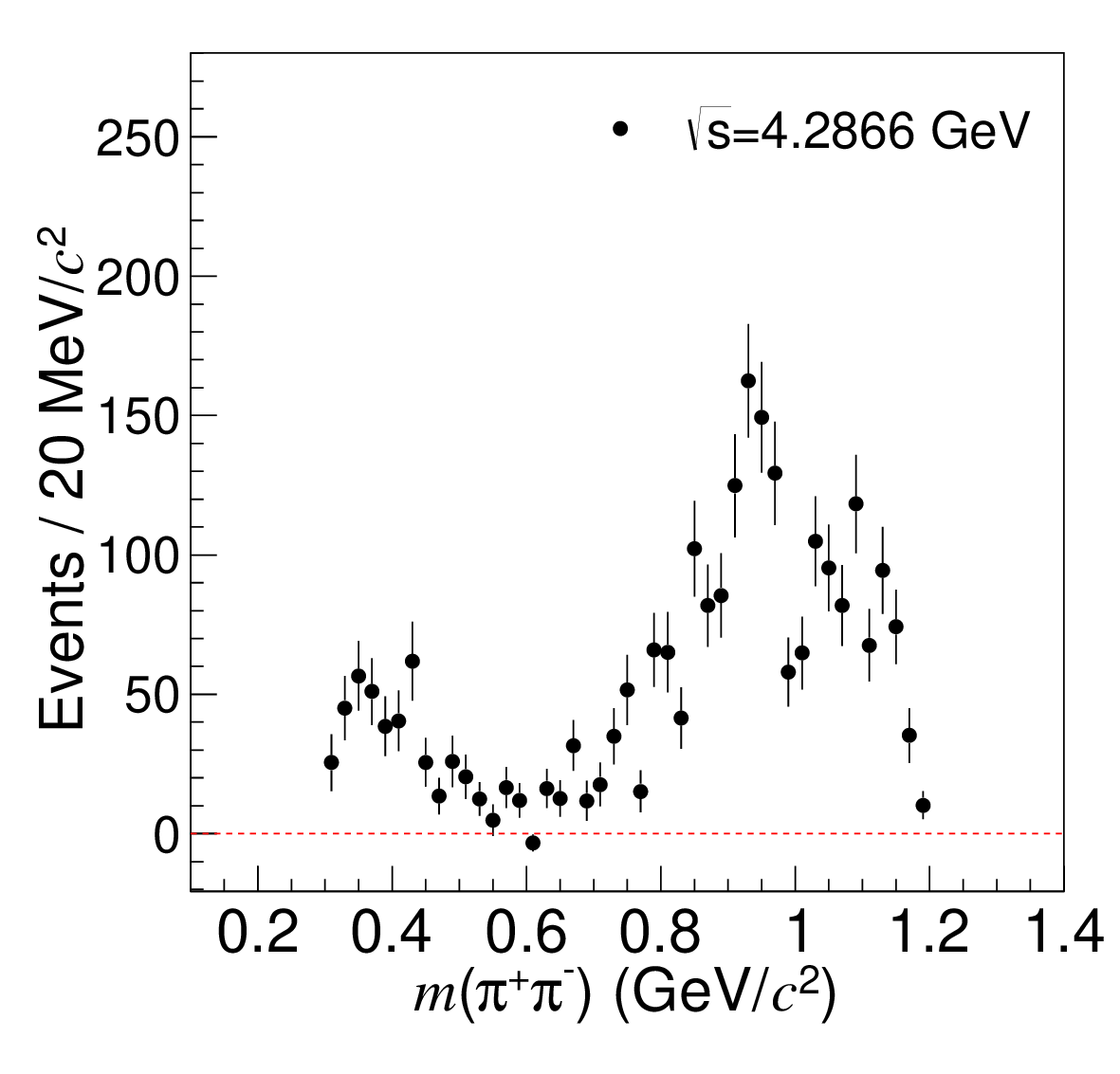}
  \includegraphics[width=0.21\textwidth, height=0.21\textwidth]{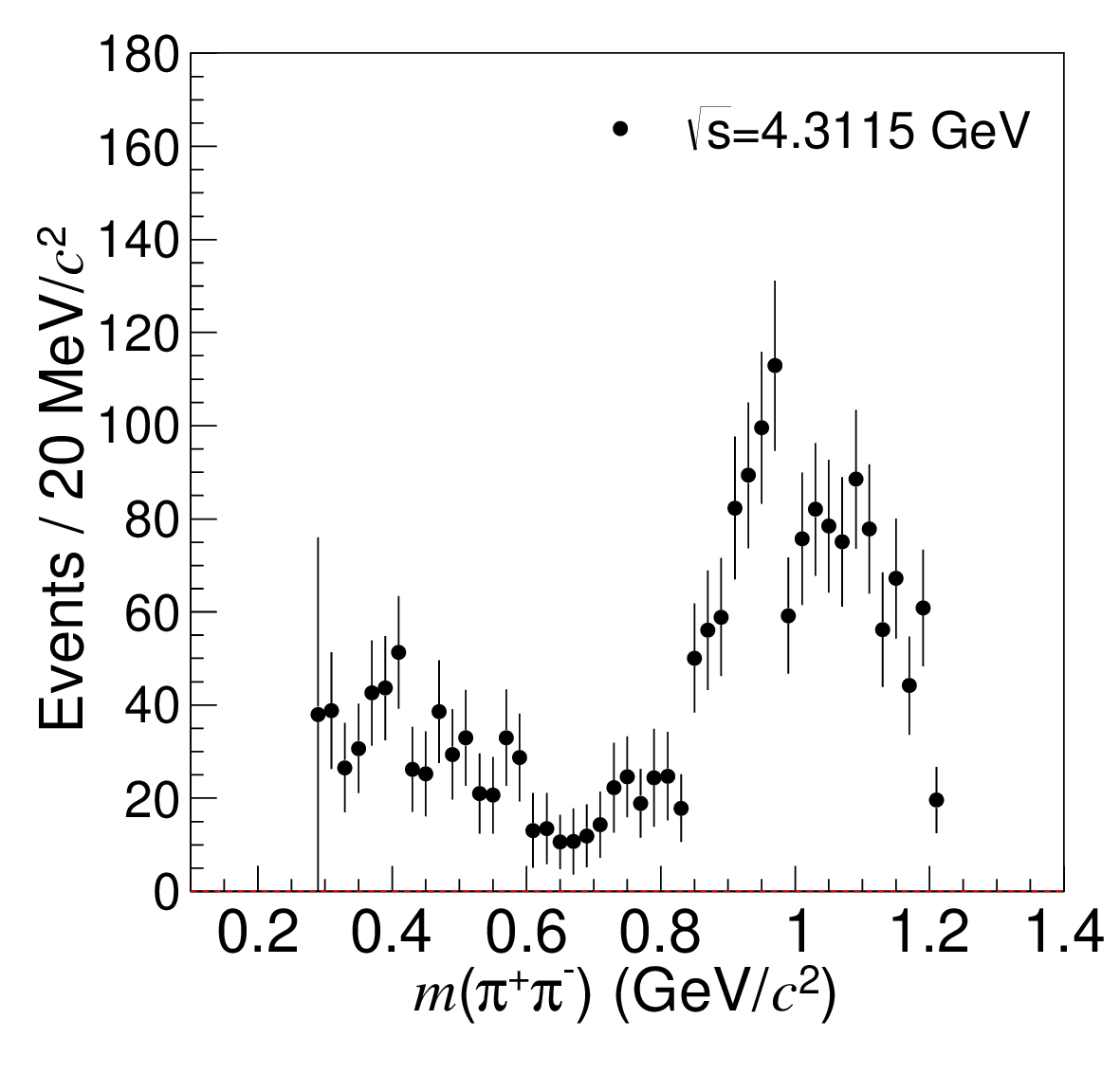}
  \includegraphics[width=0.21\textwidth, height=0.21\textwidth]{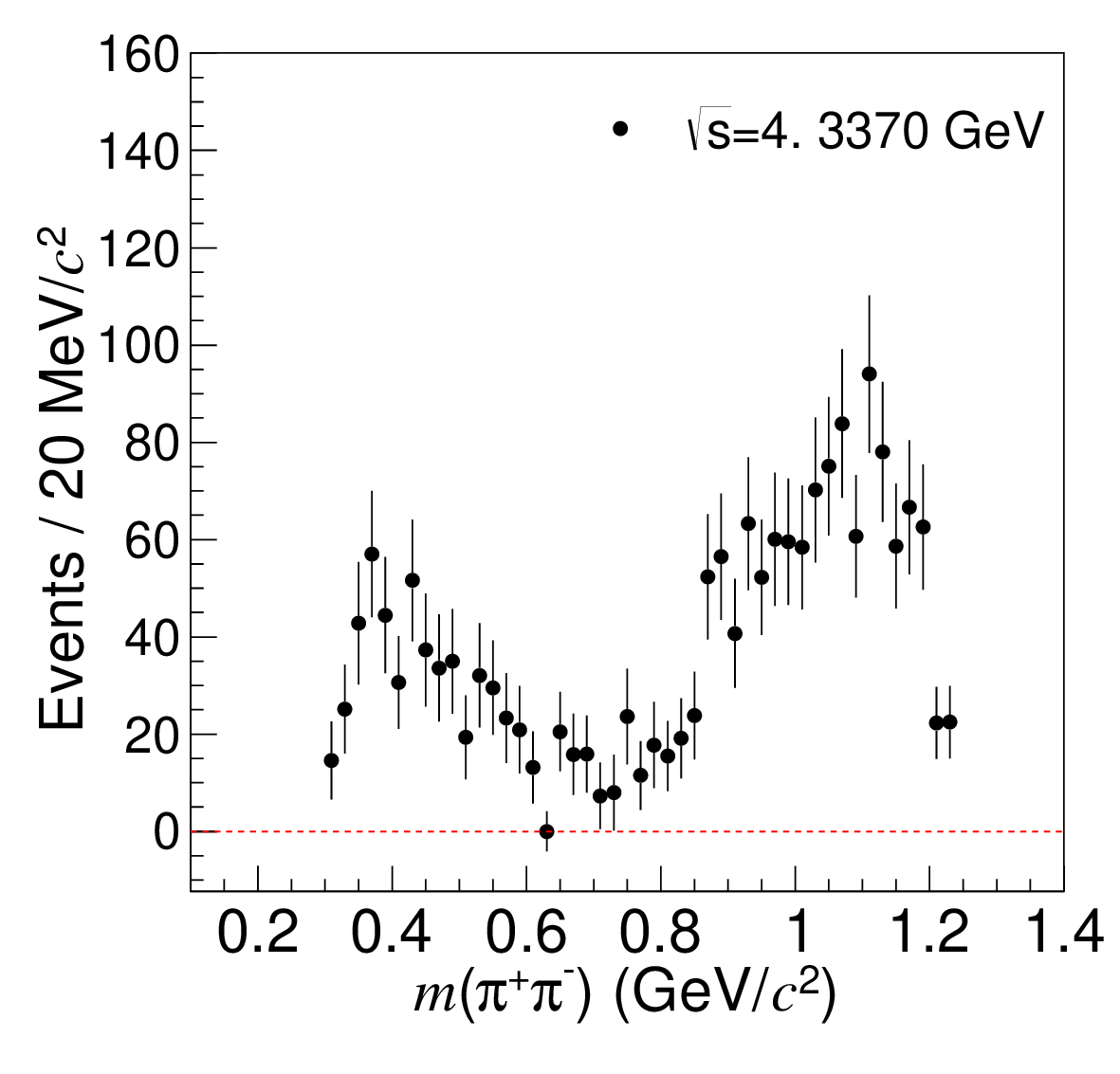}
  \includegraphics[width=0.21\textwidth, height=0.21\textwidth]{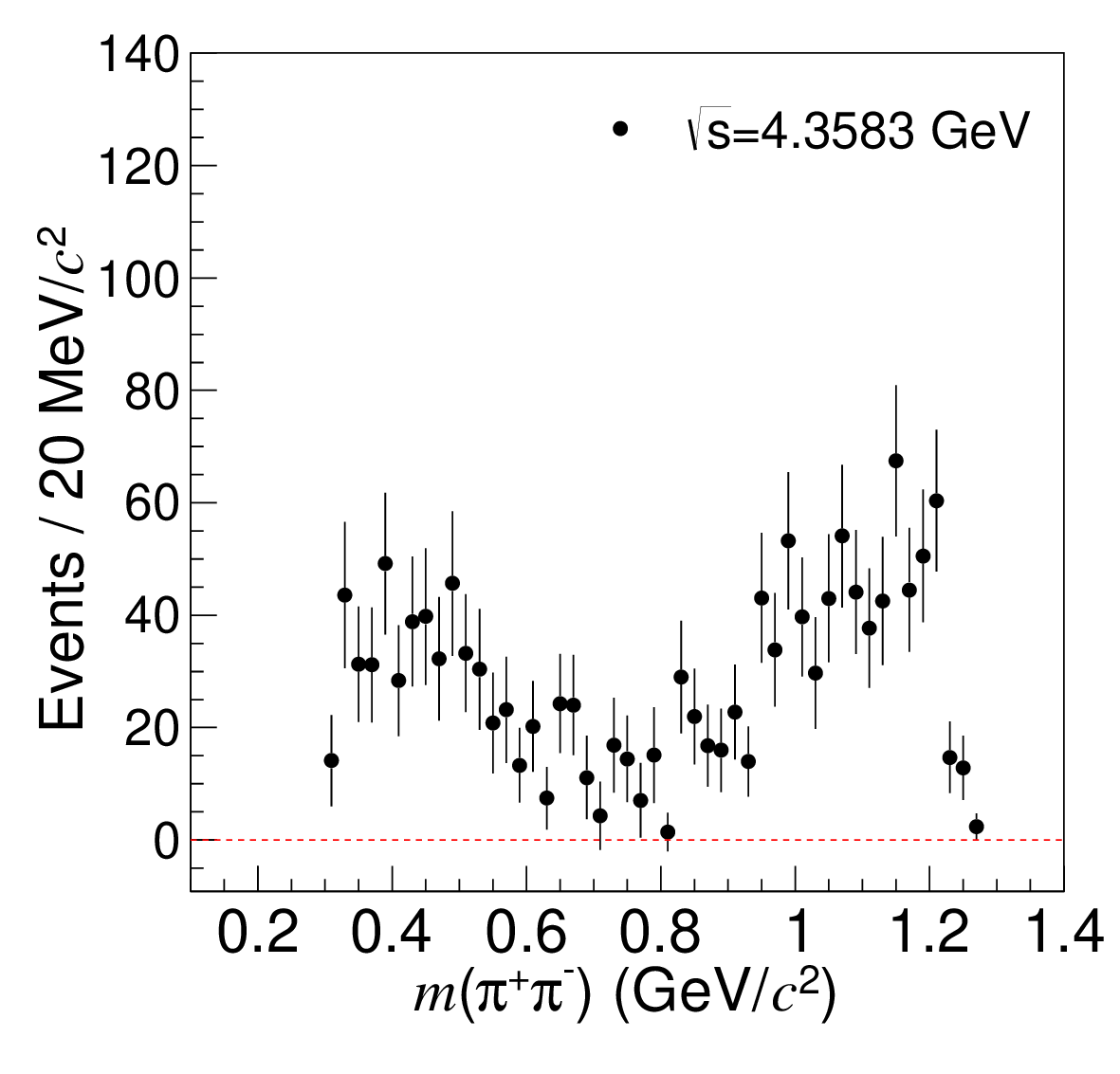}
\caption{The $m(\PP)$ distribution after the efficiency correction for the other 16 energy points except the $\sqrt{s}=$4.1780 GeV sample. The sideband events are used as an estimate of the background events and subtracted from events in the signal region.}
\label{CorrectMasspipi}
\end{figure*}

\begin{figure*}
\centering
  \includegraphics[width=0.45\textwidth, height=0.21\textwidth]{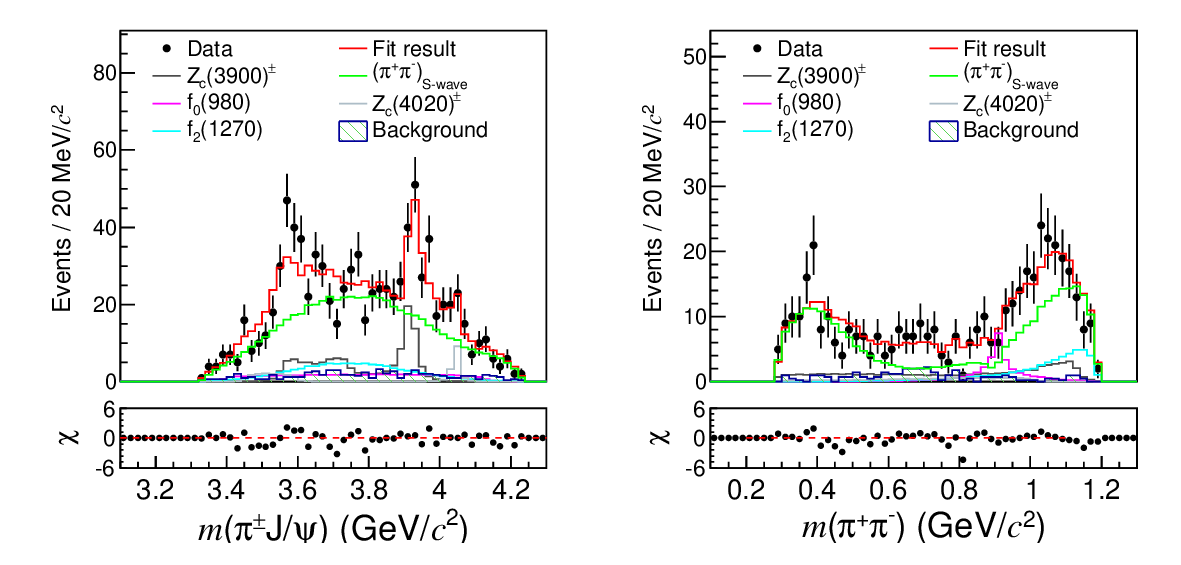}
  \includegraphics[width=0.45\textwidth, height=0.21\textwidth]{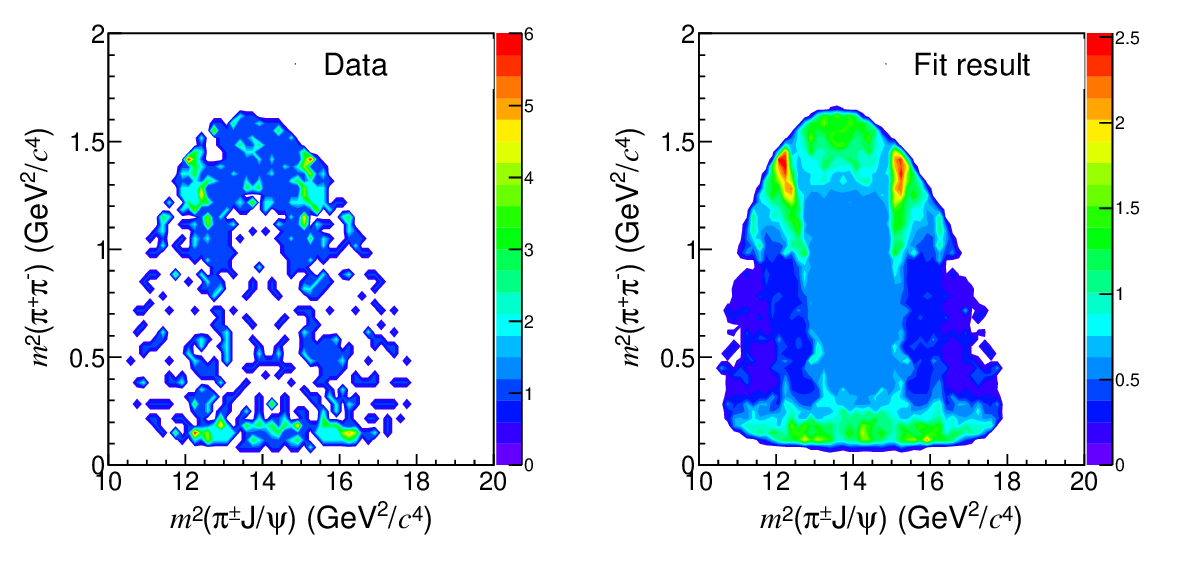}
  \includegraphics[width=0.45\textwidth, height=0.21\textwidth]{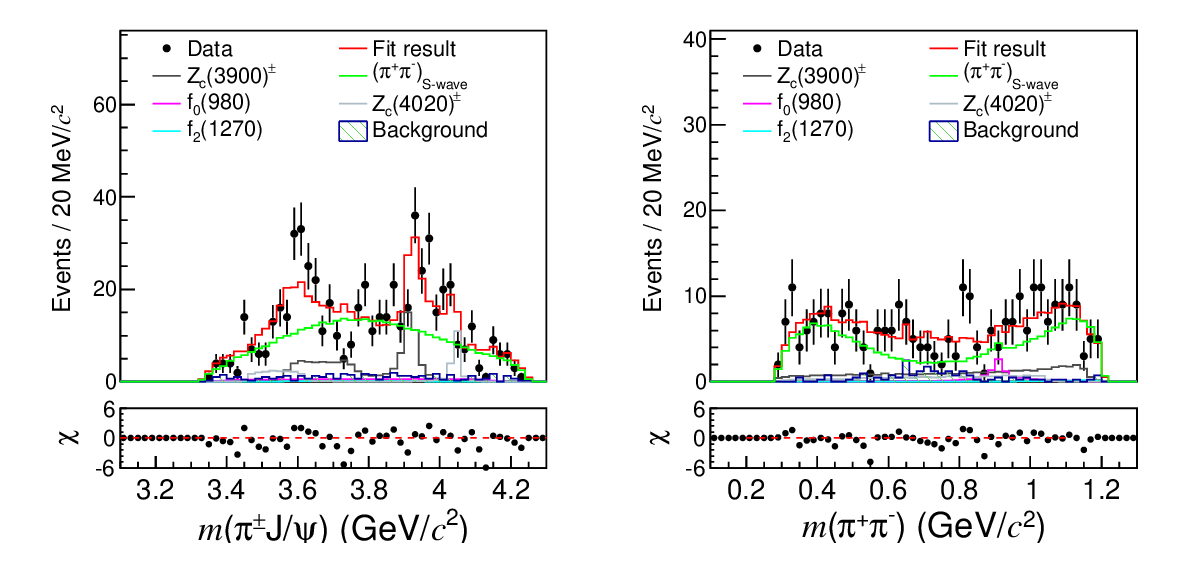}
  \includegraphics[width=0.45\textwidth, height=0.21\textwidth]{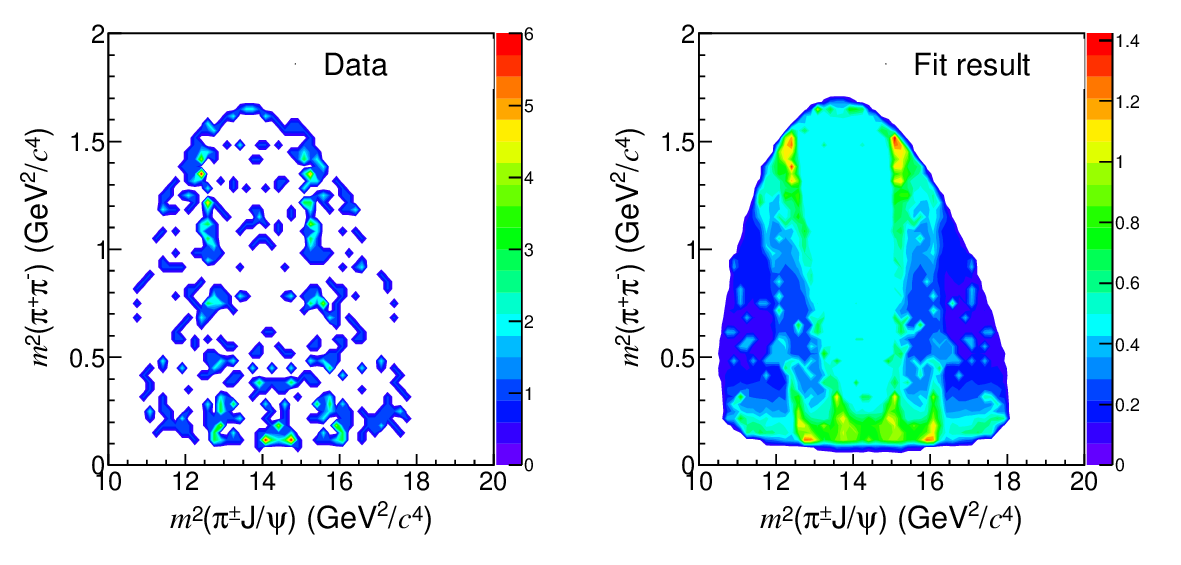}
  \includegraphics[width=0.45\textwidth, height=0.21\textwidth]{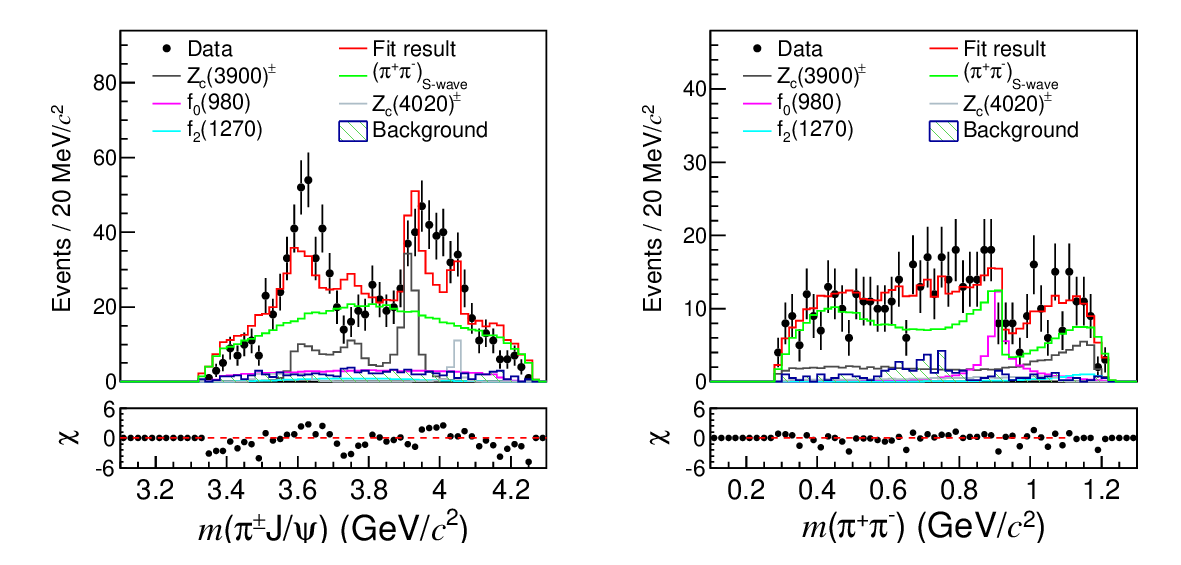}
  \includegraphics[width=0.45\textwidth, height=0.21\textwidth]{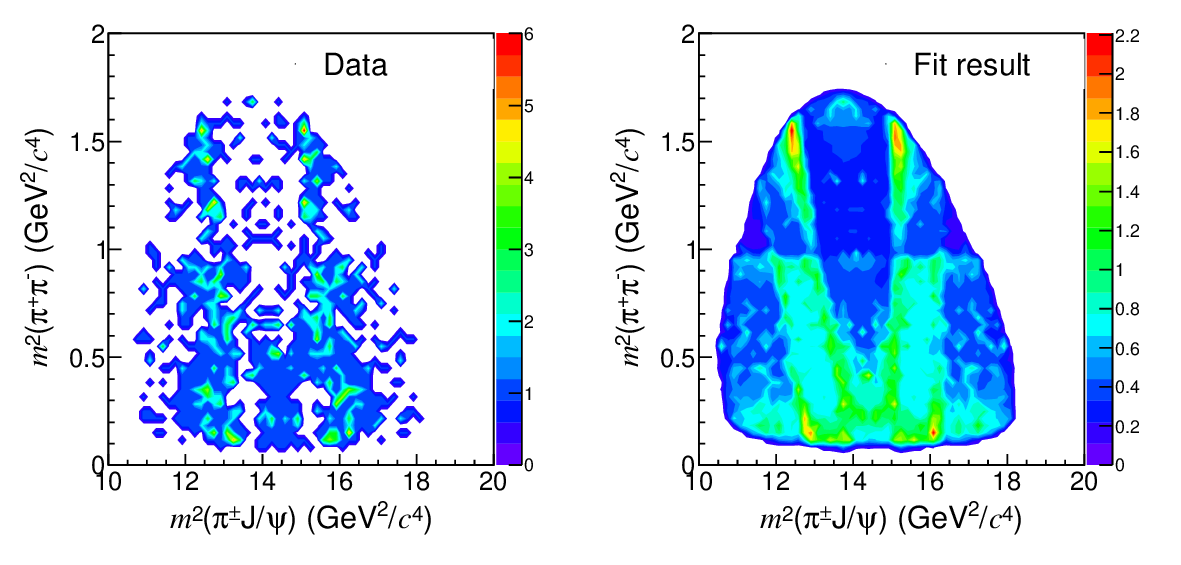}
  \includegraphics[width=0.45\textwidth, height=0.21\textwidth]{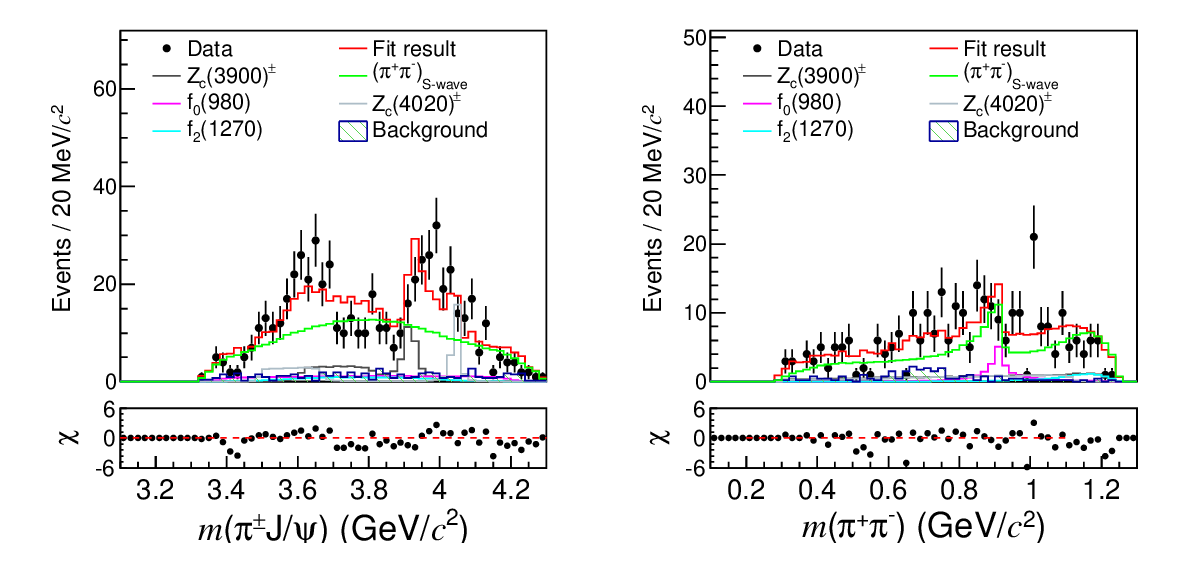}
  \includegraphics[width=0.45\textwidth, height=0.21\textwidth]{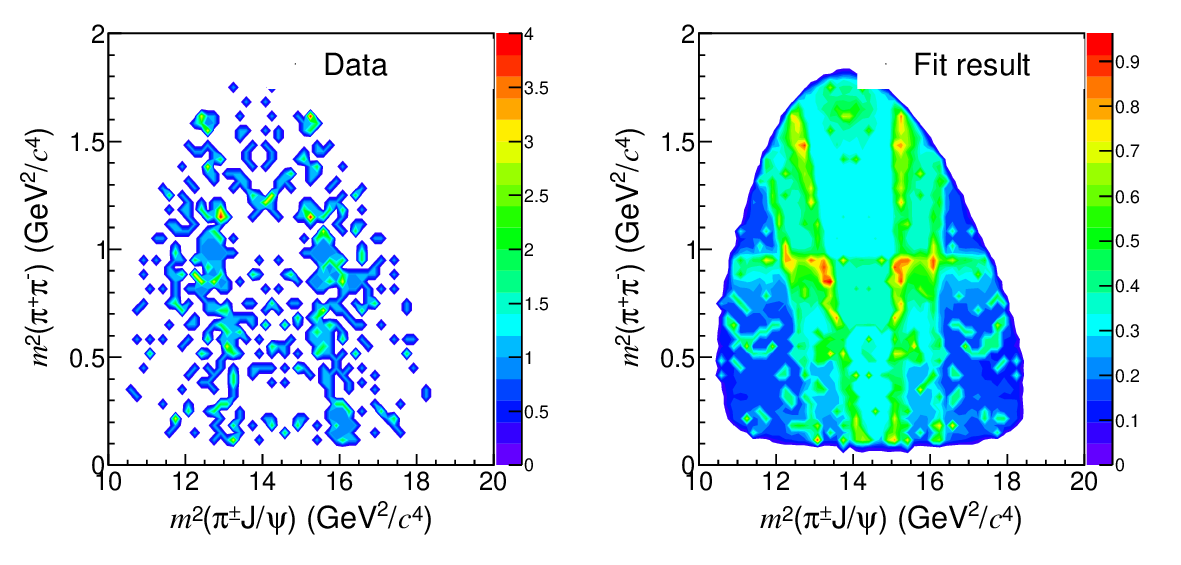}
\caption{\footnotesize{The PWA fit results with the samples at $\sqrt{s}=$4.3768 (1st row), 4.3954 (2nd row), 4.4156 (3rd row), 4.4359 (4th row) GeV.} The plots show the distributions of $m(\pi^{\pm}J/\psi)$ (1st column), $m(\pi^{+}\pi^{-})$ (2nd column), and Dalitz plots from data (3rd column) and the fit result (4th column). In addition to $\PZC$, $f_{0}(500)J/\psi$ $f_{0}(980)J/\psi$, $f_{0}(1370)J/\psi$ and $f_{2}(1270)J/\psi$, the subprocess also includes $\pi^{\pm}Z_{c}(4020)^{\mp}$. The $m(\pi^{\pm}J/\psi)$ is not well fitted.}
\label{PWAabove4360}
\end{figure*}

 Figures~\ref{CorrectMasspiJpsi} and~\ref{CorrectMasspipi} show the distributions of $m(\pi^{\pm} J/\psi)$ and $m(\PP)$ after background deduction and efficiency correction. In Figs.~\ref{CorrectMasspiJpsi} and~\ref{CorrectMasspipi}, the sideband events are used as an estimation of the background events and subtracted from events in the signal region. Figure~\ref{PWAabove4360} shows the PWA fit results of the samples with $\sqrt{s}=$4.3768, 4.3954, 4.4156, and 4.4359 GeV, and the subprocess $\pi^{\pm}Z_{c}(4020)^{\mp}$ is considered.


\par
\begin{figure*}
\centering
  \includegraphics[width=0.3\textwidth, height=0.225\textwidth]{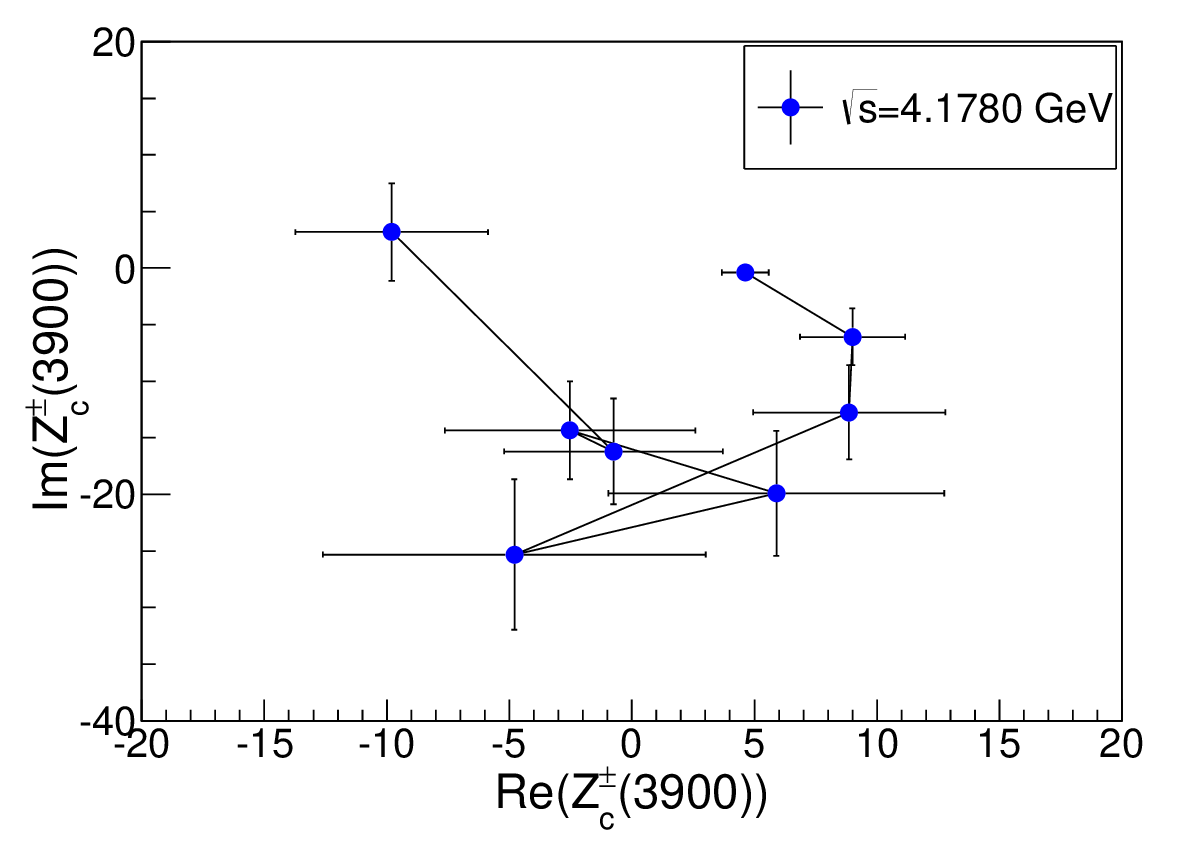}
  \includegraphics[width=0.3\textwidth, height=0.225\textwidth]{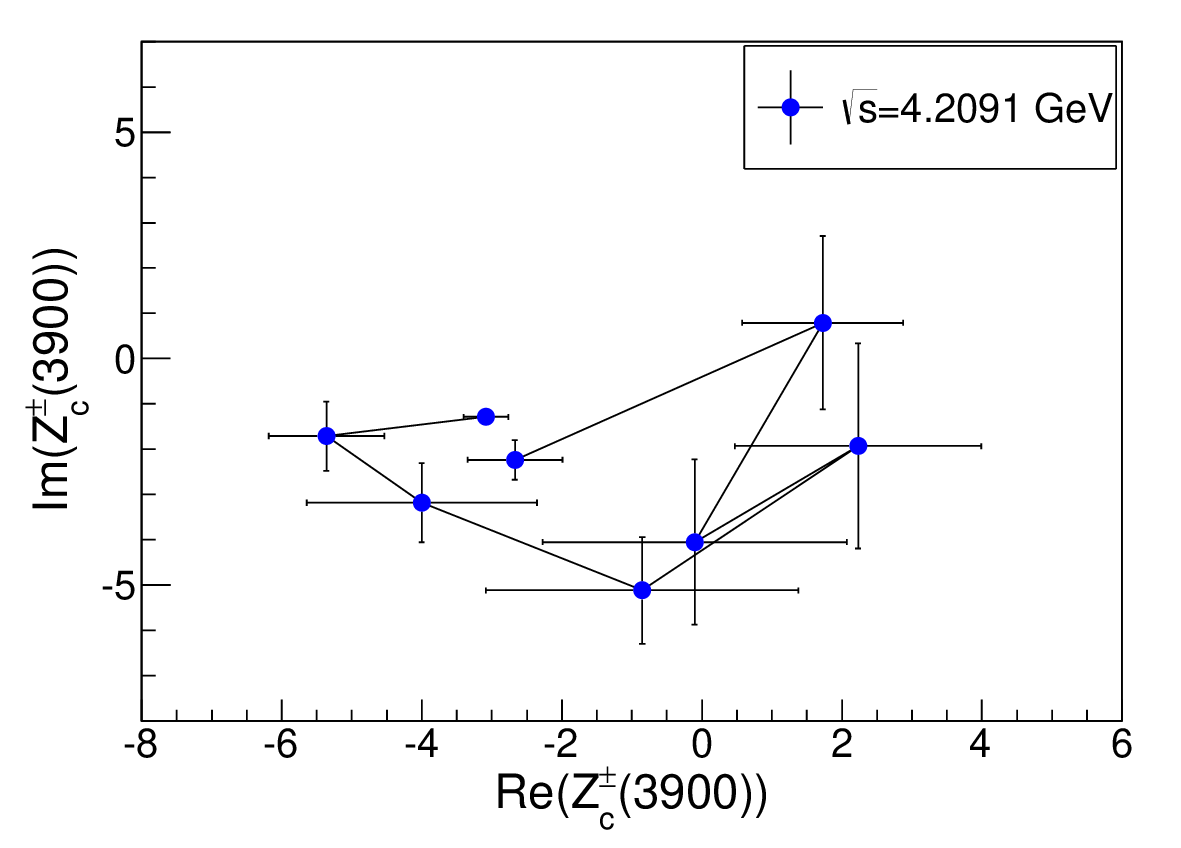}
  \includegraphics[width=0.3\textwidth, height=0.225\textwidth]{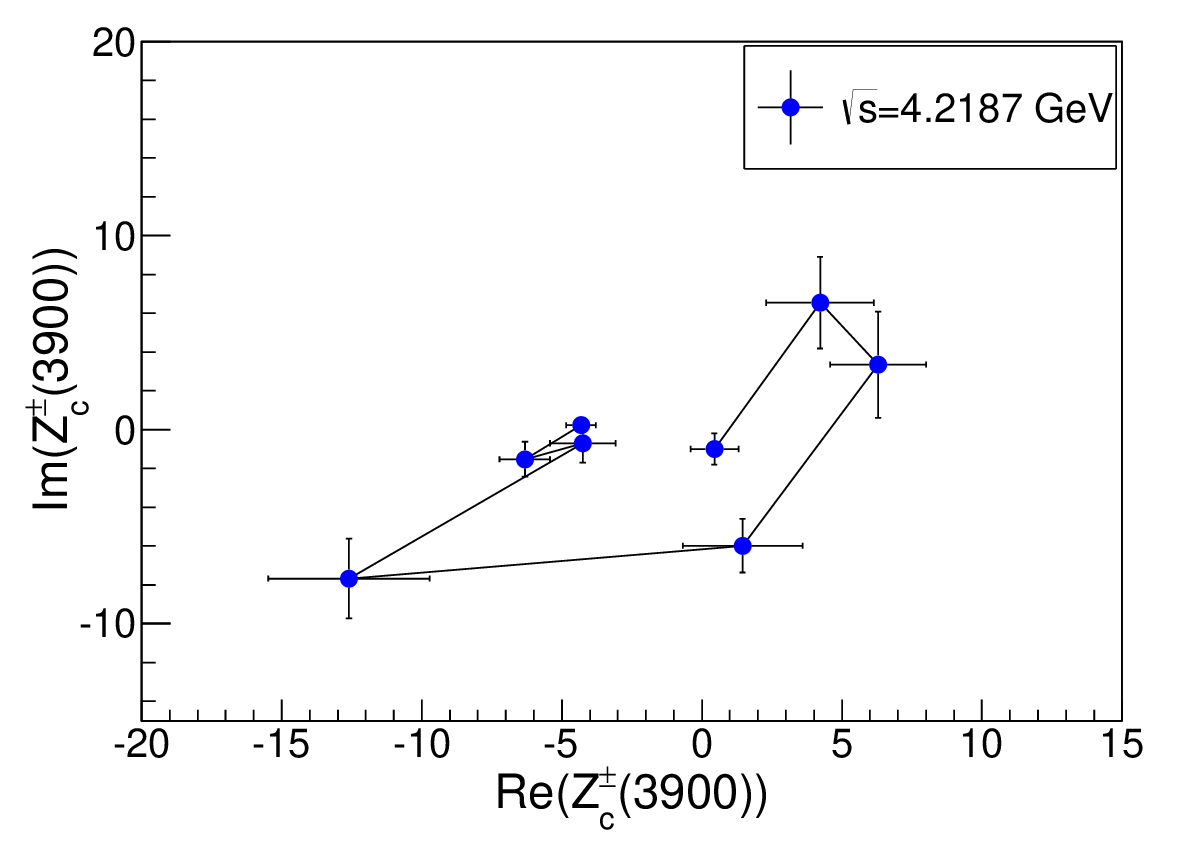}
  \includegraphics[width=0.3\textwidth, height=0.225\textwidth]{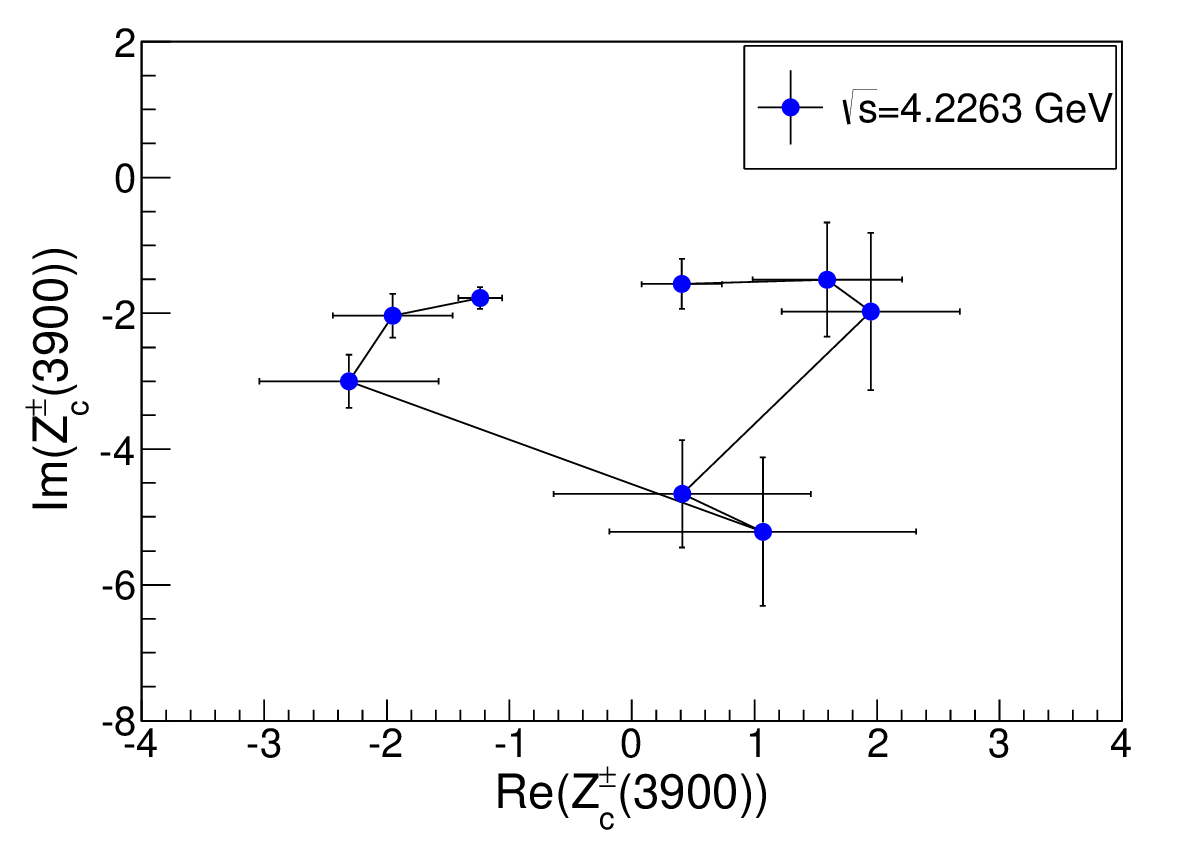}
  \includegraphics[width=0.3\textwidth, height=0.225\textwidth]{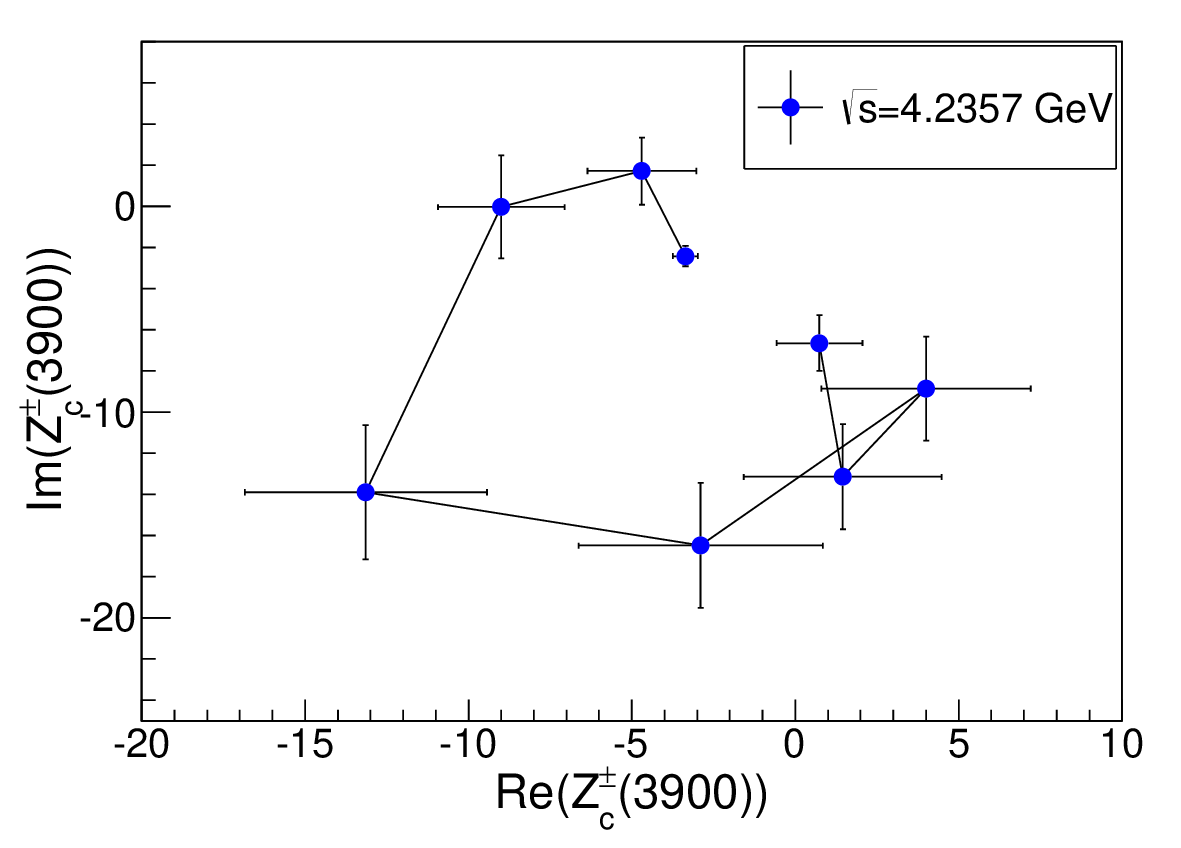}
  \includegraphics[width=0.3\textwidth, height=0.225\textwidth]{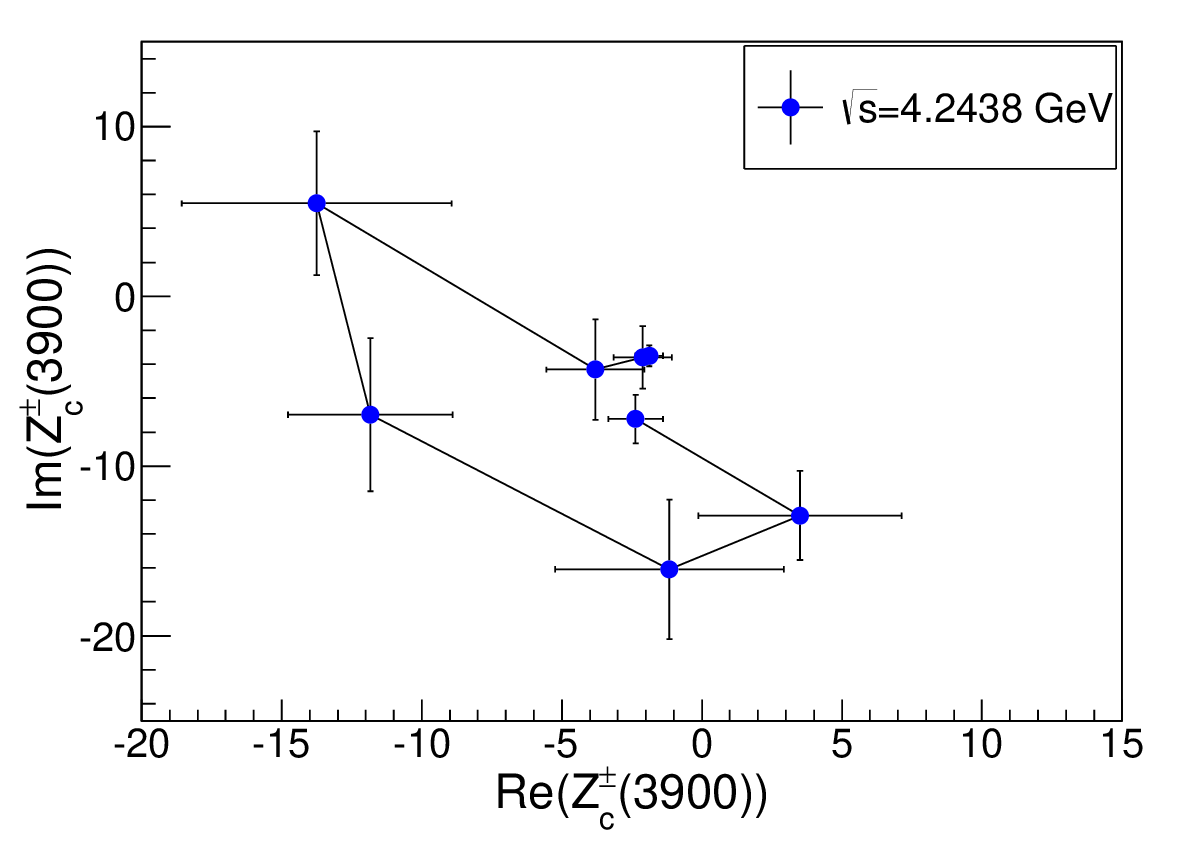}
  \includegraphics[width=0.3\textwidth, height=0.225\textwidth]{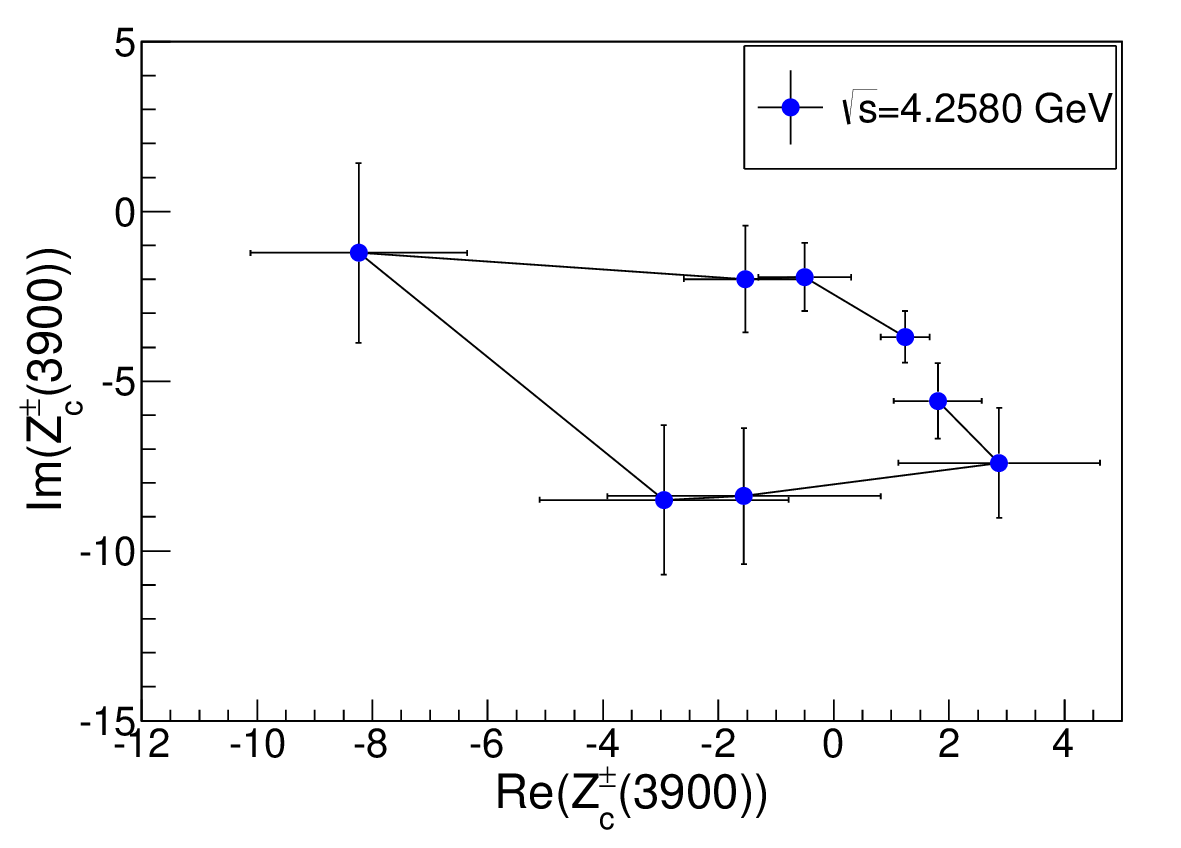}
  \includegraphics[width=0.3\textwidth, height=0.225\textwidth]{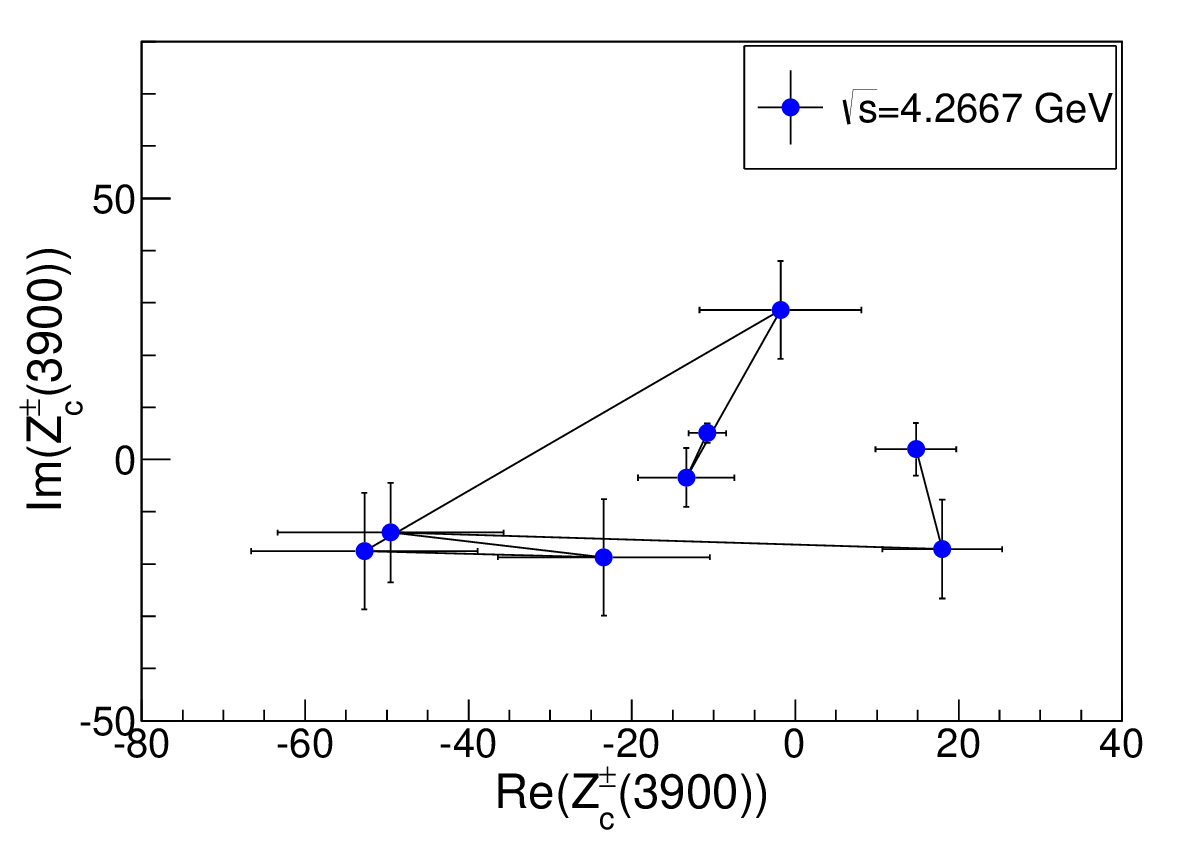}
  \includegraphics[width=0.3\textwidth, height=0.225\textwidth]{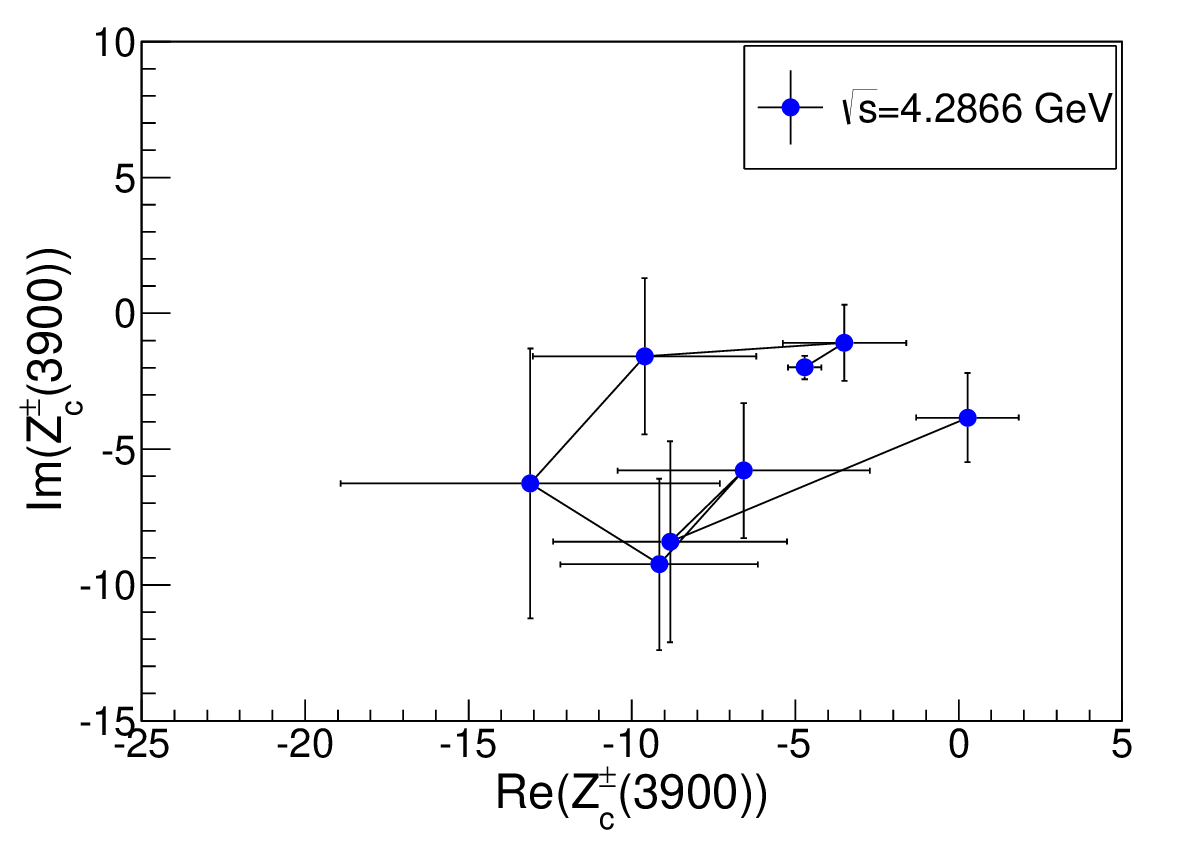}
  \includegraphics[width=0.3\textwidth, height=0.225\textwidth]{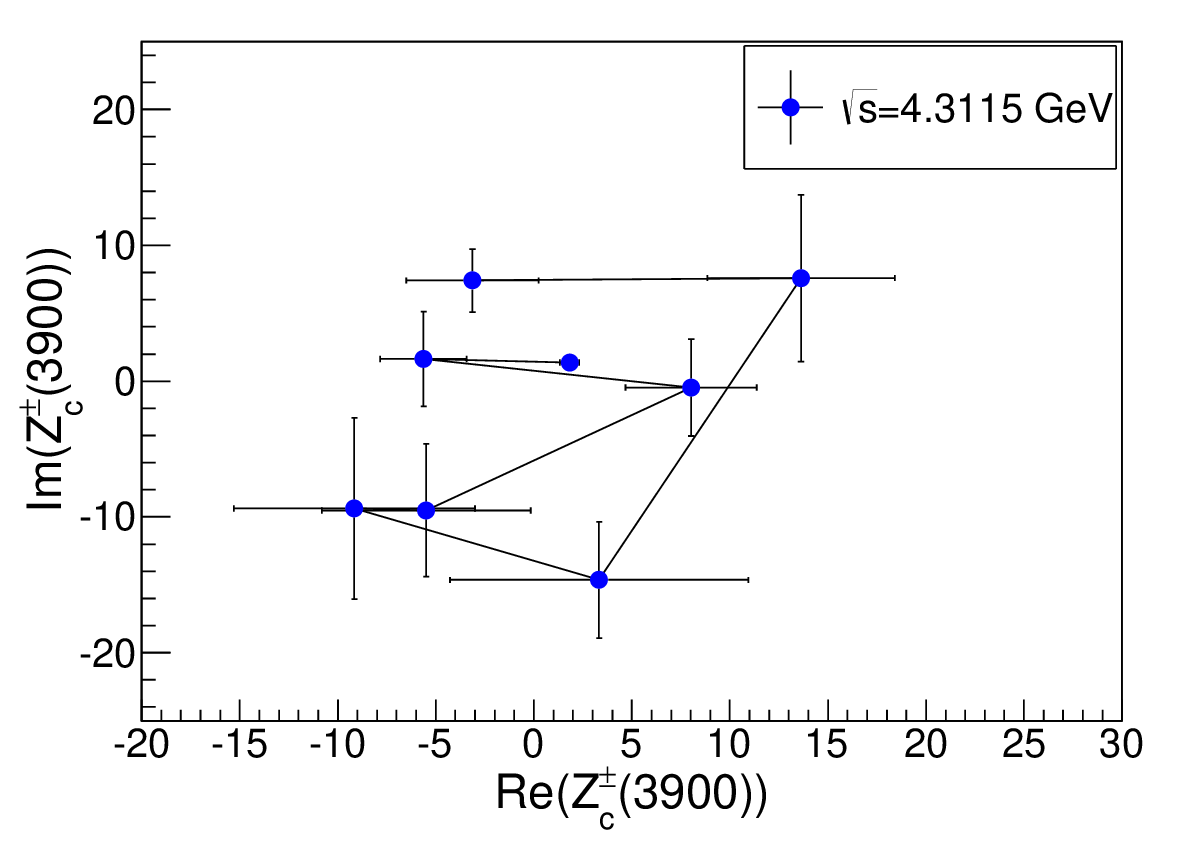}
  \includegraphics[width=0.3\textwidth, height=0.225\textwidth]{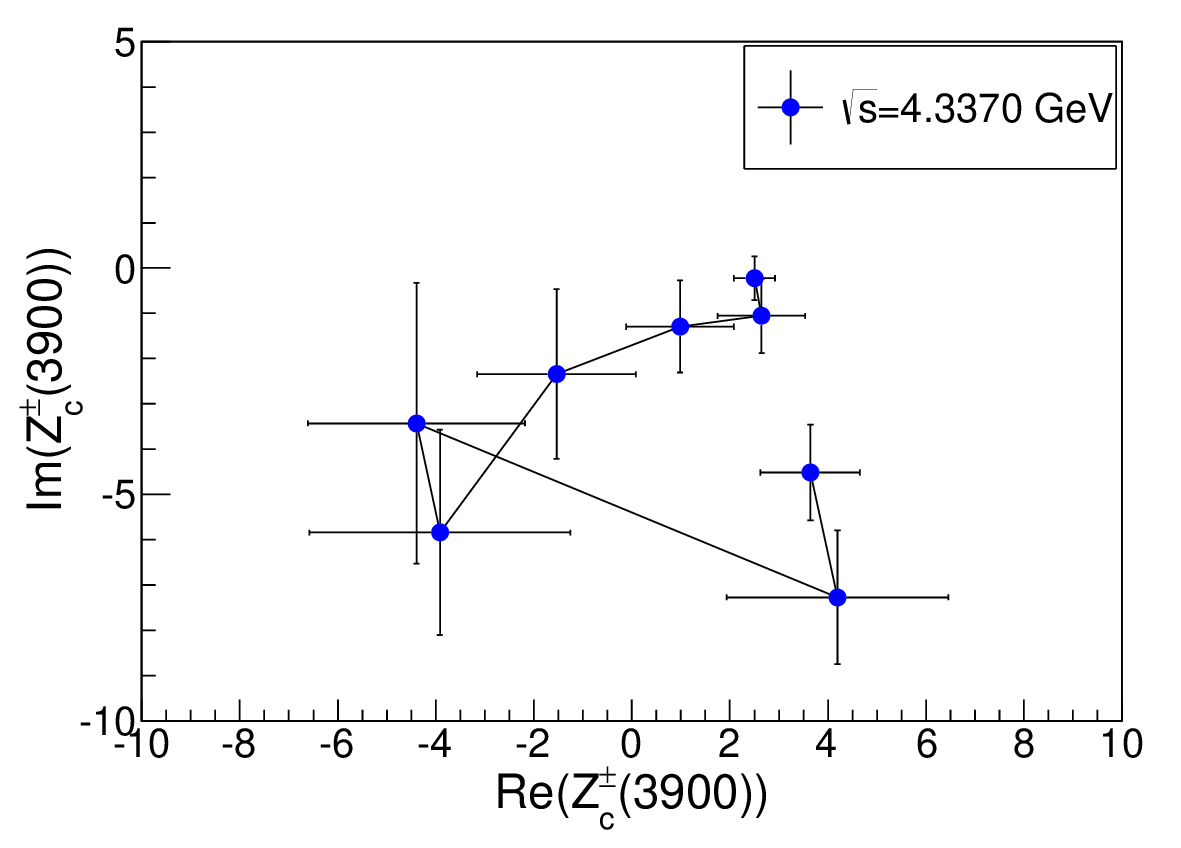}
\caption{\footnotesize{The Argand diagram with the samples at $\sqrt{s}=$4.1780, 4.2091, 4.2187, 4.2263, 4.2357, 4.2438, 4.2580, 4.2667, 4.2866, 4.3115, and 4.3370 GeV.} According to the distribution of $m(\pi^{\pm}J/\psi)$, it is divided into nine intervals: (--, 3.780), (3.780, 3.845), (3.845, 3.865), (3.865, 3.880), (3.880, 3.895), (3.895, 3.910), (3.910, 3.930), (3.930, 3.955), and (3.955, --) GeV/$c^{2}$. The two bins on the edges use the 0th and 7th pairs of parameters, and the other bins are interpolated using the neighboring parameters according to the mass of $m(\pi^{\pm}J/\psi)$. The connected line indicates the $m(\pi^{\pm}J/\psi)$ increases counterclockwise, which is expected for resonance behavior.}
\label{Agrandplot}
\end{figure*}

 Figure~\ref{Agrandplot} shows the Argand plots of 11 energy points with large statistics. The Argand plots are obtained by using a complex piecewise amplitude to describe the $Z_{c}(3900)^{\pm}$ contribution, where the complex amplitude is obtained by fitting the data and is not normalized.

\begin{table*}
\caption{The fractions (in \%) of $\pi^{\pm}Z_{c}(3900)^{\mp}$, $f_{0}(980)J/\psi$, $f_{2}(1270)J/\psi$, and $\PPS J/\psi$ corresponding to different FCN values that appeared in 10,000 random initialization fits at each energy point. Some solutions with the same FCN but with less than 100 cumulative times are not listed. The ``Nt" is the cumulative times. The $\PPS J/\psi$ is the whole of $f_{0}(500)J/\psi$, $f_{0}(980)J/\psi$, and $f_{0}(1370)J/\psi$. The $\PZC$, $f_{0}(980)J/\psi$, and $f_{2}(1270)J/\psi$ stand for $\EE\rightarrow\pi^{+}Z_{c}(3900)^{-}+c.c.\rightarrow\pi^{+}\pi^{-}J/\psi$, $f_{0}(980)(\rightarrow\pi^{+}\pi^{-})J/\psi$, and $f_{2}(1270)(\rightarrow\pi^{+}\pi^{-})J/\psi$, respectively.}
\centering
\renewcommand\arraystretch{1.2}
\renewcommand\tabcolsep{6.0pt}
\scalebox{0.9}{
\begin{tabular}{llllllll}
\hline
\hline
     $\sqrt{s}$ (GeV) & FCN  &  Nt & \makecell[c]{$\PZC$} &  \makecell[c]{$f_{0}(980)J/\psi$}    & \makecell[c]{ $f_{2}(1270)J/\psi$}   & $\PPS J/\psi$  \\
\hline
  \multirow{5}*{\makecell[c]{4.1271}} & -2335.3 & 5947 & $12.1\pm4.7$ & $53.7\pm16.8$ & $2.8\pm3.0$  & $75.1\pm11.1$  \\
                                    & -2334.9 & 1356 & $12.7\pm4.4$ & $51.0\pm15.2$ & $0.5\pm1.5$  & $81.8\pm11.0$  \\
                                    & -2334.8 & 1345 & $13.0\pm5.3$ & $47.4\pm14.8$ & $0.3\pm1.5$  & $81.7\pm11.0$  \\
                                    & -2328.3 & 316  & $11.6\pm4.5$ & $79.4\pm17.1$ & $0.5\pm1.6$  & $82.6\pm10.4$  \\
                                    & -2323.4 & 939  & $10.7\pm4.5$  & $62.4\pm19.1$& $1.9\pm2.9$  & $86.2\pm12.2$  \\
\hline
              4.1567                & -2242.1 & 10000& $17.6\pm6.3$ & $54.3\pm13.5$ & $12.3\pm4.8$ & $75.3\pm11.4$ \\
\hline
  \multirow{6}*{\makecell[c]{4.1780}} & -20981.4 & 5830 & $36.3\pm3.3$ & $36.5\pm4.9$ & $9.2\pm2.2$ & $68.4\pm4.2$  \\
                                    & -20975.8 & 1801 & $38.1\pm3.2$ & $28.5\pm5.3$  & $3.7\pm1.6$ & $73.5\pm4.3$  \\
                                    & -20939.8 & 1152 & $24.8\pm4.4$ & $40.6\pm11.5$ & $9.5\pm4.4$ & $73.0\pm6.9$  \\
                                    & -20936.3 & 109  & $22.0\pm2.1$ & $45.8\pm6.1$  & $8.2\pm3.3$ & $76.7\pm4.5$  \\
                                    & -20907.9 & 476  & $22.5\pm3.3$ & $52.3\pm6.2$  & $3.2\pm1.6$ & $71.5\pm4.5$  \\
                                    & -20895.2 & 153  & $28.7\pm2.6$ & $28.0\pm5.2$  & $19.4\pm3.2$& $64.8\pm4.6$  \\
\hline
  \multirow{7}*{\makecell[c]{4.1888}} & -4036.6 & 2704 & $30.0\pm6.7$ & $35.3\pm10.3$ & $4.3\pm3.5$ & $78.3\pm8.9$  \\
                                    & -4036.5 & 596  & $26.9\pm5.3$ & $43.2\pm8.8$   & $1.2\pm1.9$ & $79.1\pm8.5$  \\
                                    & -4034.9 & 552  & $33.7\pm7.6$ & $41.6\pm9.2$   & $0.1\pm0.3$ & $92.8\pm9.6$  \\
                                    & -4034.5 & 3380 & $15.9\pm4.1$ & $41.4\pm10.4$  & $2.1\pm2.0$ & $83.0\pm8.9$ \\
                                    & -4033.2 & 386  & $10.3\pm3.2$ & $48.2\pm9.5$   & $0.7\pm1.2$ & $85.0\pm8.8$  \\
                                    & -4033.1 & 616  & $9.2\pm2.8$  & $49.2\pm9.9$   & $0.9\pm1.6$ & $84.8\pm8.9$  \\
                                    & -4025.8 & 1588 & $25.7\pm9.7$ & $38.4\pm10.1$  & $6.0\pm4.4$ & $78.3\pm8.9$  \\
\hline
  \multirow{2}*{\makecell[c]{4.1989}} & -6497.0 & 5005 & $22.4\pm3.6$ & $34.2\pm8.4$  & $0.2\pm0.9$ & $71.3\pm6.5$  \\
                                    & -6495.1 & 4962 & $44.5\pm5.6$ & $32.0\pm7.1$  & $1.5\pm1.4$ & $69.3\pm6.6$  \\
\hline
  \multirow{4}*{\makecell[c]{4.2091}} & -12580.5 & 3283 & $16.4\pm2.3$ & $42.0\pm8.6$  & $0.7\pm0.9$ & $81.9\pm5.2$  \\
                                    & -12580.4 & 4862 & $16.2\pm2.4$ & $52.9\pm12.1$ & $0.5\pm0.8$ & $81.2\pm5.0$  \\
                                    & -12578.0 & 1142 & $13.4\pm2.1$ & $55.6\pm8.3$  & $0.4\pm0.6$ & $79.2\pm5.0$ \\
                                    & -12575.9 & 662  & $13.2\pm2.1$ & $49.2\pm7.7$  & $1.1\pm1.0$ & $80.6\pm5.1$  \\
\hline
  \multirow{3}*{\makecell[c]{4.2187}} & -20265.7 & 6780 & $16.0\pm2.3$ & $30.3\pm4.0$ & $1.5\pm1.3$ & $75.9\pm4.0$ \\
                                    & -20261.2 & 2880 & $20.2\pm2.2$ & $30.8\pm3.6$ & $0.5\pm0.5$ & $76.9\pm4.0$ \\
                                    & -20254.5 & 306  & $13.8\pm1.8$ & $33.9\pm4.1$ & $0.6\pm0.8$ & $76.4\pm4.1$ \\
\hline
  \multirow{3}*{\makecell[c]{4.2263}} & -54577.5 & 5841 & $12.8\pm1.1$ & $35.3\pm3.1$ & $2.2\pm0.9$ & $78.9\pm2.7$ \\
                                    & -54551.9 & 3200 & $15.7\pm1.2$ & $29.2\pm2.5$ & $0.5\pm0.5$ & $80.9\pm2.7$  \\
                                    & -54535.6 & 949  & $14.6\pm1.2$ & $46.1\pm4.9$ & $0.6\pm0.4$ & $79.8\pm2.6$ \\
\hline
  \multirow{3}*{\makecell[c]{4.2357}} & -24313.9 & 6716 & $12.3\pm1.6$ & $26.9\pm3.5$ & $1.4\pm1.4$ & $83.5\pm4.1$  \\
                                    & -24311.4 & 1105 & $14.0\pm1.8$ & $27.1\pm3.7$ & $2.2\pm1.6$ & $83.5\pm4.2$  \\
                                    & -24279.7 & 2154 & $26.7\pm3.0$ & $25.7\pm3.2$ & $0.4\pm0.7$ & $90.9\pm4.2$  \\
\hline
  \multirow{5}*{\makecell[c]{4.2438}} & -24622.5 & 1982 & $7.3\pm1.2$ & $33.2\pm7.7$ & $0.9\pm1.1$ & $90.4\pm4.1$  \\
                                    & -24621.5 & 3773 & $7.3\pm1.3$ & $18.9\pm2.5$ & $0.2\pm0.4$ & $91.9\pm4.1$  \\
                                    & -24614.3 & 383  & $8.6\pm1.3$ & $18.6\pm2.3$ & $0.5\pm0.6$ & $92.2\pm4.0$  \\
                                    & -24612.1 & 3840 & $8.5\pm1.3$ & $34.1\pm9.2$ & $0.4\pm0.5$ & $91.7\pm4.0$  \\
                                    & -24575.6 & 3840 & $6.7\pm1.2$ & $72.6\pm7.2$ & $0.3\pm0.4$ & $91.2\pm4.1$  \\
\hline
\hline
\end{tabular}}
\label{FCNallBW1}
\end{table*}

\clearpage

\begin{table*}
\caption{The fractions (in \%) of $\pi^{\pm}Z_{c}(3900)^{\mp}$, $f_{0}(980)J/\psi$, $f_{2}(1270)J/\psi$, and $\PPS J/\psi$ corresponding to different FCN values that appeared in 10,000 random initialization fits at each energy point. Some solutions with the same FCN but with less than 100 cumulative times are not listed. The ``Nt" is the cumulative times. The $\PPS J/\psi$ is the whole of $f_{0}(500)J/\psi$, $f_{0}(980)J/\psi$, and $f_{0}(1370)J/\psi$. The $\PZC$, $f_{0}(980)J/\psi$, $f_{2}(1270)J/\psi$ stand for $\EE\rightarrow\pi^{+}Z_{c}(3900)^{-}+c.c.\rightarrow\pi^{+}\pi^{-}J/\psi$, $f_{0}(980)(\rightarrow\pi^{+}\pi^{-})J/\psi$, and $f_{2}(1270)(\rightarrow\pi^{+}\pi^{-})J/\psi$, respectively.}
\centering
\renewcommand\arraystretch{1.2}
\renewcommand\tabcolsep{6.0pt}
\scalebox{0.9}{
\begin{tabular}{llllllll}
\hline
\hline
  $\sqrt{s}$ (GeV)   & FCN  &  Nt & \makecell[c]{$\PZC$} &  \makecell[c]{$f_{0}(980)J/\psi$}    & \makecell[c]{ $f_{2}(1270)J/\psi$}   & $\PPS J/\psi$  \\
\hline
  \multirow{4}*{\makecell[c]{4.2580}} & -28755.7 & 7958 & $8.3\pm1.2$  & $25.2\pm4.7$  & $6.8\pm2.6$  & $83.6\pm4.6$ \\
                                    & -28698.2 & 1188 & $8.3\pm1.1$  & $68.3\pm7.4$   & $0.5\pm0.7$  & $88.9\pm3.7$ \\
                                    & -28693.6 &  476 & $9.9\pm1.4$  & $91.8\pm15.2$  & $1.8\pm1.2$  & $87.2\pm3.9$ \\
                                    & -28684.9 &  366 & $10.9\pm1.4$ & $98.4\pm9.0$   & $2.1\pm1.2$  & $86.4\pm3.8$ \\
\hline
  \multirow{5}*{\makecell[c]{4.2667}} & -16248.2 & 3872 &$7.1\pm1.5$ & $8.8\pm2.4 $ & $0.1\pm0.2 $  & $94.2\pm4.9$ \\
                                    & -16234.3 & 2261 & $7.0\pm1.4$  & $44.3\pm6.0$ & $ 1.1\pm1.3$  & $91.8\pm4.9$ \\
                                    & -16223.7 & 1941 & $14.0\pm2.6$ & $38.4\pm5.7$ & $ 1.7\pm1.9$  & $95.4\pm5.3$ \\
                                    & -16203.5 & 279  & $14.5\pm2.3$ & $42.6\pm6.8$ & $ 3.1\pm2.5$  & $93.0\pm5.4$ \\
                                    & -16201.6 & 1531 & $9.9\pm1.9 $ & $41.3\pm7.0$ & $ 14.6\pm4.1$ & $77.9\pm6.1$ \\
\hline
  \multirow{9}*{\makecell[c]{4.2776}} & -4179.6 & 1425 & $13.7\pm4.2$ & $28.2\pm9.6  $ & $12.2\pm6.9$ & $85.3\pm10.8$ \\
                                     & -4177.9 & 140  & $12.7\pm3.9$ & $30.9\pm9.7  $ & $8.6\pm5.3$  & $89.4\pm10.5$ \\
                                     & -4177.6 & 1372 & $13.3\pm4.1$ & $30.1\pm9.3  $ & $8.5\pm5.9$  & $88.7\pm10.2$ \\
                                     & -4174.4 & 113  & $6.7\pm2.9 $ & $4.2\pm2.3   $ & $0.1\pm0.4$  & $93.0\pm8.4$  \\
                                     & -4173.5 & 323  & $6.3\pm2.3 $ & $5.6\pm5.0   $ & $1.3\pm2.6$  & $92.4\pm8.6$  \\
                                     & -4170.3 & 120  & $11.7\pm3.8$ & $33.3\pm10.1 $ & $3.9\pm3.8$  & $86.2\pm9.1$  \\
                                     & -4164.4 & 5221 & $14.8\pm3.7$ & $92.8\pm13.1 $ & $4.9\pm3.4$  & $85.0\pm8.2$  \\
                                     & -4162.4 & 270  & $15.2\pm3.8$ & $93.1\pm13.3 $ & $4.0\pm3.0$  & $88.0\pm8.1$  \\
                                     & -4158.0 & 186  & $11.8\pm3.5$ & $101.0\pm15.2$ & $1.9\pm2.3$  & $89.3\pm8.1$  \\
\hline
  \multirow{3}*{\makecell[c]{4.2866}} & -14157.2 & 5614 & $9.6\pm1.8$ & $6.4\pm1.7$ & $3.0\pm1.7$ & $89.9\pm5.3$ \\
                                    & -14150.9 & 1245 & $7.9\pm1.7$ & $9.2\pm2.5$ & $7.5\pm3.5$ & $84.6\pm6.2$ \\
                                    & -14148.3 & 3120 & $7.0\pm1.4$ & $8.4\pm2.1$ & $5.7\pm3.0$ & $88.0\pm5.8$ \\
\hline
  \multirow{4}*{\makecell[c]{4.3115}} & -11081.1 & 6636 & $8.4\pm1.9 $ & $5.6\pm2.5$ & $9.5\pm4.7$  & $83.8\pm7.9$ \\
                                    & -11077.7 & 1843 & $6.2\pm1.5 $ & $4.0\pm1.8$ & $4.6\pm2.8$  & $87.8\pm6.3$ \\
                                    & -11061.5 & 576  & $22.1\pm4.0$ & $6.0\pm3.0$ & $9.1\pm3.7$  & $90.9\pm7.5$ \\
                                    & -11038.1 & 400  & $13.1\pm4.0$ & $7.4\pm3.1$ & $36.0\pm7.5$ & $55.6\pm12.8$ \\
\hline
  \multirow{6}*{\makecell[c]{4.3370}} & -9396.1 & 6303 & $12.1\pm2.7$ & $2.7\pm2.2 $ & $8.7\pm5.5$ & $88.2\pm8.7$  \\
                                    & -9391.2 & 946  & $10.3\pm2.2$ & $1.3\pm1.2 $ & $1.5\pm1.5$ & $91.9\pm6.5$  \\
                                    & -9390.3 & 736  & $24.8\pm4.2$ & $0.9\pm1.1 $ & $0.8\pm1.2$ & $96.5\pm6.5$  \\
                                    & -9368.0 & 206  & $13.7\pm2.5$ & $22.9\pm8.9$ & $2.0\pm2.3$ & $89.1\pm6.6$  \\
                                    & -9352.6 & 150  & $23.1\pm3.8$ & $58.0\pm8.3$ & $1.0\pm1.3$ & $94.8\pm6.2$  \\
                                    & -9348.3 & 1635 & $7.5\pm1.9 $ & $1.5\pm1.9 $ & $5.6\pm4.1$ & $89.8\pm7.2$  \\
\hline
  \multirow{8}*{\makecell[c]{4.3583}} & -6633.9 & 1661 & $9.2\pm2.3 $ & $2.1\pm2.0 $ & $1.7\pm1.4$ & $91.9\pm7.1$  \\
                                    & -6629.1 & 3826 & $10.8\pm2.6$ & $36.0\pm8.2$ & $1.9\pm1.5$ & $92.7\pm7.3$  \\
                                    & -6627.9 & 206  & $8.3\pm2.4 $ & $1.7\pm1.7 $ & $1.0\pm1.0$ & $89.5\pm7.2$  \\
                                    & -6626.8 & 213  & $11.1\pm3.4$ & $4.5\pm3.7 $ & $8.7\pm6.0$ & $87.0\pm8.6$  \\
                                    & -6625.6 & 393  & $12.5\pm3.4$ & $38.6\pm7.6$ & $3.1\pm2.4$ & $92.8\pm7.5$  \\
                                    & -6611.8 & 233  & $8.5\pm2.5 $ & $18.4\pm9.5$ & $1.7\pm1.3$ & $90.6\pm6.9$ \\
                                    & -6611.0 & 113  & $10.6\pm3.2$ & $42.1\pm8.9$ & $2.2\pm1.6$ & $92.3\pm7.1$  \\
                                    & -6608.6 & 3107  & $7.0\pm2.4 $ & $44.4\pm9.4$ & $2.0\pm1.5$ & $93.1\pm7.2$  \\
\hline
\hline
\end{tabular}}
\label{FCNallBW2}
\end{table*}

\begin{sidewaystable*}
\caption{The fit fractions (in \%) of the subprocesses in the PWA for different energy points, calculated by using the Breit-Wigner and K-matrix methods to parametrize the $\PPS$. The $\PPS J/\psi$ is the whole of $f_{0}(500)J/\psi$, $f_{0}(980)J/\psi$, and $f_{0}(1370)J/\psi$. The $\PZC$, $f_{0}(980)J/\psi$, $f_{2}(1270)J/\psi$ stand for $\EE\rightarrow\pi^{+}Z_{c}(3900)^{-}+c.c.\rightarrow\pi^{+}\pi^{-}J/\psi$, $f_{0}(980)(\rightarrow\pi^{+}\pi^{-})J/\psi$, and $f_{2}(1270)(\rightarrow\pi^{+}\pi^{-})J/\psi$, respectively. The first uncertainty is statistical, and the second one systematic. }
\centering
\renewcommand\arraystretch{1.1}
\renewcommand\tabcolsep{5.0pt}
\scalebox{1.0}{
\begin{tabular}{llllllll}
\toprule
\toprule
  \multirow{2}*{ $\sqrt{s}$ (GeV)} & \multicolumn{4}{c}{Breit-Wigner} & \multicolumn{3}{c}{K-matrix}  \\
  \cmidrule(rl){2-5}\cmidrule(rl){6-8}
                 &  \makecell[c]{$\PZC$} &   \makecell[c]{$f_{0}(980)J/\psi$}   & \makecell[c]{ $f_{2}(1270)J/\psi$} &  \makecell[c]{$\PPS J/\psi$ }  & \makecell[c]{ $\PZC$ } & \makecell[c]{ $f_{2}(1270)J/\psi$ }  & \makecell[c]{ $\PPS J/\psi$ } \\
  \cmidrule(l){0-1}\cmidrule(rl){2-5}\cmidrule(rl){6-8}
  4.1271 & $12.4\pm4.7\pm2.4$  & $53.4\pm16.7\pm19.5$  & $2.7\pm3.0\pm2.0$  & $75.1\pm11.1\pm8.5$ & $9.9\pm3.6\pm2.2$   & $2.9\pm3.4\pm0.1$  & $73.8\pm7.0\pm1.3$   \\
  4.1567 & $17.9\pm4.7\pm2.5$  & $54.0\pm13.7\pm11.9$  & $12.2\pm4.5\pm2.8$ & $75.3\pm11.4\pm3.9$ & $15.5\pm5.4\pm2.1$  & $10.4\pm5.1\pm1.9$ & $76.5\pm13.9\pm1.2$  \\
  4.1780 & $36.6\pm3.3\pm8.2$  & $36.3\pm4.8\pm17.0$   & $9.0\pm2.1\pm11.4$ & $68.4\pm4.2\pm15.3$ & $27.4\pm4.1\pm9.0$  & $1.2\pm1.0\pm8.0$  & $82.7\pm11.4\pm14.3$ \\
  4.1888 & $30.2\pm6.9\pm5.1$  & $35.0\pm10.3\pm9.9$   & $4.1\pm3.7\pm3.7$  & $78.3\pm8.9\pm4.4$  & $30.5\pm5.9\pm0.5$  & $3.3\pm3.2\pm1.0$  & $84.9\pm4.6\pm6.6$   \\
  4.1989 & $22.6\pm3.6\pm3.6$  & $33.9\pm8.4\pm12.4$   & $0.1\pm0.9\pm1.4$  & $71.3\pm6.5\pm9.8$  & $24.6\pm4.6\pm2.2$  & $1.9\pm1.4\pm1.6$  & $75.9\pm7.8\pm4.6$   \\
  4.2091 & $13.6\pm2.1\pm2.8$  & $55.4\pm8.3\pm30.0$   & $0.3\pm0.6\pm0.7$  & $81.9\pm5.0\pm5.0$  & $11.9\pm2.2\pm1.5$  & $0.3\pm0.4\pm0.2$  & $80.0\pm5.6\pm1.9$   \\
  4.2187 & $16.2\pm2.1\pm2.6$  & $30.0\pm4.0\pm7.9$    & $1.4\pm1.2\pm1.2$  & $75.9\pm4.0\pm2.9$  & $12.6\pm1.8\pm3.4$  & $0.3\pm0.4\pm1.2$  & $79.6\pm7.4\pm3.7$   \\
  4.2263 & $13.0\pm1.1\pm2.2$  & $35.0\pm3.1\pm11.6$   & $2.1\pm0.9\pm0.9$  & $78.8\pm2.7\pm2.8$  & $10.6\pm1.1\pm2.2$  & $1.5\pm1.1\pm0.7$  & $82.5\pm4.7\pm3.6$   \\
  4.2357 & $12.5\pm1.6\pm2.4$  & $26.6\pm3.5\pm10.9$   & $1.3\pm1.4\pm0.6$  & $83.5\pm4.1\pm1.5$  & $10.0\pm1.4\pm2.3$  & $0.6\pm0.9\pm0.9$  & $85.0\pm7.4\pm1.5$   \\
  4.2438 & $7.5\pm1.2\pm2.4$   & $32.9\pm7.7\pm15.6$   & $0.8\pm1.1\pm0.6$  & $90.4\pm4.1\pm2.1$  & $5.3\pm1.1\pm2.0$   & $1.3\pm2.0\pm0.4$  & $87.6\pm2.8\pm2.8$   \\
  4.2580 & $8.6\pm1.2\pm1.7$   & $24.9\pm4.7\pm14.6$   & $6.6\pm2.6\pm3.7$  & $83.6\pm4.5\pm4.2$  & $9.0\pm1.2\pm0.7$   & $2.8\pm1.7\pm4.0$  & $82.8\pm3.1\pm0.8$   \\
  4.2667 & $7.3\pm1.5\pm1.8$   & $8.5\pm2.5\pm6.5$     & $0.1\pm0.2\pm0.4$  & $94.2\pm4.8\pm1.4$  & $8.6\pm1.3\pm1.5$   & $1.8\pm0.7\pm1.7$  & $89.3\pm8.5\pm4.9$   \\
  4.2776 & $13.9\pm4.2\pm2.2$  & $27.9\pm9.6\pm10.9$   & $12.1\pm6.9\pm5.9$ & $85.3\pm10.8\pm7.6$ & $14.1\pm6.1\pm0.4$  & $5.9\pm4.2\pm6.2$  & $85.7\pm45.2\pm0.5$  \\
  4.2866 & $9.9\pm1.8\pm2.6$   & $6.1\pm1.7\pm6.3$     & $2.9\pm1.7\pm1.6$  & $89.9\pm5.3\pm1.4$  & $7.8\pm1.9\pm1.9$   & $6.3\pm5.0\pm3.3$  & $85.1\pm14.7\pm4.8$  \\
  4.3115 & $8.6\pm1.9\pm2.6$   & $5.3\pm2.5\pm4.3$     & $9.3\pm4.8\pm5.8$  & $83.8\pm7.9\pm4.5$  & $10.8\pm2.2\pm2.4$  & $1.4\pm1.5\pm8.1$  & $87.2\pm9.2\pm3.4$   \\
  4.3370 & $12.3\pm2.7\pm3.0$  & $2.4\pm2.2\pm4.5$     & $8.6\pm5.4\pm7.9$  & $88.2\pm8.6\pm4.8$  & $12.9\pm2.6\pm0.8$  & $1.6\pm1.7\pm7.1$  & $92.2\pm6.2\pm4.1$   \\
  4.3583 & $9.4\pm2.3\pm1.8$   & $1.8\pm2.0\pm4.3$     & $1.6\pm1.4\pm0.9$  & $91.9\pm7.1\pm1.6$  & $15.1\pm3.4\pm6.0$  & $1.6\pm1.6\pm0.1$  & $95.5\pm9.8\pm3.6$   \\
\hline
\hline
\end{tabular}}
\label{totfraction}
\end{sidewaystable*}

\begin{sidewaystable*}
\caption{The cross sections (in pb) of the subprocesses in the PWA for different energy points by using the Breit-Wigner and K-matrix methods to parametrize the $\PPS$. The $\PPS J/\psi$ is the whole of $f_{0}(500)J/\psi$, $f_{0}(980)J/\psi$, and $f_{0}(1370)J/\psi$. The $\PZC$, $f_{0}(980)J/\psi$, $f_{2}(1270)J/\psi$ stand for $\EE\rightarrow\pi^{+}Z_{c}(3900)^{-}+c.c.\rightarrow\pi^{+}\pi^{-}J/\psi$, $(f_{0}(980)\rightarrow\pi^{+}\pi^{-})J/\psi$, and $(f_{2}(1270)\rightarrow\pi^{+}\pi^{-})J/\psi$ respectively. The $\mathcal{L}$ (pb$^{-1}$) is the integrated luminosity. The first uncertainty is statistical, and the second one systematic. The uncertainty of $\mathcal{L}$ is dominated by systematic uncertainty.}
\centering
\renewcommand\arraystretch{1.1}
\renewcommand\tabcolsep{5.0pt}
\scalebox{1.0}{
\begin{tabular}{lllllllll}
\toprule
\toprule
  \multirow{2}*{ $\sqrt{s}$ (GeV)} & \multirow{2}*{$\mathcal{L}$ (pb$^{-1}$)} & \multicolumn{4}{c}{Breit-Wigner} & \multicolumn{3}{c}{K-matrix}  \\
  \cmidrule(rl){3-6}\cmidrule(rl){7-9}
                 &          & \makecell[c]{$\PZC$} &   \makecell[c]{$f_{0}(980)J/\psi$}   & \makecell[c]{ $f_{2}(1270)J/\psi$} &  \makecell[c]{$\PPS J/\psi$ }  & \makecell[c]{ $\PZC$ } & \makecell[c]{ $f_{2}(1270)J/\psi$ }  & \makecell[c]{ $\PPS J/\psi$ } \\
  \cmidrule(l){0-2}\cmidrule(rl){3-6}\cmidrule(rl){7-9}
  4.1271& $390.0 \pm2.6$  &$1.5\pm0.6\pm0.3$   & $6.4\pm2.0\pm2.4$   & $0.3\pm0.4\pm0.2$  & $9.0\pm1.3\pm1.0$  & $1.2\pm0.4\pm0.3$  & $0.3\pm0.4\pm0.0$  & $8.9\pm0.8\pm0.2$  \\
  4.1567& $409.9 \pm2.7$  &$2.1\pm0.6\pm0.3$   & $6.4\pm1.6\pm1.5$   & $1.4\pm0.5\pm0.3$  & $8.9\pm1.3\pm0.5$  & $1.8\pm0.6\pm0.3$  & $1.2\pm0.6\pm0.2$  & $9.1\pm1.6\pm0.1$ \\
  4.1780& $3194.5\pm31.9$ &$4.8\pm0.4\pm1.1$   & $4.8\pm0.6\pm2.3$   & $1.2\pm0.3\pm1.5$  & $9.0\pm0.6\pm2.0$  & $3.6\pm0.5\pm1.2$  & $0.2\pm0.1\pm1.0$  & $10.9\pm1.5\pm1.9$ \\
  4.1888& $570.1 \pm2.2$  &$5.0\pm1.2\pm0.9$   & $5.8\pm1.7\pm1.7$   & $0.7\pm0.6\pm0.6$  & $13.0\pm1.5\pm0.7$ & $5.1\pm1.0\pm0.1$  & $0.5\pm0.5\pm0.2$  & $14.1\pm0.8\pm1.1$ \\
  4.1989& $524.6 \pm2.1$  &$6.5\pm1.0\pm1.1$   & $9.7\pm2.4\pm3.6$   & $0.0\pm0.2\pm0.4$  & $20.4\pm1.9\pm2.8$ & $7.0\pm1.3\pm0.6$  & $0.5\pm0.4\pm0.5$  & $21.7\pm2.2\pm1.3$ \\
  4.2091& $572.1 \pm1.8$  &$6.3\pm1.0\pm1.3$   & $25.4\pm3.8\pm13.8$ & $0.1\pm0.3\pm0.3$  & $37.6\pm2.3\pm2.3$ & $5.5\pm1.0\pm0.7$  & $0.1\pm0.2\pm0.1$  & $36.7\pm2.6\pm0.9$ \\
  4.2187& $569.2 \pm1.8$  &$11.5\pm1.5\pm1.9$  & $21.3\pm2.9\pm5.7$  & $1.0\pm0.9\pm0.9$  & $53.7\pm2.8\pm2.0$ & $8.9\pm1.3\pm2.4$  & $0.2\pm0.3\pm0.9$  & $56.4\pm5.2\pm2.6$ \\
  4.2263& $1100.9\pm7.0$  &$11.6\pm1.0\pm2.0$  & $31.3\pm2.8\pm10.4$ & $1.9\pm0.8\pm0.8$  & $70.4\pm2.4\pm2.5$ & $9.4\pm1.0\pm2.0$  & $1.4\pm1.0\pm0.6$  & $73.6\pm4.2\pm3.2$ \\
  4.2357& $530.6 \pm2.4$  &$10.1\pm1.3\pm1.9$  & $21.4\pm2.8\pm8.8$  & $1.0\pm1.1\pm0.5$  & $67.2\pm3.3\pm1.2$ & $8.1\pm1.1\pm1.9$  & $0.5\pm0.7\pm0.7$  & $68.4\pm6.0\pm1.2$ \\
  4.2438& $594.0 \pm2.7$  &$5.6\pm0.9\pm1.8$   & $24.6\pm5.7\pm11.6$ & $0.6\pm0.8\pm0.5$  & $67.5\pm3.1\pm1.5$ & $4.0\pm0.8\pm1.5$  & $0.9\pm1.5\pm0.3$  & $65.4\pm2.1\pm2.1$ \\
  4.2580& $828.4 \pm5.5$  &$5.2\pm0.7\pm1.0$   & $15.1\pm2.9\pm8.9$  & $4.0\pm1.6\pm2.2$  & $50.6\pm2.8\pm2.6$ & $5.5\pm0.7\pm0.4$  & $1.7\pm1.0\pm2.4$  & $50.1\pm1.9\pm0.5$ \\
  4.2667& $529.7 \pm3.1$  &$4.1\pm0.8\pm1.0$   & $4.7\pm1.4\pm3.7$   & $0.1\pm0.1\pm0.2$  & $52.5\pm2.7\pm0.8$ & $4.8\pm0.7\pm0.8$  & $1.0\pm0.4\pm0.9$  & $49.8\pm4.8\pm2.8$ \\
  4.2776& $175.7 \pm1.0$  &$7.4\pm2.2\pm1.3$   & $14.9\pm5.1\pm5.9$  & $6.4\pm3.7\pm3.2$  & $45.5\pm5.8\pm4.1$ & $7.5\pm3.2\pm0.2$  & $3.2\pm2.2\pm3.3$  & $45.8\pm24.1\pm0.2$\\
  4.2866& $498.5 \pm3.3$  &$5.3\pm1.0\pm1.4$   & $3.3\pm0.9\pm3.4$   & $1.5\pm0.9\pm0.9$  & $48.0\pm2.8\pm0.8$ & $4.1\pm1.0\pm1.0$  & $3.4\pm2.7\pm1.7$  & $45.5\pm7.8\pm2.5$ \\
  4.3115& $499.2 \pm3.3$  &$3.3\pm0.7\pm1.0$   & $2.0\pm1.0\pm1.7$   & $3.6\pm1.9\pm2.2$  & $32.3\pm3.0\pm1.7$ & $4.2\pm0.9\pm0.9$  & $0.5\pm0.6\pm3.1$  & $33.6\pm3.5\pm1.3$ \\
  4.3370& $511.5 \pm3.4$  &$4.3\pm0.9\pm1.1$   & $0.9\pm0.8\pm1.6$   & $3.0\pm1.9\pm2.8$  & $30.8\pm3.0\pm1.7$ & $4.5\pm0.9\pm0.3$  & $0.6\pm0.6\pm2.5$  & $32.3\pm2.2\pm1.4$ \\
  4.3583& $543.9 \pm3.6$  &$2.4\pm0.6\pm0.5$   & $0.5\pm0.5\pm1.1$   & $0.4\pm0.4\pm0.2$  & $23.6\pm1.8\pm0.4$ & $3.9\pm0.9\pm1.5$  & $0.4\pm0.4\pm0.0$  & $24.6\pm2.5\pm0.9$ \\
\hline
\hline
\end{tabular}}
\label{totCrosssection}
\end{sidewaystable*}

\begin{table*}
\caption{The interference magnitude between each subprocess at $\sqrt{s} = 4.2263$ and 4.2580 GeV. The $Z_{c}(3900)^{\pm}$, $f_{0}(500)$, $f_{0}(980)$, $f_{0}(1370)$ and $f_{2}(1270)$ represent $\PZC$, $f_{0}(500)J/\psi$ $f_{0}(980)J/\psi$, $f_{0}(1370)J/\psi$ and $f_{2}(1270)J/\psi$ subprocesses, respectively. The corrections in Table~\ref{MWCorrection} are not considered here.(In \%)}
\centering
\renewcommand\arraystretch{1.3}
\renewcommand\tabcolsep{4.0pt}
\begin{tabular}{llllll}
\hline
\hline
  $4.2263$ GeV  & $Z_{c}(3900)^{\pm}$   & $f_{0}(500)$    & $f_{0}(980)$     & $f_{0}(1370)$    &  $f_{2}(1270)$  \\
\hline
  $Z_{c}(3900)^{\pm}$  &   $12.8\pm1.1$  & $-6.9\pm3.8$  & $-6.5\pm4.4$   & $18.3\pm8.2$   & $1.1\pm2.2$   \\
  $f_{0}(500)$   &                   & $24.1\pm2.5$      & $-6.4\pm5.8$   & $-62.1\pm6.8$  & $0.1\pm3.7$   \\
  $f_{0}(980)$   &                   &                   & $35.3\pm3.1$   & $-14.2\pm7.8$  & $0.1\pm4.8$   \\
  $f_{0}(1370)$  &                   &                   &                & $102.1\pm5.4$  & $0.1\pm7.6$   \\
  $f_{2}(1270)$  &                   &                   &                &                & $2.3\pm0.9$   \\
\hline
\hline
  $4.2580$ GeV  &  $Z_{c}(3900)^{\pm}$   & $f_{0}(500)$    & $f_{0}(980)$    & $f_{0}(1370)$    &  $f_{2}(1270)$  \\
\hline
  $Z_{c}(3900)^{\pm}$  &  $8.3\pm1.2$   & $-7.0\pm5.1$ & $-4.6\pm6.6$   & $11.6\pm12.8$  &  $1.4\pm4.4$  \\
  $f_{0}(500)$   &                   & $29.9\pm3.5$    & $-7.6\pm9.1$   & $-67.9\pm11.0$ &  $0.1\pm6.0$  \\
  $f_{0}(980)$   &                   &                 & $25.2\pm4.8$   & $-1.65\pm12.5$ &  $0.1\pm8.2$  \\
  $f_{0}(1370)$  &                   &                 &                & $105.8\pm8.8$  &  $0.1\pm12.2$ \\
  $f_{2}(1270)$  &                   &                 &                &                &  $6.8\pm2.6$  \\
\hline
\hline
\end{tabular}
\label{InterferenceTerm4230}
\end{table*}

\begin{table*}
\caption{The $\chi^{2}/ndf$ for $m(\pi^{\pm}J/\psi)$, $m(\pi^{+}\pi^{-})$, and the Dalitz distribution of $m^{2}(\pi^{\pm}J/\psi)$ vs $m^{2}(\pi^{+}\pi^{-})$ in the fit. }
\centering
\renewcommand\arraystretch{1.2}
\renewcommand\tabcolsep{8.0pt}
\begin{tabular}{l llc}
\hline
\hline
  $\sqrt{s}$ (GeV) & $m(\pi^{\pm}J/\psi)$ & $m(\pi^{+}\pi^{-})$ & $m^{2}(\pi^{\pm}J/\psi)$ vs $m^{2}(\pi^{+}\pi^{-})$\\
\hline
  4.1271    &     1.05     &   1.05    &  1.16 \\
  4.1567    &     1.29     &   0.82    &  0.82 \\
  4.1780    &     1.48     &   1.24    &  1.71 \\
  4.1888    &     0.86     &   1.05    &  1.09 \\
  4.1989    &     1.47     &   1.47    &  1.42 \\
  4.2091    &     1.05     &   1.20    &  1.46 \\
  4.2187    &     1.48     &   1.43    &  1.21 \\
  4.2263    &     1.72     &   1.77    &  1.47 \\
  4.2357    &     0.87     &   0.69    &  1.03 \\
  4.2438    &     1.71     &   1.25    &  1.23 \\
  4.2580    &     1.05     &   1.25    &  1.20 \\
  4.2667    &     1.60     &   1.47    &  1.15 \\
  4.2776    &     0.77     &   1.03    &  1.02 \\
  4.2866    &     1.75     &   1.56    &  1.18 \\
  4.3115    &     1.13     &   0.80    &  1.39 \\
  4.3368    &     1.49     &   0.98    &  1.75 \\
  4.3583    &     1.21     &   0.96    &  1.38 \\
\hline
\hline
\end{tabular}
\label{chisqvsndf1}
\end{table*}

\begin{thebibliography}{90}

\bibitem{BESIII:2013ris}
M.~Ablikim \textit{et al.} (BESIII Collaboration),
\href{https://journals.aps.org/prl/abstract/10.1103/PhysRevLett.111.019901}
{Phys. Rev. Lett. \textbf{110}, 252001  (2013)}.

\bibitem{Belle:2013yex}
Z.~Q.~Liu \textit{et al.} (Belle Collaboration),
\href{https://journals.aps.org/prl/abstract/10.1103/PhysRevLett.110.252002}
{Phys. Rev. Lett. \textbf{110}, 252002 (2013)}.

\bibitem{Xiao:2013iha}
T.~Xiao \textit{et al.},
\href{https://www.sciencedirect.com/science/article/abs/pii/S0370269313008484}
{Phys. Lett. B \textbf{727}, 366 (2013)}.


\bibitem{BESIII:2017bua}M.~Ablikim \textit{et al.} (BESIII Collaboration),
\href{https://journals.aps.org/prl/abstract/10.1103/PhysRevLett.119.072001}
{Phys. Rev. Lett. \textbf{119}, 072001 (2017)}.

\bibitem{Pilloni:2016obd}
A.~Pilloni \textit{et al.} (JPAC Collaboration),
\href{https://inspirehep.net/literature/1505197}
{Phys. Lett. B \textbf{772} (2017)}.

\bibitem{Danilkin:2020kce}
I.~Danilkin, D.~A.~S.~Molnar, and M.~Vanderhaeghen,
\href{https://inspirehep.net/literature/1793248}
{Phys. Rev. D \textbf{102} (2020)}.


\bibitem{Wang:2013cya}Q.~Wang, C.~Hanhart, and Q.~Zhao,
\href{https://journals.aps.org/prl/abstract/10.1103/PhysRevLett.111.132003}
{Phys. Rev. Lett. \textbf{111}, 132003 (2013)}.

\bibitem{Li:2013xia}G.~Li,
\href{https://link.springer.com/article/10.1140/epjc/s10052-013-2621-5}
{Eur. Phys. J. C \textbf{73}, 2621 (2013)}.

\bibitem{Cui:2013yva}C.~Y.~Cui {\it et al.},
\href{https://iopscience.iop.org/article/10.1088/0954-3899/41/7/075003/meta}
{J. Phys. G \textbf{41}, 075003 (2014)}.

\bibitem{Voloshin:2013dpa}M.~B.~Voloshin,
\href{https://journals.aps.org/prd/abstract/10.1103/PhysRevD.87.091501}
{Phys. Rev. D \textbf{87}, 091501 (2013)}.


\bibitem{Braaten:2013boa}E.~Braaten,
\href{https://inspirehep.net/literature/1236093}
{Phys. Rev. Lett. \textbf{111}, 162003 (2013)}.


\bibitem{Wilbring:2013cha}E.~Wilbring {\it et al.},
\href{https://www.sciencedirect.com/science/article/abs/pii/S0370269313006989}
{Phys. Lett. B \textbf{726}, 326 (2013)}.

\bibitem{Guo:2017jvc}
F.~K.~Guo, C.~Hanhart, U.~G.~Mei\ss{}ner, Q.~Wang, Q.~Zhao, and B.~S.~Zou,
\href{https://journals.aps.org/rmp/abstract/10.1103/RevModPhys.94.029901}
{Rev. Mod. Phys. \textbf{94}, 029901 (2022)}.

\bibitem{Wang:2013vex}
Z.~G.~Wang and T.~Huang,
\href{https://inspirehep.net/literature/1257741}
{Phys. Rev. D \textbf{89}, 054019 (2014)}


\bibitem{Cleven:2013mka}M.~Cleven {\it et al.},
\href{https://inspirehep.net/literature/1257632}
{Phys. Rev. D \textbf{90}, 074039 (2014)}

\bibitem{Giron:2020fvd} J.~F.~Giron and R.~F.~Lebed,
\href{https://inspirehep.net/literature/1784109}
{Phys. Rev. D \textbf{101}, 074032 (2020)}.


\bibitem{Ablikim:2020pzw} M.~Ablikim (BESIII Collaboration),
\href{https://inspirehep.net/literature/1793431}
{Phys. Rev. D \textbf{102}, 012009 (2020)}.

\bibitem{MartinezTorres:2009xb} A.~Martinez Torres {\it et al.},
\href{https://journals.aps.org/prd/abstract/10.1103/PhysRevD.80.094012}
{Phys. Rev. D \textbf{80}, 094012 (2009)}.


\bibitem{Coito:2019cts} S.~Coito and F.~Giacosa,
\href{https://inspirehep.net/literature/1721820}
{Acta Phys. Pol. B \textbf{51}, 1713 (2020)}.

\bibitem{Gong:2016jzb} Q.~R.~Gong, J.~L.~Pang, Y.~F.~Wang, and H.~Q.~Zheng,
\href{https://inspirehep.net/literature/1506405}
{Eur. Phys. J. C \textbf{78}, 276 (2018)}.

\bibitem{Wigner:1946zz}
E.~P.~Wigner,
\href{https://inspirehep.net/literature/45426}
{Phys. Rev. \textbf{70}, 15 (1946)}.

\bibitem{Aitchison:1972ay}
I.~J.~R.~Aitchison,
\href{https://inspirehep.net/literature/76474}
{Nucl. Phys. A \textbf{189}, 417 (1972)}.

\bibitem{Anisovich:2002ij}V.~V.~Anisovich and A.~V.~Sarantsev,
\href{https://inspirehep.net/literature/586064}
{Eur. Phys. J. A \textbf{16}, 229 (2003)}.

\bibitem{Peters:2004qw} K.~J.~Peters,
\href{https://inspirehep.net/literature/666255}
{Int. J. Mod. Phys. A \textbf{21}, 5618 (2006)}.


\bibitem{BESIII:2022qal} M.~Ablikim \textit{et al.} (BESIII Collaboration),
\href{https://journals.aps.org/prd/abstract/10.1103/PhysRevD.106.072001}
{Phys. Rev. D \textbf{106}, 072001 (2022)}.


\bibitem{bes3} M.~Ablikim {\it et al.} (BESIII Collaboration),
\href{https://www.sciencedirect.com/science/article/abs/pii/S0168900209023870}
{Nucl. Instrum. Methods Phys. Res., Sect. A \textbf{614}, 345 (2010)}.

\bibitem{bepcii} C.~H.~Yu {\it et al.}, \textit{Proceedings of IPAC2016, Busan, Korea (2016)}; Report No.
\href{https://accelconf.web.cern.ch/ipac2018/papers/thpaf085.pdf}
{JACoW~IPAC2016-TUYA01}.


\bibitem{etof}
 X.~Li {\it et al.},
 \href{https://link.springer.com/article/10.1007/s41605-017-0014-2}
 {Radiat. Detect. Technol. Methods} {\bf 1}, 13 (2017);
 Y.~X.~Guo {\it et al.},
 \href{https://link.springer.com/article/10.1007/s41605-017-0012-4}
 {Radiat. Detect. Technol. Methods} {\bf 1}, 15 (2017);
 P.~Cao {\it et al.},
 \href{https://www.sciencedirect.com/science/article/abs/pii/S0168900219314068}
{ Nucl.\ Instrum.\ Methods Phys. Res., Sect. A {\bf 953}, 163053 (2020)}.


\bibitem{BESIII:2022xii}
M.~Ablikim \textit{et al.} (BESIII Collaboration),
\href{https://arxiv.org/abs/2203.03133}
{Chin. Phys. C \textbf{46} 113002(2022)}.

\bibitem{BESIII:2020eyu}
M.~Ablikim {\it et al.} (BESIII Collaboration),
\href{https://iopscience.iop.org/article/10.1088/1674-1137/ac1575/meta}
{Chin. Phys. C \textbf{45} 103001 (2021)}.

\bibitem{GEANT4} S.~Agostinelli {\it et al.} (GEANT4 Collaboration),
\href{https://d1wqtxts1xzle7.cloudfront.net/43621608/GEANT4_A_Simulation_toolkit20160311-26489-1lx8o16-with-cover-page-v2.pdf?Expires=1639225873&Signature=Ebsse6mbaHTWHECEvkzqKcakMk2xMhx0hMYfskR-xTrK0PpRplBg2FPL-T0H~RWSFt3Np0jTc5kM5aQKuXjTf-1DTMgU2HyuRgkA443wlSpaKP3bPpWftD4-q~FVMCim9v9BBw20XhANFQiaElwAXI54~R4LspqUVztWzXL~FJWS4MotG4gGQFWIonDTX6-5OX875cPf2ZCE0UtBkbm~UEXQZ6FRULyo94WfivGDhDHpBXvDzvaMx6aFD7XyLIdCtA1xMjz0~UI-pKWBwXDrXmfrI-Ks1YL1uDq3QDn~SyVqsNUvVph~mLvUtPg~uQ~--DRstdOZvL55toEpJPit9A__&Key-Pair-Id=APKAJLOHF5GGSLRBV4ZA}
{Nucl.\ Instrum.\ Methods Phys. Res., Sect. A \textbf{506}, 250 (2003)}.

\bibitem{boost} Z. Y. Deng {\it et al.},
\href{http://hepnp.ihep.ac.cn/article/id/283d17c0-e8fa-4ad7-bfe3-92095466def1}
{Chin. Phys. C \textbf{30}, 371 (2006)}.

\bibitem{EvtGen} D.~J.~Lange,
\href{https://www.sciencedirect.com/science/article/abs/pii/S0168900201000894}
{Nucl.~Instrum. Methods Phys. Res., Sect. A \textbf{462}, 152 (2001)}; R.~G.~Ping,
\href{http://hepnp.ihep.ac.cn/article/doi/10.1088/1674-1137/32/8/001}
{Chin. Phys. C \textbf{32}, 599 (2008)}.

\bibitem{Jadach:2000ir}
S.~Jadach, B.~F.~L.~Ward and Z.~Was,
\href{https://journals.aps.org/prd/abstract/10.1103/PhysRevD.63.113009}
{Phys. Rev. D \textbf{63}, 113009 (2001)}.

\bibitem{photos} E.~Richter-Was,
\href{https://www.sciencedirect.com/science/article/abs/pii/037026939390062M}
{Phys. Lett. B \textbf{303}, 163 (1993)}.

\bibitem{Berends:1986if}
F.~A.~Berends, P.~H.~Daverveldt and R.~Kleiss,
\href{https://doi.org/10.1016/0010-4655(86)90114-1}
{Comput. Phys. Commun. \textbf{40}, 271 (1986).}

\bibitem{Berends:1986ig}
F.~A.~Berends, P.~H.~Daverveldt and R.~Kleiss,
\href{https://doi.org/10.1016/0010-4655(86)90115-3}
{Comput. Phys. Commun. \textbf{40}, 285 (1986).}

\bibitem{PDG}
S.~Navas \textit{et al.} [Particle Data Group],
\href{https://journals.aps.org/prd/abstract/10.1103/PhysRevD.110.030001}
{Phys. Rev. D \textbf{110}, 030001 (2024)}.


\bibitem{Chung:1997jn}S.~U.~Chung,
\href{https://inspirehep.net/literature/448883}
{Phys. Rev. D \textbf{57}, 431 (1998)}; \href{https://inspirehep.net/literature/362925}{\textbf{48} (1993)}; \href{https://inspirehep.net/literature/362925}{48, 1225 (1993)}.
\bibitem{Chung:2007nn}S.~U.~Chung and J.~Friedrich,
\href{https://inspirehep.net/literature/768176}
{Phys. Rev. D \textbf{78}, 074027 (2008)}.

\bibitem{LHCb:2015yax}R.~Aaij \textit{et al.} (LHCb Collaboration),
\href{https://inspirehep.net/literature/1382595}
{Phys. Rev. Lett. \textbf{115}, 072001 (2015)}.

\bibitem{pwa}M.~Shepherd, J.~Stevens, R.~Mitchell, M.~Albrecht, B.~Grube, A.~Austregesilo, N.~D.~Hoffman, and N.~H\"usken,
\href{https://zenodo.org/records/10961168}
{10.5281/zenodo.10961168}.

\bibitem{BES:2004twe}M.~Ablikim \textit{et al.} (BES Collaboration),
\href{https://inspirehep.net/literature/663288}
{Phys. Lett. B \textbf{607}, 243 (2005)}.



\bibitem{BaBar:2008inr}B.~Aubert \textit{et al.} (\textit{BABAR} Collaboration),
\href{https://inspirehep.net/literature/783326}
{Phys. Rev. D \textbf{78}, 034023 (2008)}.

\bibitem{LHCb:2024yzj}R.~Aaij \textit{et al.} (LHCb Collaboration),
\href{https://inspirehep.net/literature/2851668}
{Phys. Rev. Lett. \textbf{134}, 101802 (2025)}.


\bibitem{BESIII:2013qmu}M.~Ablikim \textit{et al.} (BESIII Collaboration),
\href{https://inspirehep.net/literature/1256939}
{Phys. Rev. Lett. \textbf{112}, 022001 (2014)}.

\bibitem{DataAvailability} M.~Ablikim \textit{et al.} (BESIII Collaboration),
\href{https://www.hepdata.net/record/ins2922807}
{Inspire Record 2922807 DOI 10.17182/hepdata.166173}.

\end{thebibliography}
\end{document}